\documentclass[aps,rmp,longbibliography,unsortedaddress,twocolumn,nofootinbib]{revtex4-2}
\usepackage{amsmath,amssymb,amsfonts}
\usepackage{graphicx}
\usepackage{hyperref}
\usepackage{bm}
\usepackage{xcolor}
\usepackage{braket}

\setcitestyle{numbers}

\newcommand{\dd}{\mathrm{d}}
\newcommand{\ii}{\mathrm{i}}
\newcommand{\GF}{G_\mathrm{F}}
\newcommand{\thetaW}{\theta_W}
\newcommand{\QW}{Q_W}

\newcommand{\sigmav}{\bm{\sigma}}
\newcommand{\muN}{\mu_N}
\newcommand{\alphav}{\bm{\alpha}}
\newcommand{\Mn}{\mathcal{M}}
\newcommand{\Fn}{\mathcal{F}}

\begin{document}

\title{Probing the electroweak structure of nuclei with rare atoms and molecules}

\author{Silviu-Marian Udrescu}
\email{sudrescu@princeton.edu}
\affiliation{\mbox{Department of Physics and Astronomy, Johns Hopkins University, Baltimore, MD 21210, USA}}
\affiliation{\mbox{Department of Physics, Princeton University, Princeton, NJ 08544, USA}}

\author{Antoine Belley}
\email{abelley@mit.edu}
\affiliation{\mbox{Massachusetts Institute of Technology, Cambridge, Massachusetts 02139, USA}}

\author{Jason D. Holt}
\email{jholt@triumf.ca}
\affiliation{\mbox{TRIUMF 4004 Wesbrook Mall, Vancouver BC V6T 2A3, Canada}}
\affiliation{\mbox{Department of Physics, McGill University, Montr\'eal, QC H3A 2T8, Canada}}

\author{Gilad Perez}
\email{gilad.perez@weizmann.ac.il}
\affiliation{\mbox{Department of Particle Physics and Astrophysics, Weizmann Institute of Science, Rehovot 761001, Israel}}

\author{Ronald F. Garcia Ruiz} 
\email{rgarciar@mit.edu}
\affiliation{\mbox{Massachusetts Institute of Technology, Cambridge, Massachusetts 02139, USA}}

\begin{abstract}
Precision experiments of atoms and molecules have become a powerful probe of the electroweak structure of atomic nuclei and of physics beyond the Standard Model. We review how the interaction between a nucleus and its surrounding bound electrons can be exploited to precisely measure the electromagnetic, parity-violating, and CP-violating properties of nuclei and their fundamental constituents. We focus on rare, unstable isotopes, surveying the experimental techniques and facilities developed in recent years that have extended these measurements to the most exotic regions of the nuclear chart. Recent advances in the precision control and interrogation of single molecules, together with direct laser excitation of nuclear transitions, are opening new frontiers in nuclear and particle physics. At the same time, progress in nuclear theory, machine learning, and high-performance computing is strengthening the connection between our microscopic description of nature and laboratory observables. In many cases, the precision with which nuclear and particle physics properties can be extracted is now limited not by experiment, but by the molecular, atomic, or nuclear theory required to interpret the measurements. This challenge presents a major opportunity for combined theoretical and experimental advances that will enable future discoveries.
\end{abstract}

\date{\today}

\maketitle

\setcounter{tocdepth}{5}

\raggedbottom
\tableofcontents

\section{Introduction}
\label{sec:intro}
 
In recent years, experiments with atoms and molecules have reached a precision at which the tiny energy shifts produced by different fundamental forces, each far smaller than the electronic binding energy, can be resolved and disentangled from one another \cite{safronova2018,yang2023laser}. These capabilities, developed through decades of advances in laser spectroscopy, ion and atom trapping, and quantum-state control, have turned atoms and molecules into sensitive probes of physics at scales far removed from the atomic one: the internal structure of the nucleus ($\sim$~fm), the dynamics of the strong force ($\sim$~GeV), and the possible existence of new particles and interactions at the TeV scale and beyond \cite{safronova2018,yang2023laser,demille2024quantum,Jad25}.
 
The reason is that every atomic and molecular energy level depends, through the
electron--nucleus interaction, on a hierarchy of nuclear and particle-physics
properties, each producing a characteristic fractional shift of the electronic
energy scale, typically of order $\sim 1~\mathrm{eV}$. The finite spatial extent
of the nuclear charge distribution gives rise to field shifts ranging from the
kHz to GHz scale, corresponding to fractional shifts of roughly
$10^{-6}$--$10^{-10}$ of the electronic energy. Nuclear magnetic dipole and
electric quadrupole moments generate hyperfine splittings of comparable
magnitude. At still smaller scales, the parity-violating weak interaction
between electrons and neutrons mixes states of opposite parity, producing
effects at or below the few-parts-in-$10^{13}$ level. CP-violating interactions
are expected to produce fractional energy shifts more than three orders of
magnitude smaller than those induced by parity-violating interactions. Each effect carries information about a distinct aspect of nuclear and particle physics \cite{udrescu2021isotope,yang2023laser}, the distribution of charge and magnetization \cite{wilkins2023observation,Bissell2026HyperfineAnomaly}, the weak-charge distribution dominated by neutrons \cite{lackenby2018weak}, or CP violating moments such as the nuclear Schiff moment \cite{chupp2019electric}, offering probes complementary to those at large-scale colliders \cite{pospelov2005electric,feng2013naturalness,Jad25}.
 
The experimental progress has been accompanied by advances in nuclear theory of comparable scope. Nuclear models such as the nuclear shell-model and nuclear density functional theory have already been providing theoretical results for those quantities for years~\cite{caurier2005theshellmodel,bender2003self,schunck2019energy}. \textit{Ab initio} methods based on chiral effective field theory now allow the properties of nuclei in the medium mass range (charge radii, electromagnetic moments, rates of beta decay, and electroweak matrix elements needed to interpret experiments) to be computed with systematically improvable approximations and quantifiable uncertainties \cite{machleidt2011chiral,hu2022ab,hammer2015three}.  A consistent derivation of nuclear forces and electroweak currents within the same chiral EFT framework enables controlled error estimation and reduces the reliance on phenomenological adjustments.  These developments have advanced our ability to connect the Standard Model Lagrangian to measured atomic and molecular transition frequencies, with controlled theoretical precision at every step.
The precision frontier in low-energy physics is no longer limited by experimental techniques alone.  Increasingly, the limiting factor is our ability to compute the relevant nuclear matrix elements with controlled uncertainties.  Developing nuclear theories capable of matching the precision of modern experimental measurements remains a central challenge and opportunity for the field.
 
The past decade has seen these developments come together.  On the experimental side, laser spectroscopy at radioactive beam facilities such as ISOLDE-CERN, FRIB (US), ANL (US),  TRIUMF (Canada), GSI (Germany), GANIL (France), IGISOL (Finland),  RIKEN (Japan), and
emerging rare-isotope facilities in Asia such as HIRFL/HIAF (China) and RAON (Korea), has extended measurements of nuclear charge radii and electromagnetic
moments to hundreds of short-lived isotopes far from stability, revealing
structural phenomena that challenge existing nuclear models
\cite{yang2023laser,Brinson2026AlChargeRadii,Gus25}. Molecular experiments searching for the electron electric dipole moment have reached sensitivities three orders of magnitude beyond what was achieved with atoms \cite{acme2018improved,roussy2023improved}, probing CP violating physics at energy scales exceeding tens of TeV.  Recent spectroscopic measurements of radioactive molecules containing short-lived, deformed nuclei have opened a new experimental direction \cite{garcia2020spectroscopy,udrescu2024precision,arrowsmith2023opportunities,AthanasakisKaklamanakis2025AcF,Jad25,conn2025production}.  And the recent direct laser excitation of the $^{229}$Th nuclear isomer \cite{tiedau2024,zhang2024nature} has established nuclear clock technology as a viable platform for searches for ultralight dark matter and new forces coupling to the nuclear sector.
 
This review presents a consistent framework for connecting atomic and molecular
observables to the electroweak structure of nuclei, while highlighting the
advances in nuclear theory needed to interpret these measurements. We emphasize
the common theoretical structure by using a multipole expansion of the
nucleus--electron interaction that organizes observables, from charge radii to
symmetry-violating nuclear moments, into a single factorized form, and by highlighting the
complementary roles of atoms and molecules as experimental platforms.

The review is organized as follows. Section~\ref{sec:ch2}
explains the interaction Hamiltonian between the nucleus and the
electrons, showing how electromagnetic, weak, and CP violating nuclear
properties arise as distinct terms in a unified multipole expansion,
each factorizing into a nuclear operator multiplied by an electronic
field operator. Section~\ref{sec:nuclear_theory} surveys the current
status of nuclear theory, from phenomenological approaches to
\textit{ab initio} methods based on chiral effective field theory.
Section~\ref{sec:EM_properties} addresses the electromagnetic structure
of the nucleus as probed by atoms and molecules, including nuclear
charge radii and electromagnetic moments, and discusses how these
symmetry-conserving observables serve both as probes of nuclear many-body methods and as benchmarks to guide our understanding of the nuclear force. These observables are critical to test the reliability of theoretical models needed to interpret
symmetry-violation experiments. Section~\ref{sec:symm_violation} is
devoted to parity-violating properties, including the nuclear weak
charge, the anapole moment, and the weak quadrupole moment, as well as
to CP violating properties such as Schiff and magnetic quadrupole
moments, the electron electric dipole moment, and the atomic and
molecular experiments that target them. Section~\ref{sec:Th229}
highlights the $^{229}$Th nuclear isomer as a unique case study in
which nuclear clock technology enables searches for ultralight dark
matter, temporal variations of fundamental constants, and new forces, while Sec. \ref{sec:beta_decay} discusses tests of CKM matrix unitarity using precision $\beta$-decay measurements. Finally, Section~\ref{sec:conclusion} summarizes the current status of the field and discusses promising future directions.

\section{Connecting the properties of nuclei with the structure of atoms
and molecules}
\label{sec:ch2}

The energy levels of an atom or a molecule can be sensitive to the electromagnetic, weak, and CP violating properties of the nucleus. The interaction Hamiltonian between the nucleus and the surrounding electrons can be expressed as a multipole expansion, with each term expressed as the product between a \emph{nuclear quantity} (a multipole moment of the charge, current, or weak charge distribution) and an \emph{electronic quantity} with the value of the corresponding electromagnetic or weak field (or its spatial derivatives) produced by the electrons at the nuclear site. This factorization is what makes atoms and molecules precision tools. The electronic factors can be computed using atomic and molecular theory, or extracted experimentally, while the nuclear moments are extracted from the measured energy splittings or transition amplitudes.

We introduce below the formalism for the general case of a diatomic molecule, which contains all the relevant physics. The molecular case naturally includes the atomic one as a limit, but also introduces additional degrees of freedom (vibration, rotation, opposite-parity doublets) that qualitatively enhance the sensitivity to complementary nuclear properties. The extension to polyatomic molecules can be done by accounting for the extra degrees of freedom (e.g., new vibrational and rotational modes).

\subsection{The Hamiltonian of a diatomic molecule}
\label{sec:full_H}

Consider a diatomic molecule composed of two nuclei, labeled 1 and 2, with $Z_1$ and $Z_2$ protons and $N_1$ and $N_2$ neutrons, surrounded by $Z_1 + Z_2$ electrons. The Hamiltonian describing the energy level of this system can be written as:

\begin{equation}
    H = H_{el} + H_N + H_{eN}^{(\text{EM})} + H_{eN}^{(\text{W})} + H_{eN}^{(\text{CPV})}\,,
    \label{eq:H_total}
\end{equation}
where $H_{el}$ includes the kinetic energy of the electrons, their Coulomb interaction with each other and with the two (point-like) nuclei, as well as the Coulomb interaction between the two nuclei. This latter term is added for convenience into $H_{el}$, as it represents just an energy shift for a given internuclear distance, $R$ \cite{brown2003rotational}. The next term is given by:
\begin{equation}
    H_N = -\frac{\hbar^2}{2\mu R^2}\frac{\partial}{\partial R}\left(R^2\frac{\partial}{\partial R}\right) + \frac{\hbar^2}{2\mu R^2}(\mathbf{J}-\mathbf{L}-\mathbf{S})^2,
\end{equation}
and it contains the vibrational and rotational energy of the molecular frame containing the two nuclei. Here $\mu$ is the reduced molecular mass, $\mathbf{S}$ and $\mathbf{L}$ are the electron spin and orbital angular momentum, while $\mathbf{J}$ is the total angular momentum of the molecule (we are ignoring the nuclear spin for now). 

In order to solve the Schr\"odinger equation for the $H_{el} + H_N$ Hamiltonian, a perturbative approach is used in practice. The first step commonly taken is the Born--Oppenheimer (BO) approximation \cite{born2000quantum}, which amounts to ignoring all couplings between different electronic levels due to the nuclear motion. This implies that the electrons are able to follow the nuclear motion adiabatically and thus the electronic eigenstates of the system do not change in time. We can therefore write the molecular wavefunctions as a product between an electronic, a vibrational, and a rotational wavefunction: $\psi_{mol}~=~\psi_{el}\times \psi_{vib}\times \psi_{rot}$. The presence of electronic and nuclear spin, finite nuclear size corrections, weak interaction, and CP violating physics ($H_{eN}^{(\text{EM})} + H_{eN}^{(\text{W})} + H_{eN}^{(\text{CPV})}$) can be added as a perturbation on top of this picture in a systematic way using an effective Hamiltonian approach \cite{brown2003rotational}. 

\subsubsection{The electronic Hamiltonian.}
This produces the dominant energy contribution to the molecular system, obtained by solving the Schr\"odinger equation (SE):
\begin{equation}
H_\text{el}(R) \psi_{el} = E_{el}(R)\psi_{el}.
\label{eq:Hel}
\end{equation}
This equation is solved for each internuclear distance, providing a potential energy curve $E_{el}(R)$. The characteristic energy scale is $E_\text{el} \sim 1$--$10$~eV.

\subsubsection{The vibrational Hamiltonian.}
Under the BO approximation, one can assume that the nuclei are moving inside the potential created by the electrons, $E_{el}(R)$, within each electronic manifold, $\psi_{el}$. In this case, the relevant Hamiltonian is given by:
\begin{equation}
H_\text{vib} = -\frac{\hbar^2}{2\mu R^2}\frac{\partial}{\partial R}\left(R^2\frac{\partial}{\partial R}\right) + E_\text{el}(R).
\label{eq:Hvib}
\end{equation}
Near the equilibrium distance $R_e$, the potential $E_\text{el}(R)$ is approximately harmonic, and the vibrational eigenstates are labeled by the quantum number $v = 0, 1, 2, \ldots$, with energies:
\begin{equation}
E_v = \omega_e(v + \tfrac{1}{2}) - \omega_e x_e (v + \tfrac{1}{2})^2 + \ldots\,,
\label{eq:Ev}
\end{equation}
where $\omega_e$ is the harmonic vibrational frequency and $\omega_e x_e$ is the leading anharmonicity correction.  The vibrational frequency scales as $\omega_e \propto \mu^{-1/2}$, giving a characteristic energy $E_\text{vib} \sim 10^{-2}$--$10^{-1}$~eV.  The dependence of $\omega_e$ on the nuclear mass $M$ and, through $E_\text{el}(R)$, on the nuclear charge radius (which modifies $V_{eN}$ inside the nuclear volume) represents the origin of the vibrational contributions to the molecular isotope shift \cite{udrescu2021isotope}.

\subsubsection{The rotational Hamiltonian.}
The end-over-end rotation of the molecule is described by the Hamiltonian:
\begin{equation}
H_{rot} = \frac{\hbar^2}{2\mu R^2}(\mathbf{J}-\mathbf{L}-\mathbf{S})^2,
\label{eq:Hrot}
\end{equation}
with the rotational energy levels being given by:

\begin{equation}
    E_{rot} \approx B_v J(J+1),
\end{equation}
with $B_v = \hbar^2/(2\mu \langle R^{2}\rangle_v)$ being the rotational constant for vibrational level $v$. The characteristic scale for this energy is $E_\text{rot} \sim 10^{-5}$--$10^{-3}$~eV.  The rotational constant depends on the nuclear mass through $\mu$ and on the nuclear charge radius through the equilibrium distance $R_e$ (which shifts slightly with the electron density inside the nucleus), contributing to the rotational part of the molecular isotope shift \cite{udrescu2021isotope}.

A feature of the rotational structure that is of central importance for searches for P-violating and CP violating interactions discussed below is that adjacent rotational levels have opposite parity. As these levels are separated by energies on the order of $\Delta E_\pm \sim 10^{-5}$--$10^{-3}$~eV and thus much smaller than the electronic energy spacing usually found in atoms ($\sim 1$ eV), the sought-for parity-violating signals, which scale as $1/\Delta E_\pm$, can be significantly enhanced.  Moreover, in certain molecular electronic states (e.g., $^2\Pi_{1/2}$, $^3\Delta_1$), molecules can display pairs of opposite-parity states, within the same rotational level, with even smaller splittings, $\Delta E_\pm \lesssim 10^{-6}$~eV \cite{brown2003rotational}.  These small opposite-parity gaps play a major role in amplifying symmetry-violating effects, as will be shown in Secs.~\ref{sec:W_expansion} and~\ref{sec:CPV_expansion}.

\subsubsection{Nuclear-structure and fundamental-interaction corrections.} 
The remaining three terms in Eq.~(\ref{eq:H_total})---$H_{eN}^{(\text{EM})}$, $H_{eN}^{(\text{W})}$, and $H_{eN}^{(\text{CPV})}$---represent the electromagnetic, P-violating, and CP violating interactions between the electrons and nuclei (with the exception of the EM interaction between electrons and point-like nuclei, which is already included in $E_{el}$ above). These effects are usually small compared to the ones previously discussed, ranging from $\delta E/E_\text{el} \sim 10^{-5}$, for finite nuclear size effects, down to $< 10^{-16}$ for CP violating effects. Yet, they encode the nuclear and possible new-physics information that is the subject of this review.

\begin{figure}
\centering
\includegraphics[width=\linewidth]{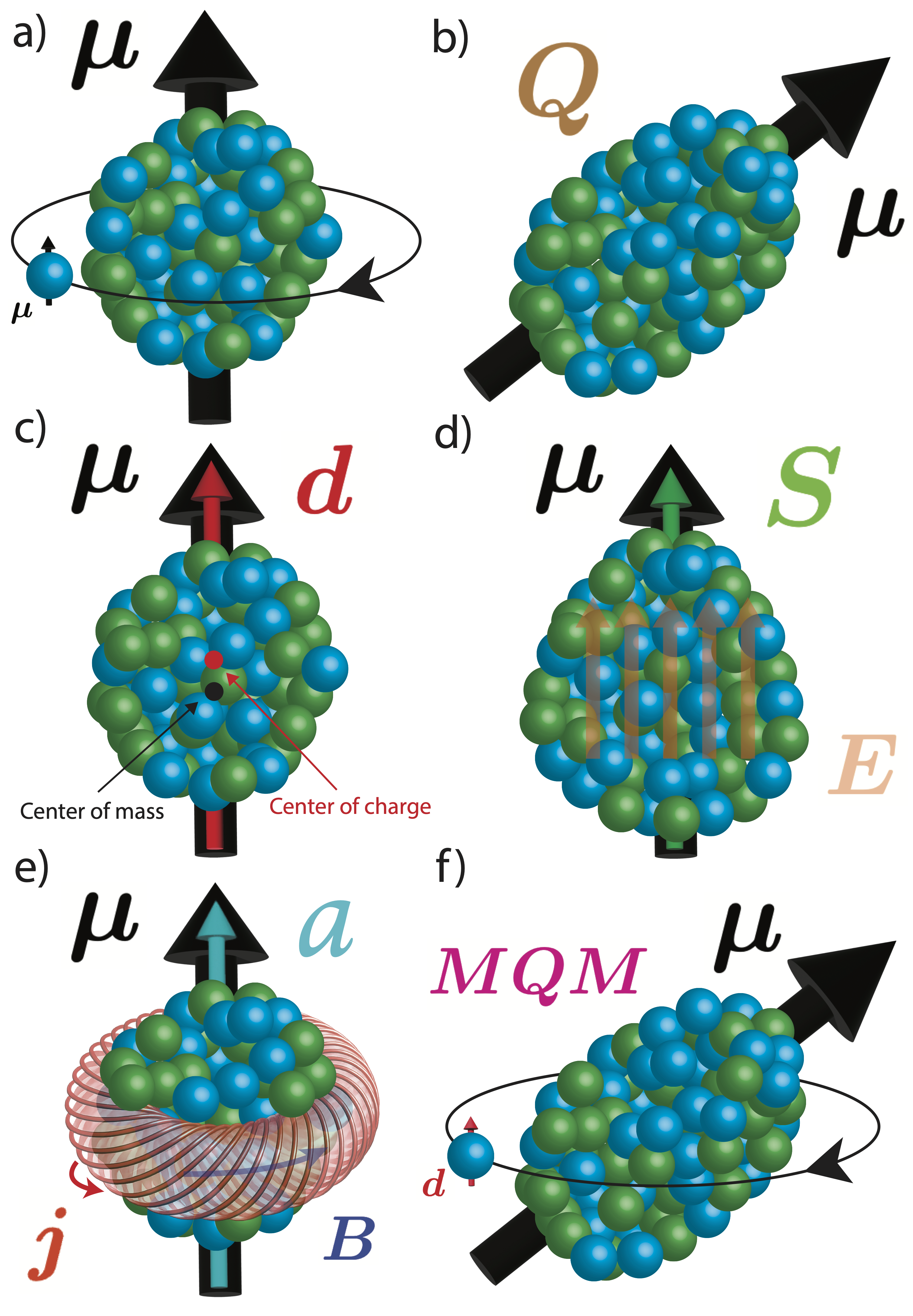}
\caption{Nuclear multipole moments accessible through precision spectroscopy of atoms and molecules.  Blue and green spheres represent protons and neutrons; the black arrow indicates the nuclear spin direction. The non-zero nuclear size is visible in each case.
(a)~The magnetic dipole moment $\bm{\mu}$ and its spatial distribution over the nuclear volume; the current loop indicates the magnetization distribution that gives rise to the Bohr--Weisskopf correction. 
(b)~The electric quadrupole moment $\mathbf{Q}$, reflecting the anisotropy of the proton distribution in a deformed nucleus.
(c)~The nuclear electric dipole moment $\mathbf{d}$ (red arrow), arising from a displacement between the center of mass and the center of charge. In neutral atoms and molecules, this is not a direct observable, being screened by the electron cloud.
(d)~The nuclear Schiff moment $\mathbf{S}$, the residual charge asymmetry that survives electrostatic screening (Schiff's theorem). Classically, the interaction with the electron cloud that penetrates the nucleus is due to a uniform electric field $\mathbf{E}$ (orange arrows) created inside the nucleus. In octupole-deformed (pear-shaped) nuclei, this moment can be enhanced by up to three orders of magnitude compared to non-deformed isotopes.
(e)~The nuclear anapole moment $\mathbf{a}$, a toroidal electromagnetic current distribution generated by parity-violating admixtures in the nuclear wavefunction, producing a magnetic field $\mathbf{B}$ confined within the nucleus.  Although it arises from the weak nucleon--nucleon interaction, the electron couples to it electromagnetically (see Sec.~\ref{sec:anapole}).
(f)~The nuclear magnetic quadrupole moment, MQM ($P$-odd, $T$-odd), arising from the interplay of nucleon electric dipole moments $\mathbf{d}$ and the nuclear magnetization; unlike the Schiff moment, the MQM is not subject to electrostatic screening.}
\label{fig:multipoles}
\end{figure}

The hierarchy of energy scales in the molecular Hamiltonian can be roughly summarized as:
\begin{equation}
E_\text{el} \gg E_\text{vib} \gg E_\text{rot} \gg E_\text{hfs} \approx E_\text{FS} \gg E_\text{PV} \gg E_\text{CPV}\,,
\label{eq:hierarchy_mol}
\end{equation}
spanning roughly $16$ orders of magnitude from $\sim 1$~eV to $\sim 10^{-16}$~eV. Here $E_\text{hfs}$ is the hyperfine energy and $E_\text{FS}$ represents corrections due to finite nuclear size. This vast dynamic range is what makes molecules such powerful spectroscopic laboratories: the gross structure (electronic, vibrational, rotational) provides a well-understood scaffold of energy levels, within which the tiny nuclear-structure and fundamental-interaction effects can be isolated through their unique signatures: isotope dependence (charge radius); spin dependence (electromagnetic moments); parity violation (weak charge, anapole); and simultaneous parity and time-reversal violation (electric dipole moments, Schiff moments, magnetic quadrupole moments) \cite{yang2023laser,arrowsmith2023opportunities}.
Below we expand our discussion for each of the interaction terms: $H_{eN}^{(\text{EM})}$, $H_{eN}^{(\text{W})}$, and $H_{eN}^{(\text{CPV})}$ as multipole expansions, and identify the resulting contributions to the molecular energy levels.

\subsection{The electromagnetic nucleus--electron interaction: multipole expansion}
\label{sec:EM_expansion}

\subsubsection{General framework}
Each nucleus $\alpha$ ($\alpha = 1,2$) has a charge density $\rho_\alpha(\mathbf{r})$ and a current density $\mathbf{j}_\alpha(\mathbf{r})$.  The electromagnetic interaction between the nucleus and the surrounding electrons is given by:
\begin{equation}
H_{eN,\alpha}^{(\text{EM})} = \int \dd^3 r\, \rho_\alpha(\mathbf{r})\, \Phi_e(\mathbf{r}) - \int \dd^3 r\, \mathbf{j}_\alpha(\mathbf{r}) \cdot \mathbf{A}_e(\mathbf{r})\,,
\label{eq:HeN_EM}
\end{equation}
where $\Phi_e(\mathbf{r})$ and $\mathbf{A}_e(\mathbf{r})$ are the scalar and vector potentials generated by all the electrons and the other nucleus at the position $\mathbf{r}$ inside nucleus $\alpha$.  These electronic fields can be expanded in a Taylor series about the center of nucleus $\alpha$ (taken as $\mathbf{r} = 0$):
\begin{align}
\Phi_e(\mathbf{r}) &= \Phi_e(0) + r_i\, \partial_i \Phi_e(0) + \tfrac{1}{2}\, r_i r_j\, \partial_i\partial_j \Phi_e(0) + \ldots\,, \nonumber \\
A_{e,k}(\mathbf{r}) &= A_{e,k}(0) + r_i\, \partial_i A_{e,k}(0) + \ldots\,,
\label{eq:Taylor}
\end{align}
where all derivatives are evaluated at the nuclear center.  Substituting into Eq.~(\ref{eq:HeN_EM}) and collecting terms according to their transformation properties under rotations---i.e., decomposing into irreducible spherical tensors---yields the multipole expansion:
\begin{equation}
H_{eN,\alpha}^{(\text{EM})} = \sum_{\lambda} \Mn_\alpha^{(\lambda)} \cdot \Fn_e^{(\lambda)},
\label{eq:multipole_master}
\end{equation}
where $\Mn_\alpha^{(\lambda)}$ is the nuclear multipole moment of rank $\lambda$, and $\Fn_e^{(\lambda)}$ is the corresponding electronic field operator (or its derivatives) evaluated at the position of nucleus $\alpha$.   Every electromagnetic interaction between electrons and the nucleus inside atoms and molecules is a specific term in the sum of equation~(\ref{eq:multipole_master}).

Some explicit terms in this expansion, which will be discussed in more detail further in this review, are:
\begin{align}
H_{eN,\alpha}^{(\text{EM})} = & \underbrace{\frac{Ze^2}{6\epsilon_0}\, \langle r^2 \rangle \, \rho_e(0)}_{\text{finite-size correction}} \nonumber \\[4pt]
& - \underbrace{\bm{\mu} \cdot \mathbf{B}_e(0)}_{\text{magnetic dipole}} \nonumber \\[4pt]
& - \underbrace{\tfrac{1}{6}\, Q_{ij}\, \partial_i E_{e,j}(0)}_{\text{electric quadrupole}} \nonumber \\[4pt]
& - \underbrace{ec\mathbf{a} \cdot \boldsymbol{\alpha}\delta(r)}_{\text{anapole (if $P$-violated inside the nucleus)}} \nonumber \\[4pt]
& - \underbrace{\mathbf{d} \cdot \mathbf{E}_e(0)}_{\text{electric dipole (if $P$,$T$-violated inside the nucleus)}} \nonumber \\[4pt]
& - \underbrace{4\pi\mathbf{S} \cdot \nabla\delta(r)}_{\text{Schiff moment (if $P$,$T$-violated inside the nucleus)}} \nonumber \\[4pt]
& - \underbrace{\tfrac{1}{6}\, M_{ij}\, \partial_i B_{e,j}(0)}_{\text{magnetic quadrupole (if $P$,$T$-violated inside the nucleus)}} + \ldots
\label{eq:EM_terms}
\end{align}
Here $\mathbf{E}_e(0) = -\nabla\Phi_e(0)$ and $\mathbf{B}_e(0) = \nabla \times \mathbf{A}_e(0)$ are the electric and magnetic fields produced by the electrons at the nuclear center, $\rho_e(0) = |\psi_e(0)|^2$ is the electronic charge density at the nucleus, and $\boldsymbol{\alpha}$ is the vector of Dirac alpha matrices.  The nuclear moments appearing in each term are illustrated in Fig.~\ref{fig:multipoles}.  We discuss them in turn next.

\subsubsection{The finite-size correction and the nuclear charge radius}
\label{sec:term_FS}

The first term arises from the fact that the nuclear charge $Ze$ is not concentrated at a point but is distributed over a volume of radius $R_N \sim 1$--$7$~fm \cite{yang2023laser}.  The electronic potential $\Phi_e(\mathbf{r})$ varies slightly across this volume; the leading correction beyond the point-charge result involves the second moment of the nuclear charge distribution. To derive this, we write the difference between the actual nuclear--electron interaction and the point-charge interaction as:
\begin{equation}
\delta H_\text{FS} = \int \dd^3 r\, \rho_c(\mathbf{r})\, \Phi_e(\mathbf{r}) - Ze\,\Phi_e(0)\,.
\label{eq:deltaH_FS}
\end{equation}
Expanding $\Phi_e(\mathbf{r})$ about $\mathbf{r} = 0$ and using the normalization $\int \rho_c\,\dd^3 r = Ze$ and the fact that $\nabla^2\Phi_e(0) = -\rho_e(0)/\epsilon_0$, the leading correction due to nuclear size is:
\begin{equation}
\delta H_\text{FS} = \frac{Ze^2}{6\epsilon_0}\, \langle r^2 \rangle_\text{ch}\, |\psi_e(0)|^2\,,
\label{eq:FS_shift}
\end{equation}
where
\begin{equation}
\langle r^2 \rangle_\text{ch} = \frac{1}{Ze}\int \rho_c(\mathbf{r})\, r^2\, \dd^3 r
\label{eq:r2_def}
\end{equation}
is the mean-square charge radius of the nucleus.  This is the product between a nuclear parameter, $\langle r^2 \rangle_\text{ch}$, and an electronic quantity, $|\psi_e(0)|^2$. $\langle r^2 \rangle_\text{ch}$ encodes the spatial extent of the nuclear charge distribution, which is determined by the balance between the attractive nuclear force, the Pauli exclusion principle, and the Coulomb repulsion among protons \cite{yang2023laser}.

\paragraph{\textbf{Isotope shifts in atoms.}}  The absolute value of $\delta H_\text{FS}$ is not directly observable (it is absorbed into the definition of the electronic energy).  However, the \emph{change} of this term between two isotopes $A$ and $A'$ of the same element is observable as the \emph{field shift} component of the isotope shift:
\begin{equation}
\delta\nu_\text{FS}^{A,A'} = F \cdot \delta\langle r^2 \rangle^{A,A'}\,,
\label{eq:IS_FS}
\end{equation}
where $\delta\langle r^2\rangle^{A,A'} = \langle r^2\rangle_{A'} - \langle r^2\rangle_A$ is the nuclear quantity and
\begin{equation}
F = \frac{Ze^2}{6\epsilon_0}\, \Delta|\psi_e(0)|^2
\label{eq:F_def}
\end{equation}
is the electronic factor, given by the change in electron density at the nucleus between the two electronic levels involved in the transition \cite{king2013isotope}. 

The total isotope shift also receives a \emph{mass shift} contribution from the nuclear kinetic energy term $T_\text{nuc}$.  In the center-of-mass frame, the nuclear recoil contributes:
\begin{equation}
H_\text{recoil} = \frac{1}{2M}\left(\sum_i p_i^2 + \sum_{i\neq j} \mathbf{p}_i \cdot \mathbf{p}_j\right),
\label{eq:recoil}
\end{equation}
where $M$ is the nuclear mass and $\mathbf{p}_i$ are the electron momenta.  The resulting mass shift between two isotopes is:
\begin{equation}
\delta\nu_\text{MS}^{A,A'} = \frac{M_A - M_{A'}}{M_A M_{A'}}(K_\text{NMS} + K_\text{SMS})\,,
\label{eq:MS}
\end{equation}
where $K_\text{NMS} = m_e \nu$ (the normal mass shift, calculable from the transition frequency) and $K_\text{SMS}$ (the specific mass shift, requiring many-body atomic theory) are purely electronic quantities \cite{king2013isotope}.  The total isotope shift is therefore:
\begin{equation}
\delta\nu_\text{IS}^{A,A'} = \underbrace{F \cdot \delta\langle r^2 \rangle^{A,A'}}_{\text{nuclear size}} + \underbrace{\frac{M_A - M_{A'}}{M_A M_{A'}}(K_\text{NMS} + K_\text{SMS})}_{\text{nuclear mass}}\,.
\label{eq:IS_total}
\end{equation}
This is again a sum of factorized terms: nuclear quantities ($\delta\langle r^2\rangle$, masses) multiplied by electronic factors ($F$, $K$).

\paragraph{\textbf{Higher radial moments.}}  Extending the Taylor expansion of $\Phi_e(\mathbf{r})$ to higher orders introduces corrections proportional to $\langle r^4\rangle$, $\langle r^6\rangle$, etc.:
\begin{equation}
\delta\nu_\text{FS}^{A,A'} = F\Big(\delta\langle r^2\rangle^{A,A'} + a\,\delta\langle r^4\rangle^{A,A'} + b\,\delta\langle r^6\rangle^{A,A'} + \ldots\Big),
\label{eq:FS_higher}
\end{equation}
where $a$ and $b$ are functions of $Z$ and the electronic wave functions.  The $\langle r^4\rangle$ moment probes the nuclear surface diffuseness, providing information beyond the overall size \cite{reinhard2020beyond}. Such higher-order radial moments, as well as higher-order corrections to the mass shifts, and polarizability effects appear as subleading contributions to the isotope shift effect (see Sec. \ref{sec:IS_newphysics} for further discussion). 

\paragraph{\textbf{Isotope shifts in molecules.}}  In a diatomic molecule, the electron density at the nucleus depends on the internuclear distance, $R$.  Consequently, the electronic factors become functions of $R$, and the field shift contributes not only to the electronic energy but also to the vibrational and rotational constants.  For a rovibrational level $(v, J)$, the molecular isotope shift, to leading order in $J$ and $\nu$, takes the form \cite{knecht2012nuclear,almoukhalalati2016nuclear,udrescu2021isotope}:
\begin{align}
\delta E^{A,A'}(v,J) = \; & \Big(f_{00}\,\delta\langle r^2\rangle^{A,A'} + \delta_{00}\tfrac{\Delta M}{M'}\Big) \nonumber \\
& + \Big(f_{10}\,\delta\langle r^2\rangle^{A,A'} + \delta_{10}\tfrac{\Delta M}{M'}\Big)\left(\tfrac{\mu}{\mu'}\right)^{1/2}(v+\tfrac{1}{2}) \nonumber \\
& + \Big(f_{01}\,\delta\langle r^2\rangle^{A,A'} + \delta_{01}\tfrac{\Delta M}{M'}\Big)\tfrac{\mu}{\mu'}\, J(J+1)\,,
\label{eq:mol_IS}
\end{align}
where $\mu$, $\mu'$ are the reduced masses of the two molecules, and $f_{ij}$, $\delta_{ij}$ are electronic parameters that encode the $R$-dependence of the electron density at the nucleus and the mass-shift sensitivity, respectively.  The factorization persists: each term is a nuclear quantity times an electronic quantity, but now the electronic quantities carry information about the molecular potential energy surface.

\subsubsection{The magnetic dipole interaction and the nuclear magnetic moment}
\label{sec:term_mu}

The second term in Eq.~(\ref{eq:EM_terms}),
\begin{equation}
H_\mu = -\bm{\mu} \cdot \mathbf{B}_e(0)\,,
\label{eq:Hmu}
\end{equation}
describes the interaction of the nuclear magnetic dipole moment $\bm{\mu}$ with the magnetic field $\mathbf{B}_e(0)$ produced by the electrons at the nuclear center (Fig.~\ref{fig:multipoles}a).  For a nucleus with spin $I \geq 1/2$, we write $\bm{\mu} = g_I \muN \mathbf{I}$, where $g_I$ is the nuclear $g$-factor, $\muN = e\hbar/(2m_p)$ is the nuclear magneton, and $\mathbf{I}$ is the nuclear spin operator.  Since the electronic field at the nucleus is proportional to the electronic angular momentum, $\mathbf{B}_e(0) \propto \mathbf{J}$, the interaction takes the standard hyperfine form:
\begin{equation}
H_\mu = A_\text{hfs}\, \mathbf{I} \cdot \mathbf{J}\,,
\label{eq:Hhfs}
\end{equation}
where the magnetic hyperfine constant is:
\begin{equation}
A_\text{hfs} = \underbrace{\mu/I}_{\text{nuclear}} \cdot \underbrace{B_e(0)/J}_{\text{electronic}}\,.
\label{eq:Ahfs}
\end{equation}
This interaction splits each electronic fine-structure level (or rotational level, in the case of a molecule) into hyperfine sublevels labeled by the total angular momentum $F$, with energies \cite{yang2023laser,arrowsmith2023opportunities}:
\begin{equation}
E_F^\text{(mag)} = \frac{A_\text{hfs}}{2}\big[F(F+1) - I(I+1) - J(J+1)\big]\,.
\label{eq:EF}
\end{equation}
From the measured splittings, $A_\text{hfs}$ can be extracted.  If $B_e$ is known (from atomic theory or independent experiments), the nuclear magnetic moment $\mu$ can be determined.

\paragraph{\textbf{The Bohr--Weisskopf correction.}}  In the derivation above, we assumed that the nuclear magnetization is treated as concentrated at a single point.  In reality, the magnetization is distributed over the entire nuclear volume (Fig.~\ref{fig:multipoles}a), described by a distribution function $f_M(\mathbf{r})$ with $\bm{\mu}(\mathbf{r}) = \mu\, f(\mathbf{r})$ and $\int f(\mathbf{r})d^3 r = 1$.  The electronic field $\mathbf{B}_e(\mathbf{r})$ varies inside the nucleus, and the exact interaction is given by:
\begin{equation}
H_\mu = -\int \mathbf{B}_e(\mathbf{r}) \cdot \dd\bm{\mu}(\mathbf{r})\,.
\label{eq:BW}
\end{equation}
The deviation from the point-dipole result is called the Bohr--Weisskopf effect, and it is sensitive to the spatial distribution of nuclear currents \cite{bohr1950influence,skripnikov2020nuclear,yang2023laser,wilkins2023observation}.

\subsubsection{The electric quadrupole interaction and the nuclear shape}
\label{sec:term_Q}

The next term in Eq.~(\ref{eq:EM_terms}),
\begin{equation}
H_Q = -\frac{1}{6}\, Q_{ij}\, \partial_i E_{e,j}(0),
\label{eq:HQ}
\end{equation}
describes the interaction of the nuclear electric quadrupole moment $Q_{ij}$ with the electric field gradient (EFG) $V_{ij} = \partial_i E_{e,j}(0)$ produced by the electrons at the nucleus (Fig.~\ref{fig:multipoles}b).  This term exists only for nuclei with spin $I \geq 1$.

The quadrupole interaction produces a contribution to the hyperfine energies given by \cite{yang2023laser}:
\begin{equation}
E_F^\text{(quad)} = B_\text{hfs}\, \frac{\frac{3}{4}K_H(K_H+1) - I(I+1)J(J+1)}{2I(2I-1)J(2J-1)}\,,
\label{eq:EF_quad}
\end{equation}
where $K_H \equiv F(F+1) - I(I+1) - J(J+1)$, and $B_\text{hfs}~=~eQ\,V_{zz}$ is the quadrupole hyperfine constant, with $Q$ being the spectroscopic nuclear quadrupole moment. 
 The matrix element of $H_Q$ vanishes for electronic states with total angular momentum $J < 1$.  This means that for atoms with a $J = 1/2$ ground state (for example, all alkali atoms), the nuclear quadrupole moment cannot be measured from ground-state hyperfine splittings.

 Unlike atoms, in a diatomic molecule, the electron orbital angular momentum is not a good quantum number \cite{brown2003rotational}, which means that, in general, ground electronic levels in molecules have a non-zero sensitivity to the electric quadrupole hyperfine interaction.  Moreover, the EFG at the nucleus in a molecule can be relatively large, reflecting the breaking of spherical symmetry present in atoms, and the different rotational levels in the ground electronic and vibrational state are long-lived \cite{brown2003rotational}.  This makes molecules the preferred systems for measuring nuclear quadrupole moments, especially for elements where the atomic ground state has $J < 1$.  An example of such a situation is the measurement of the quadrupole moments of $^{39,40,41}$K through the hyperfine structure of potassium monofluoride, KF \cite{kello1998quadrupole}.

\paragraph{\textbf{Quadrupole shifts}}
\label{sec:qshifts}
The finite nuclear size also generates penetration corrections to the
higher-rank interactions themselves: just as the monopole acquires the field
shift $\propto\langle r^2\rangle$ and the magnetic dipole the Bohr--Weisskopf correction, the quadrupole interaction acquires an analogous rank-2 penetration
correction, the \emph{quadrupole shift} \cite{Koch2010QS}. Arising from the penetration of $s$
and relativistic $p_{1/2}$ electrons into the nuclear volume, it is the
quadrupole analog of the field shift~\cite{Koch2010QS,Rose2012toymodels} and can
be expressed as a nuclear \emph{quasi-quadrupole moment}
\begin{equation}
  \tilde{Q} \;\propto\; \int \rho_c(r)\,(3z^2-r^2)\,r^2 \,\dd^3 r
  \;\propto\; \langle r^4 Y_{20}\rangle ,
  \label{eq:quasiQ}
\end{equation}
which carries an extra $r^2$ relative to the ordinary quadrupole moment
$Q\propto\langle r^2 Y_{20}\rangle$, multiplied by an electronic factor set by
the curvature of the electron density at the nucleus,
$n_{zz}\propto(\partial_z^2-\tfrac{1}{3}\nabla^2)\,n_e(0)$, the rank-2 analog of
the contact density $|\psi_e(0)|^2$. The measured quadrupole coupling constant
is therefore the sum of the point-nucleus interaction and the quadrupole shift,
\begin{equation}
  \nu_Q \;\approx\; \nu_{\mathrm{QI}} + \nu_{\mathrm{QS}}
  \;=\; \frac{1}{h}\!\left(eQV_{zz}
        - \tfrac{1}{140}\,e\tilde{Q}\,n_{zz}\right) ,
  \label{eq:nuQS}
\end{equation}
and, because $\nu_{\mathrm{QS}}$ shares the symmetry of $\nu_{\mathrm{QI}}$, the
two are indistinguishable in any single measurement. Reliable values require
fully relativistic calculations with an explicit finite nucleus. The effect scales with nuclear charge
($\nu_{\mathrm{QS}}/\nu_{\mathrm{QI}}\propto Z^4$) and is predicted to be a
$0.1$--$1\%$ effect for heavy elements ($Z\gtrsim 60$), where it can reach the
$\sim\!100$~kHz level ~\cite{Koch2010QS}.

While $\nu_{\mathrm{QS}}$ cannot be isolated in a single experiment, it can be
extracted through the \emph{quadrupole anomaly} $\delta$, defined in direct
analogy to the hyperfine (Bohr--Weisskopf) anomaly. For two isotopes in the same electronic environment, the ratio of quadrupole splittings deviates from the ratio
of quadrupole moments as:
\begin{equation}
  \frac{\nu_{Q,1}}{\nu_{Q,2}} = \frac{Q_1}{Q_2}\,(1+\delta), \qquad
  \delta \;\propto\; \frac{n_{zz}}{V_{zz}}\delta \langle r^2 \rangle^{1,2},
  \label{eq:Qanomaly}
\end{equation}
where $\delta \langle r^2 \rangle^{1,2}$ is the nuclear charge radii difference between the two isotopes of interest. Comparing this ratio for the same
isotope pair measured in two different molecules, for instance $^{223,227}$Ra in
RaF and RaI can provide an experimental
observable that is nonzero because of electron penetration into the
anisotropic nuclear charge distribution~\cite{Koch2010QS,Rose2012toymodels}.
This anomaly is sensitive to the deformation-weighted moment
$\langle r^4 Y_{20}\rangle$, and has not been observed yet. As with the higher radial moments above, molecules are a natural platform for exploring this effect.

\subsubsection{The nuclear anapole moment}
\label{sec:anapole}

The electromagnetic multipole expansion also contains $P$-odd nuclear moments, which vanish in a nucleus with a parity-eigenstate wavefunction but become nonzero if parity-violating physics is present \cite{Zeldovich1958,bouchiat1991nuclear,khriplovich1991}.  The leading such moment is the \emph{nuclear anapole moment}, $\mathbf{a}$ (Fig.~\ref{fig:multipoles}e), which can be written in terms of the ordinary electromagnetic current density, $\mathbf{j}_N(\mathbf{r})$, as:
\begin{equation}
\mathbf{a} = -\pi \int \dd^3 r\, r^2\, \mathbf{j}_N(\mathbf{r})\,.
\label{eq:anapole_def}
\end{equation}
Note that this is the \emph{same} nuclear current density that produces the magnetic dipole moment, $\bm{\mu}~=~\frac{1}{2}\int \mathbf{r}~\times~\mathbf{j}_N\,\dd^3 r$. In a nuclear state with well-defined parity, the expectation value of the anapole moment vanishes identically \cite{flambaum1984nuclear,flambaum1997anapole,hao2020nuclear}. However, due to the presence of parity-violating nucleon--nucleon interactions $H_\text{PV}^{NN}$ (mediated by weak pion exchange), opposite-parity nuclear states are mixed into the ground-state wavefunction \cite{hao2020nuclear}:
\begin{equation}
|\Psi\rangle = |\Psi_0\rangle + \sum_n \frac{\langle n | H_\text{PV}^{NN} | \Psi_0\rangle}{E_0 - E_n}\, |n\rangle\,,
\label{eq:PV_admixture}
\end{equation}
leading to a non-zero anapole moment in nuclei with $I>0$. Thus, the anapole moment is an \emph{electromagnetic} quantity, a moment of $\mathbf{j}_N$, evaluated in a nuclear state that is not a parity eigenstate due to the \emph{weak} interaction between nucleons.

Physically, the parity-violating admixtures generate a toroidal current distribution inside the nucleus (Fig. \ref{fig:multipoles}e), which produces a magnetic vector potential $\mathbf{A}_\text{ana}(\mathbf{r})$ that is nonzero only within the nuclear volume.  The electron interacts with this vector potential through the ordinary electromagnetic coupling $-e\,\alphav \cdot \mathbf{A}_\text{ana}(\mathbf{r})$ \cite{flambaum1985enhancement,flambaum1980p,haxton2001atomic}. 

For a nucleus with spin $I$, the anapole moment points along $\mathbf{I}$, and the resulting interaction Hamiltonian takes the factorized form \cite{flambaum1984nuclear,flambaum1985enhancement,safronova2018}:
\begin{equation}
H_\text{anapole} = \underbrace{\kappa_a}_{\substack{\text{nuclear}\\\text{(anapole moment)}}} \times \underbrace{\frac{\GF}{\sqrt{2}}\,\frac{\alphav \cdot \mathbf{I}}{I(I+1)}\,\rho_\text{nuc}(r)}_{\substack{\text{electronic}\\\text{(EM field inside nucleus)}}}\,,
\label{eq:Hanapole_factor}
\end{equation}
where $\kappa_a$ is a dimensionless constant proportional to the anapole moment, and $\rho_\text{nuc}(r)$ is the nuclear density. The contact character arises because the anapole vector potential is confined to the nuclear volume.

The anapole interaction (Eq.~(\ref{eq:Hanapole_factor})) has the same operator structure, $\alphav \cdot \mathbf{I}\,\rho_\text{nuc}(r)$, as the direct $Z^0$-exchange nuclear-spin-dependent (NSD) weak interaction (the $C_2$ coupling discussed in Sec.~\ref{sec:W_expansion}) \cite{safronova2000high}.  However, the two effects have distinct physical origins: the anapole moment is an electromagnetic interaction with a parity-violating nuclear current (Eq.~(\ref{eq:anapole_def})), while the $C_2$ term arises from the weak interaction involving the nucleon axial current and the electron vector current.  The two cannot be separated in a single atomic measurement.  However, the latter effect is roughly constant with the atomic number, while the anapole moment scales as $A^{2/3}$. Thus, in heavy nuclei, the total NSD parity-violating signal is primarily a measurement of the anapole moment, and thus of the parity-violating nucleon--nucleon interaction inside the nucleus \cite{safronova2018,hao2020nuclear}.
The anapole moment has been measured only once, in $^{133}$Cs \cite{wood1997}, with the extracted constraints on hadronic parity violation being in tension with those derived from accelerator-based experiments \cite{haxton2013hadronic}. Although in molecules the NSD-PV effect can be strongly enhanced by the small opposite-parity splittings discussed in Sec.~\ref{sec:mol_enhance}, it has not yet been observed and remains the subject of ongoing investigations 
\cite{altuntacs2018demonstration,norrgard2019nuclear,karthein2023electroweak,blanchard2023using,kogel2025laser}.

\subsubsection{The nuclear electric dipole moment}

If parity and time-reversal are violated, the nucleus can acquire a non-zero electric dipole moment (EDM), as shown in Fig. \ref{fig:multipoles}c. This can be either due to the EDMs of the constituent protons and neutrons or due to PT (or, equivalently, CP) violating interactions between nucleons \cite{sushkov1984possibility,spevak1997,engel2025nuclear}. The EDM can be viewed as a displacement between the center of mass and the center of charge of the nucleus. However, for a neutral and non-relativistic system of point particles interacting purely electrostatically, the induced EDM of the system is identically zero: the internal rearrangement of charges exactly cancels the effect of a constituent EDM on any external measurement, known as the Schiff theorem \cite{schiff1963measurability} (see Sec. \ref{sec:appendix_schiff_theorem}). While the nuclear EDM itself is screened in a neutral atom, the finite size of the nucleus prevents perfect cancellation, leaving a residual electrostatic potential that can interact with the atomic electrons, whose leading effect is characterized by the nuclear Schiff moment.

\subsubsection{The nuclear Schiff moment}
\label{sec:schiff_mom}
The Schiff theorem can be evaded because the nucleus is not a point particle, so the
electron screening of the nuclear EDM is incomplete \cite{schiff1963measurability}.
The leading unscreened effect is characterized by the nuclear Schiff moment
(Fig.~\ref{fig:multipoles}d), whose charge contribution is given by
\cite{sushkov1984possibility,engel2025nuclear}:
\begin{equation}\label{eq:Schiff}
    \mathbf{S}^\text{ch} = \frac{|e|}{10}\left(\int\rho_\text{ch}(\mathbf{r})\,r^2\,\mathbf{r}\,\dd^3r
    \;-\;\frac{5}{3}\,\langle r^2\rangle_\text{ch}
    \int\rho_\text{ch}(\mathbf{r})\,\mathbf{r}\,\dd^3r\right),
\end{equation}
where $\langle r^2\rangle_\text{ch}$
is the mean squared charge radius defined in Eq. \ref{eq:r2_def} (see Sec. \ref{sec:appendix_schiff_moment}) and $\rho_\text{ch}(\mathbf{r})$ is the nuclear charge density. The second integral,
$\mathbf{d}_0 = \int\rho(\mathbf{r})\,\mathbf{r}\,\dd^3r$, is the dipole moment of
the charge distribution in units of length, related to the nuclear EDM by
$\mathbf{d} = |e|\,\mathbf{d}_0$, while the first is the second moment of the dipole
distribution. Both integrals vanish unless parity and time-reversal symmetry are
violated inside the nucleus, primarily through a PT violating nucleon--nucleon
interaction. The terms of order $R_N$ cancel between them, leaving $\mathbf{S}$ of
order $R_N^3$. The intrinsic EDMs $\mathbf{d}_i = D_i\boldsymbol{\sigma}_i$ of the
nucleons themselves contribute an additional piece \cite{engel2025nuclear},
\begin{equation}\label{eq:SchiffNucleon}
    \mathbf{S}^{n} = \frac{1}{6}\sum_{i=1}^{A}
    \left(r_i^2 - \langle r^2\rangle_\text{ch}\right)\mathbf{d}_i\,,
\end{equation}
in which the sum runs over all $A$ nucleons, so that the neutron EDM enters the
atomic observable despite the neutron carrying no net charge. Because Schiff
screening suppresses this contribution relative to free-neutron measurements, the nucleon-EDM piece is subdominant in the interpretation of atomic and molecular experiments, which instead probe mainly Eq.~\eqref{eq:Schiff}. Naively, the induced atomic EDM is suppressed by $(R_N/R_A)^2 \approx 10^{-8}$--$10^{-9}$ relative to the unscreened nuclear EDM, but the relativistic concentration of electrons near heavy nuclei softens this suppression to roughly $10^{-3}$ \cite{engel2025nuclear}. 

The PT odd part of the resulting nuclear electrostatic potential reduces to a contact
term, $\bar{\phi}_{PT}(\mathbf{r}) =~4\pi\,\mathbf{S}~\cdot~\nabla\delta^3(\mathbf{r})$
with $\mathbf{S}~=~\mathbf{S}^\text{ch} + \mathbf{S}^n$, so that the interaction with the electrons is expressed as the product
\cite{safronova2018,arrowsmith2023opportunities}:
\begin{equation}
H_\text{Schiff} = -4\pi\, \mathbf{S} \cdot \nabla\rho_e(\mathbf{r})\big|_{r=0}\,,
\label{eq:HSchiff}
\end{equation}
where $\nabla\rho_e(\mathbf{r})\big|_{r=0}$ is the gradient of the electronic charge
density at the nucleus. A non-zero value of this nuclear property has never been measured in an atom or
molecule and its observation, given current experimental sensitivities, would be a
clear sign of new CP violating physics beyond the SM. Again, the existence of closely
spaced levels of opposite parity in molecules lead to large enhancements in the
sensitivity to this effect compared to similar atomic experiments
\cite{safronova2018,arrowsmith2023opportunities}.

\paragraph{\textbf{Enhancement by nuclear octupole deformation.}}

The nuclear Schiff moment can be significantly enhanced in nuclei with strong octupole (pear-shaped) deformation \cite{arrowsmith2023opportunities,auerbach1996collective,Jad25, sheline1983evidence,gaffney2013studies,butler2020evolution, butler2016octupole,butler1996intrinsic}.  Such nuclei possess near-degenerate parity doublets--pairs of states with the same spin but opposite parity, separated by a very small energy gap $\Delta E_\pm$.  

Assuming the nucleus possesses intrinsic quadrupole ($\beta_2$) and octupole ($\beta_3$) deformations, a large nuclear Schiff moment can be generated, expressed as \cite{auerbach1996collective}:
\begin{equation}
S \propto \frac{\langle +|V_\text{PT}|-\rangle}{\Delta E_\pm}\, \beta_2 \beta_3^2\, Z A^{2/3}\,,
\label{eq:Schiff_enhance}
\end{equation}
where $V_\text{PT}$ is the CP violating nucleon-nucleon potential, containing the new physics of interest, and $|+\rangle$, $|-\rangle$ are the members of the parity doublet, one of which is the ground nuclear state. The enhancement arises thus both from the large octupole deformation, $\beta_3$, present in the numerator of this expression, and from the small energy denominator, $\Delta E_\pm$, which can be as low as $\sim 50$~keV in nuclei like $^{225}$Ra \cite{sheline1983evidence,andersen1989measurements}, or even below 100 eV for $^{229}$Pa \cite{Ahmad2015Pa229}.  The resulting enhancement can reach factors of more than $10^3$ compared to spherical nuclei \cite{arrowsmith2023opportunities}.

When an octupole-deformed nucleus is placed inside a polar molecule, the nuclear enhancement combines with the molecular enhancement from small opposite-parity doublet splittings and full polarizability (factor $\sim 10^2$--$10^3$), yielding a total sensitivity improvement of more than five orders of magnitude relative to a spherical nucleus in an atom \cite{safronova2018,arrowsmith2023opportunities}.  Using molecules containing such octupole-deformed isotopes is an emerging and active area of research, with great promise for the future of CP violation searches \cite{udrescu2021isotope,fan2021optical,udrescu2024precision,wilkins2023observation,garcia2020spectroscopy,arrowsmith2023opportunities,conn2025production,Jad25}.

\paragraph{\textbf{Electron EDM}}

While this review is focused on nuclear properties, the exciting advances in the use of molecules for electron EDM measurements have provided major motivation and guidance for progress toward measurements of symmetry-violating nuclear properties \cite{acme2018improved,roussy2023improved}.
Excellent reviews of the theoretical background
and of past, ongoing, and future electron EDM searches exist in the literature, and we point the interested reader to them
\cite{pospelov2005electric,chupp2019electric,safronova2018,vutha2010search,leanhardt2011high}. For the electron, Schiff's theorem is evaded not
by finite size but by the relativistic character of the electronic motion, most pronounced in the strong Coulomb field near a heavy nucleus, which enhances rather than screens the atomic EDM
\cite{sandars1965electric,sandars1966enhancement,khriplovich2012cp,commins2007electric,safronova2018}. Because polar molecules can be fully polarized in modest laboratory fields, exposing the unpaired electron to effective internal fields of tens of GV/cm, they set the
most stringent bounds on the electron EDM, probing energy scales comparable to or exceeding those accessible at high-energy facilities such as the LHC
\cite{acme2018improved,roussy2023improved}. Importantly, however, the PT odd
frequency shift measured in these paramagnetic systems is not sensitive to $d_e$ alone: it receives a second, nucleus-dependent contribution from the CP violating scalar--pseudoscalar electron--nucleon interaction, which is discussed in detail in Sec. \ref{sub:cp_e_N}.

\subsubsection{The nuclear magnetic quadrupole moment}
The lowest-order PT violating magnetic moment of the nucleus is the magnetic
quadrupole moment (MQM), $M_{ij}$ (Fig.~\ref{fig:multipoles}f). It interacts with the gradient of the electron magnetic field at the nucleus:
\begin{equation}
    H_\text{MQM} = -\frac{1}{6}\,M_{ij}\,\partial_i B_{e,j}(0)\,.
\end{equation}
As a moment of the nuclear current rather than the charge distribution, the MQM is not subject to Schiff screening and carries no compensating subtraction analogous to the second term of Eq.~\eqref{eq:Schiff}. Its contribution to atomic and molecular EDMs for non-octupole-deformed nuclei can exceed that of the Schiff moment \cite{sushkov1984possibility,lackenby2018mqm}. It requires nuclear spin $I \geq 1$ and electronic states with nonzero angular momentum \cite{flambaum2014time,lackenby2018mqm}.

Reliable nuclear many-body calculations for the MQM are not yet available in the literature.  Calculations for the collective MQM with state-of-the-art nuclear many-body methods (see  Sec.~\ref{sec:abinitio})
is thus a well-posed opportunity for nuclear theory, with direct impact on the
interpretation of molecular experiments.

At the single-nucleon level, the MQM operator is given by
\cite{sushkov1984possibility,khriplovich2012cp,dalton2023enhanced}
\begin{equation}
\label{eq:MQMoperator}
\begin{split}
\hat{M}^{\nu}_{kn} = \frac{e}{2m_N}\Bigg[
&3\mu_\nu\left(r_k\sigma_n + \sigma_k r_n
- \frac{2}{3}\,\delta_{kn}\,\boldsymbol{\sigma}\cdot\mathbf{r}\right) \\
&+ 2q_\nu\left(r_k l_n + l_k r_n\right)
\Bigg],
\end{split}
\end{equation}
where $\nu = p,n$, and $\mu_\nu$ and $q_\nu$ are the nucleon magnetic moment and
charge. 
Existing estimates of nuclear MQMs rely on oversimplified assumptions about nuclear
structure, based either on single independent nucleons
\cite{sushkov1984possibility,flambaum2014time} or on simple collective nuclear models
\cite{flambaum2022mqm}. In quadrupole-deformed
nuclei, the MQM acquires a collective character, with the single-nucleon MQM contribution being proportional to the product of the spin and orbital angular momentum projections on the deformation axis, a quantity that is even under time reversal, so the two members of each nucleon pair contribute with the same sign, and the open-shell contributions add coherently. Sums over the open shells in deformed nuclear models predict an order-of-magnitude enhancement over the single-particle value in heavy deformed nuclei \cite{lackenby2018mqm}. This enhancement, however, assumes sharp orbital occupations, neglecting pairing correlations and configuration mixing, and treats the PT odd interaction in the zero-range limit. In octupole-deformed nuclei, the parity-doublet mixing that produces collective Schiff moments can also generate an MQM, given by \cite{flambaum2022mqm}
\begin{equation}
M \approx
\frac{2I-1}{2I+3}\,
\langle\beta_3^2\rangle\,
\frac{ eV}{E_+ - E_-}\,
\mu\,\eta\, e\,\mathrm{fm}^2,
\end{equation}
where $\mu$ is the
nuclear magnetic moment in nuclear magnetons; and $\eta$ is the dimensionless
strength constant of the PT odd nucleon--nucleus potential.
As the intrinsic MQM is a moment of the magnetization, which cancels between
paired nucleons, only the unpaired nucleon contributes. The
octupole mechanism, therefore is suggested to enhance the MQM mainly through the small energy
denominator of the doublet, without the collective enhancement that benefits the Schiff moment.

\subsubsection{Higher electromagnetic terms}

The multipole expansion in Eq. \ref{eq:EM_terms} continues with the electric and magnetic octupole interactions (requiring $I \geq 3/2$), the electric and magnetic hexadecapole interactions (requiring $I \geq 2$), and so on.  Each term has the same factorized structure:
\begin{equation}
H^{(\lambda)} = \Mn^{(\lambda)} \cdot \Fn_e^{(\lambda)}\,,
\label{eq:Hlambda}
\end{equation}
where $\Mn^{(\lambda)}$ is a nuclear multipole of rank $\lambda$ and $\Fn_e^{(\lambda)}$ is the electronic field derivative of order $\lambda$ at the nucleus.  The effects decrease rapidly with $\lambda$: each successive term is suppressed by a factor of order $R_N/a_0$, where $R_N$ is the nuclear size, while $a_0$ is the Bohr radius, so that, in practice, only the monopole (charge radius), dipole, and quadrupole moments are routinely relevant for ongoing and planned experiments.  The magnetic octupole has been observed in a few atoms \cite{gerginov2003observation,gerginov2009observation,lewty2012spectroscopy,singh2013observation,de2022precision,li2022re}; higher moments remain elusive.  Molecules, with their richer level structure, offer a promising pathway to reach them.

\begin{table}
\caption{Nuclear moments organized by their whether they violate ($\checkmark$) or not ($\times$) parity ($P$) and/or time reversal ($T$). The Min. $I$ column shows the smallest spin required for the moment to be non-zero according to angular momentum selection rules.  The column $Origin$, indicates the physical origin behind the moment: electromagnetic (EM), P (PV) or CP violating (CPV) physics.}
\label{tab:moments}
\begin{ruledtabular}
\begin{tabular}{lccccl}
\textrm{Moment} & $P$ & $T$ & \textrm{Min.\ $I$} & \textrm{Origin} \\
\colrule
Charge radius $\langle r^2\rangle$ & $\times$ & $\times$ & 0 & EM \\
Mag.\ dipole $\boldsymbol{\mu}$ & $\times$ & $\times$ & 1/2 & EM \\
Elec.\ quadrupole $\boldsymbol{Q}$ & $\times$ & $\times$ & 1 & EM \\
\colrule
Anapole $\mathbf{a}$ & $\checkmark$ & $\times$ & 1/2 & PV \\
\colrule
Electric\ dipole $\boldsymbol{d}$ & $\checkmark$ & $\checkmark$ & 1/2 & CPV \\
Schiff moment $\boldsymbol{S}$ & $\checkmark$ & $\checkmark$ & 1/2 & CPV \\
Mag.\ quadrupole $\boldsymbol{M}$ & $\checkmark$ & $\checkmark$ & 1 & CPV \\
\end{tabular}
\end{ruledtabular}
\end{table}

\subsubsection{Discrete symmetries and the electromagnetic expansion}

In the electromagnetic expansion, the existence of a non-zero nuclear moment can indicate the violation of a given fundamental symmetry of nature, which can allow us to better understand the physics of the SM, or explore new physics beyond it \cite{safronova2018,arrowsmith2023opportunities}. In Table~\ref{tab:moments} we present the nuclear moments discussed above, whether their presence indicates a violation of the parity ($P$) and/or time reversal ($T$) symmetry, the minimal nuclear spin required for the moment to be non-zero, according to angular momentum selection rules, as well as the intranuclear origin of the respective moment.

\subsection{The weak nucleus--electron interaction: multipole expansion}
\label{sec:W_expansion}

In the preceding section, we expressed the electromagnetic interaction between the nucleus and the electrons as a multipole expansion of the nuclear charge density $\rho_N(\mathbf{r})$ and current density $\mathbf{j}_N(\mathbf{r})$ coupled to the electronic electromagnetic fields at the nuclear site. Even if this interaction is P- and T-conserving, it is still able to reveal P- and T-violations manifested inside the nucleus.  

We now turn to the weak (P-violating) interaction between the electrons and the nucleons, mediated by a $Z^0$-boson exchange.  This interaction admits a parallel multipole expansion to the one above, but with the nuclear \emph{weak-charge} and \emph{weak-current} replacing the electromagnetic ones, and with different electronic operators replacing the electromagnetic fields and their derivatives.

\subsubsection{The nuclear weak densities}
\label{sec:weak_densities}

At energies far below the $Z^0$ boson mass ($m_Z \approx 91$~GeV), the weak interaction between an electron and a nucleon is a contact interaction proportional to the Fermi constant $\GF$.  The relevant parity-violating, time-reversal-conserving ($P$-odd, $T$-even) electron-nucleon interaction Hamiltonian is \cite{langacker1995precision,khriplovich1980parity}:

\begin{equation}
\begin{aligned}
H_\text{PV}
= \frac{G_F}{2\sqrt{2}} \Big\{&
\underbrace{-C_1\, \bar{N}\gamma^\mu N}_{\text{nucleon vector}}
\underbrace{\bar{e}\gamma_\mu\gamma_5 e}_{\text{electron axial}}
\\
&+
\underbrace{C_2\, \bar{e}\gamma^\mu e}_{\text{electron vector}}
\underbrace{\bar{N}\gamma_\mu\gamma_5 N}_{\text{nucleon axial}}
\Big\}\,.
\end{aligned}
\label{eq:HPV_relativistic}
\end{equation}
Here $N$ and $e$ denote the nucleon and electron Dirac field operators,
$\gamma^\mu$ and $\gamma_5$ are Dirac matrices, and $C_1$ and $C_2$ are
dimensionless coupling constants that depend on the nucleon species
($N=p,n$), with tree-level Standard Model values given in
Eqs.~(\ref{eq:C1_values}) and~(\ref{eq:C2_values}). If we take the non-relativistic limit for the nucleons, we can make the following simplifications \cite{khriplovich1980parity}:
\begin{align}
\bar{N}\gamma^0 N &\;\to\; N^\dagger N = \rho_N(\mathbf{r})\,, \nonumber \\
\bar{N}\gamma^i N &\;\to\; \frac{1}{2m_N}\, N^\dagger(\overleftarrow{\nabla} - \overrightarrow{\nabla})^i N \approx 0 \quad \text{(to leading order)}\,,\nonumber \\
\bar{N}\gamma^0\gamma_5 N &\;\to\; 0 \quad\text{(suppressed by $v_N/c$)}\,, \nonumber \\
\bar{N}\gamma^i\gamma_5 N &\;\to\; N^\dagger \sigma^i N = s_N^i(\mathbf{r})\,,
\label{eq:NR_nucleon}
\end{align}
where $\rho_N(\mathbf{r})$ is the nucleon number density and $s_N^i(\mathbf{r})$ is the nucleon spin density.  Summing over all protons and neutrons in the nucleus, we can define the nuclear weak-charge density, $\rho_W(\mathbf{r})$, and the weak-current density, $\mathbf{j}_W(\mathbf{r})$:

\paragraph{\textbf{The nuclear weak-charge density.}} Analogous to the electromagnetic charge density $\rho_c$, it can be expressed as \cite{safronova2018,khriplovich1980parity}:
\begin{equation}
\rho_W(\mathbf{r}) \equiv 2C_{1,p}\, \rho_p(\mathbf{r}) + 2C_{1,n}\, \rho_n(\mathbf{r}),
\label{eq:rhoW}
\end{equation}
where $\rho_p(\mathbf{r})$ and $\rho_n(\mathbf{r})$ are the proton and neutron number densities (each normalized to $Z$ and $N$ respectively), and the coupling constants at tree level in the Standard Model are:
\begin{align}
C_{1,p} &= \tfrac{1}{2}(1 - 4\sin^2\thetaW) \approx 0.04\,, \nonumber \\
C_{1,n} &\approx -\tfrac{1}{2}\,.
\label{eq:C1_values}
\end{align}
The nuclear weak-charge density plays the same role in the weak expansion as $\rho_c(\mathbf{r})$ plays in the electromagnetic expansion.  The main difference being the coupling constants. While the electromagnetic charge density weights protons with charge $+e$ and neutrons with charge $0$, the weak-charge density weights protons with $C_{1,p} \approx 0.04$ (suppressed given that $\sin^2\thetaW \approx 1/4$) and neutrons with $C_{1,n} \approx -1/2$ \cite{safronova2018,navas2024review}.  Therefore, the weak interaction is {predominantly a probe of the neutron distribution}, complementary to the electromagnetic interaction, which probes only protons.

\paragraph{\textbf{The nuclear weak-current density.}} Analogous to the electromagnetic current density $\mathbf{j}_N$, it can be expressed as \cite{safronova2018,khriplovich1980parity}:
\begin{equation}
\mathbf{j}_W(\mathbf{r}) \equiv C_{2,p}\, \mathbf{s}_p(\mathbf{r}) + C_{2,n}\, \mathbf{s}_n(\mathbf{r}),
\label{eq:jW}
\end{equation}
where $\mathbf{s}_p(\mathbf{r})$ and $\mathbf{s}_n(\mathbf{r})$ are the proton and neutron spin densities, and:
\begin{align}
C_{2,p} &= g_A\,\tfrac{1}{2}(1 - 4\sin^2\thetaW) \approx 0.05\,, \nonumber \\
C_{2,n} &\approx -g_A\,\tfrac{1}{2}(1 - 4\sin^2\thetaW) \approx -0.05\,,
\label{eq:C2_values}
\end{align}
with $g_A \approx 1.27$, the nucleon axial coupling constant.  This density is the weak analog of $\mathbf{j}_N(\mathbf{r})$. While the electromagnetic current density describes the flow of electric charge and magnetization inside the nucleus, giving rise to the nuclear electromagnetic moments, the weak-current density describes the distribution of nucleon spins weighted by their weak axial couplings.  The rank-1 moment of $\mathbf{j}_W$ gives the NSD weak interaction, which has the same operator structure as the anapole interaction but a different physical origin (see Sec.~\ref{sec:anapole}).

\subsubsection{The electronic weak-field operators}
\label{sec:electronic_weak}

In the electromagnetic case, the nuclear densities couple to the electromagnetic fields produced by the electron: $\rho_c$ couples to the scalar potential $\Phi_e$, and $\mathbf{j}_N$ couples to the vector potential $\mathbf{A}_e$.  In the weak case, the nuclear weak densities couple to different electronic operators.  From Eq.~(\ref{eq:HPV_relativistic}):

\paragraph{\textbf{\texorpdfstring{$\rho_W$}{rhoW} couples to the electronic axial-vector density:}}
\begin{equation}
\Phi_W^{(e)}(\mathbf{r}) \equiv \frac{\GF}{2\sqrt{2}}\, \psi_e^\dagger(\mathbf{r})\, \gamma_5\, \psi_e(\mathbf{r})\,.
\label{eq:PhiW}
\end{equation}
This is the ``weak scalar potential'' produced by the electrons at position $\mathbf{r}$.  It is odd under parity, in contrast to the electromagnetic $\Phi_e$, which is even.

\paragraph{\textbf{\texorpdfstring{$\mathbf{j}_W$}{jW} couples to the electronic vector current:}}
\begin{equation}
\mathbf{A}_W^{(e)}(\mathbf{r}) \equiv \frac{\GF}{2\sqrt{2}}\, \psi_e^\dagger(\mathbf{r})\, \alphav\, \psi_e(\mathbf{r})\,,
\label{eq:AW}
\end{equation}
where $\alphav = \gamma^0\bm{\gamma}$ is the electron velocity operator.  This is the ``weak vector potential'', a vector even under parity, analogous to the electromagnetic $\mathbf{A}_e$.

With these identifications, the full weak interaction between the electrons and a single nucleus can be written in a form that is structurally identical to the electromagnetic case (Eq.~(\ref{eq:HeN_EM})):
\begin{equation}
H_{eN}^{(\text{W})} = \int \dd^3 r\, \rho_W(\mathbf{r})\, \Phi_W^{(e)}(\mathbf{r}) - \int \dd^3 r\, \mathbf{j}_W(\mathbf{r}) \cdot \mathbf{A}_W^{(e)}(\mathbf{r}).
\label{eq:HW_master}
\end{equation}
The main differences with respect to the electromagnetic interaction are: (i)~the nuclear densities ($\rho_W$, $\mathbf{j}_W$) have different coupling constants from the electromagnetic ones ($\rho_c$, $\mathbf{j}_N$), reflecting the different $Z^0$-boson couplings to protons and neutrons; and (ii)~the electronic operators ($\Phi_W^{(e)}$, $\mathbf{A}_W^{(e)}$) have different parity transformation properties from their electromagnetic counterparts ($\Phi_e$, $\mathbf{A}_e$), reflecting the parity-violating nature of the weak force.

\subsubsection{The weak multipole expansion}
\label{sec:weak_multipole}

We now expand the electronic weak-field operators about the nuclear center $\mathbf{r} = 0$, as previously done for the electromagnetic fields:
\begin{align}
\Phi_W^{(e)}(\mathbf{r}) &= \Phi_W^{(e)}(0) + r_i\,\partial_i \Phi_W^{(e)}(0) + \tfrac{1}{2}\, r_i r_j\, \partial_i\partial_j \Phi_W^{(e)}(0) + \ldots\,, \nonumber \\[4pt]
A_{W,k}^{(e)}(\mathbf{r}) &= A_{W,k}^{(e)}(0) + r_i\,\partial_i A_{W,k}^{(e)}(0) + \ldots\,.
\label{eq:Taylor_W}
\end{align}
Substituting into Eq.~(\ref{eq:HW_master}) and collecting terms by multipole rank yields:
\begin{equation}
H_{eN}^{(\text{W})} = \sum_\lambda \Mn_W^{(\lambda)} \cdot \Fn_{W,e}^{(\lambda)},
\label{eq:weak_multipole_master}
\end{equation}
where $\Mn_W^{(\lambda)}$ is the rank-$\lambda$ nuclear \emph{weak multipole moment} and $\Fn_{W,e}^{(\lambda)}$ is the corresponding electronic weak-field operator evaluated at the nuclear center.  The explicit terms can be expressed as:
\begin{align}
H_{eN}^{(\text{W})} = \; & \underbrace{\QW \cdot \Phi_W^{(e)}(0)}_{\substack{\text{weak monopole} \\ \text{(nuclear weak charge)}}} \nonumber \\[6pt]
& + \underbrace{Q_{W,ij}\, \partial_i\partial_j \Phi_W^{(e)}(0)}_{\substack{\text{weak quadrupole}}} \nonumber \\[6pt]
& + \underbrace{\text{direct } \text{$C_2$} \text{ NSD} \ \text{interaction}}_{\substack{\text{weak-current dipole}}} \nonumber \\[6pt]
& + \ldots
\label{eq:weak_terms}
\end{align}
Below, we discuss each of these terms.

\subsubsection{The nuclear weak charge (weak monopole)}

This is the weak analog of $Ze \cdot \Phi_e(0) $ plus finite-size correction. For the weak interaction, the zeroth-order term in the expansion is:
\begin{equation}
H_{W0} = \left(\int \dd^3 r\, \rho_W(\mathbf{r})\right) \Phi_W^{(e)}(0) \equiv \QW \cdot \Phi_W^{(e)}(0)\,,
\label{eq:HW0}
\end{equation}
where the {nuclear weak charge} is the weak monopole \cite{safronova2018}:
\begin{equation}
\QW = \int \dd^3 r\, \rho_W(\mathbf{r}) = 2\left(ZC_{1,p} + NC_{1,n}\right).
\label{eq:QW_def}
\end{equation}
In analogy with $\int \rho_c\, \dd^3 r = Ze$ for the electromagnetic case, the integral of the weak-charge density gives the total weak charge of the nucleus.  Using the SM values of $C_{1,p}$ and $C_{1,n}$:
\begin{equation}
\QW \approx Z(1 - 4\sin^2\thetaW) - N \approx -N\,.
\label{eq:QW_approx}
\end{equation}
Writing out the electronic operator explicitly, we get:
\begin{equation}
H_{W0} = \underbrace{\QW}_{\substack{\text{nuclear}\\\text{(weak charge)}}} \times \underbrace{\frac{\GF}{2\sqrt{2}}\gamma_5\,\rho_\text{nuc}(r)}_{\substack{\text{electronic}\\\text{(weak ``field'' at nucleus)}}}\,,
\label{eq:Q_W_formula}
\end{equation}
where $\rho_\text{nuc}(r)$ is the nuclear density appearing because the contact interaction probes the electronic wave function inside the nuclear volume.

\paragraph{\textbf{Effect on energy levels.}}  $H_{W0}$ mixes electronic states of {opposite parity}.  For two such parity eigenstates states $|\psi_+\rangle$ and $|\psi_-\rangle$, we therefore obtain \cite{safronova2018}:
\begin{equation}
\langle \psi_- | H_{W0} | \psi_+ \rangle = \underbrace{k_{\text{PV}}}_{\text{electronic}} \cdot \underbrace{\QW}_{\text{nuclear}}\,,
\label{eq:MPV}
\end{equation}
where $k_{\text{PV}} = \frac{\GF}{2\sqrt{2}} \int \psi_-^\dagger \gamma_5 \psi_+ \rho_\text{nuc}(r)\,\dd^3 r$.  This is a nuclear-spin-independent (NSI) parity-violating effect.  Since the integrand is non-zero only inside the nuclear volume ($r \lesssim R$), and the relevant $s_{1/2}$ and $p_{1/2}$ electron densities at the nucleus scale as $Z^3$, the electronic factor scales as $k_{\text{PV}} \propto Z^3$ \cite{safronova2018,arrowsmith2023opportunities}.

\paragraph{\textbf{The weak finite-size correction: the neutron skin.}}
Just as the electromagnetic monopole has a finite-size correction proportional to $\langle r^2 \rangle$, the weak monopole also has a finite-size correction related to the {weak-charge distribution}, in particular to the fact that the proton, $\rho_p(\mathbf{r})$, and neutron $\rho_n(\mathbf{r})$ densities are not identical inside a nucleus. Accounting for this difference, the weak charge gets a correction relative to the above, given by \cite{fortson1990nuclear,pollock1992atomic}:

\begin{equation}
\delta Q_W = 2NC_{1,n}\left(1-\frac{q_n}{q_p}\right),
\label{eq:HW0_FS}
\end{equation}
where $q_{n,p} \equiv \int\rho_{n,p}(r)f(r)d^3r$, and $f(r)$ is the radial electronic wavefunction inside the nucleus. This correction can be related to the  difference between the neutron and proton root-mean-square radii, also known as the neutron skin:

\begin{equation}
\Delta R_{np} = \sqrt{\langle r^2\rangle_n} - \sqrt{\langle r^2\rangle_p}\,.
\label{eq:nskin}
\end{equation}
This is the weak analog of the electromagnetic charge radius. While the isotope shift probes $\delta\langle r^2\rangle$ (the change in the {proton} distribution between isotopes), the weak finite-size correction probes $\Delta R_{np}$ (the difference between the {neutron} and proton distributions within a single isotope).  A precision measurement of the NSI PV amplitude can therefore provide information about the neutron density distribution inside the nucleus \cite{fortson1990nuclear,pollock1992atomic,arrowsmith2023opportunities}.

\subsubsection{The direct NSD weak interaction}

This is the weak analog of the magnetic dipole interaction. For the weak interaction, the rank-1 term from the weak-current density $\mathbf{j}_W$ gives the  nuclear-spin-dependent weak interaction \cite{safronova2018,arrowsmith2023opportunities}:
\begin{equation}
H_{W1} = \frac{\GF}{\sqrt{2}}\left[ZC_{2,p}\,\rho_p(r)\,\alphav \cdot \sigmav_p + NC_{2,n}\,\rho_n(r)\,\alphav \cdot \sigmav_n\right].
\label{eq:HW1_direct}
\end{equation}
This originates from the $Z^0$-exchange interaction between the electron and the nucleus. The electron vector current ($\bar{e}\gamma^\mu e$) couples directly to the nucleon axial current ($\bar{N}\gamma_\mu\gamma_5 N$).  Its operator structure, $\alphav \cdot \mathbf{I}\,\rho_\text{nuc}(r)$, is similar to that of the anapole interaction (Eq.~(\ref{eq:Hanapole_factor})), but its origin is different as the anapole is an electromagnetic coupling to a PV nuclear current, while this term is a weak coupling to the nuclear spin.

\begin{table*}[t]
\centering
\caption{Electromagnetic and weak nuclear quantities and their corresponding electronic fields by rank.}
\begin{tabular}{ccccc}
\hline\hline
\textbf{Rank} & \textbf{EM nuclear quantity} & \textbf{EM electronic field} & \textbf{Weak nuclear quantity} & \textbf{Weak electronic field} \\
\hline
$\lambda = 0$ & $\int \rho_c = Ze$ & $\Phi_e(0)$ & $\int \rho_W = Q_W$ & $\Phi_W^{(e)}(0)$ \\[2pt]
$\lambda = 0$ & $\langle r^2\rangle_\text{ch}$ & $\nabla^2\Phi_e(0)$ & $\Delta R_{np}$ & $\nabla^2\Phi_W^{(e)}(0)$ \\[2pt]
$\lambda = 1$ & $\bm{\mu}$ & $\mathbf{B}_e(0)$ & $C_2$ NSD interaction & $\mathbf{A}_W^{(e)}(0)$ \\[2pt]
$\lambda = 2$ & $Q_{ij}$ & $\nabla_i E_{e,j}(0)$ & $Q_{W,ij}^{(2)}$ & $\nabla_i\nabla_j\Phi_W^{(e)}(0)$ \\[2pt]
\hline\hline
\end{tabular}
\label{tab:em-weak-quantities}
\end{table*}

\subsubsection{The nuclear weak quadrupole moment}

This is the weak analog of the electric quadrupole interaction. In the weak interaction, the rank-2 term in the expansion of $\rho_W(\mathbf{r})$ gives the nuclear weak quadrupole moment \cite{flambaum2016enhancing,flambaum2017effect,lackenby2018weak}.  Expanding the weak-charge density in spherical harmonics:
\begin{equation}
\rho_W(\mathbf{r}) = \rho_W^{(0)}(r) + \sum_{q=-2}^{2} (-1)^q\, \rho_W^{(2,q)}(r)\, Y_{2,q}(\hat{r}) + \ldots
\label{eq:rhoW_multipole}
\end{equation}
where $\rho_W^{(0)}$ is the monopole (giving $\QW$) and $\rho_W^{(2,q)}$ are the quadrupole components:
\begin{equation}
\rho_W^{(2,q)}(r) = C_{1,p}\, \rho_p^{(2,q)}(r) + C_{1,n}\, \rho_n^{(2,q)}(r)\,.
\label{eq:rhoW2}
\end{equation}
The resulting contribution to the Hamiltonian is:
\begin{equation}
H_{W2} = Q_{W,ij}\, \nabla_i\nabla_j \Phi_W^{(e)}(0),
\label{eq:HW2}
\end{equation}
where the nuclear weak quadrupole moment is:
\begin{equation}
Q_{W,ij} = \int \dd^3 r\, \rho_W(\mathbf{r})\left(3r_i r_j - r^2 \delta_{ij}\right).
\label{eq:QW2_def}
\end{equation}
The electromagnetic quadrupole moment $Q_{ij}$ probes the quadrupole deformation of the {proton} distribution.  The weak quadrupole moment $Q_{W,ij}$ probes the quadrupole deformation of the {neutron} distribution (since $C_{1,n} \gg C_{1,p}$).

A summary comparing the electromagnetic and weak multipole expansions is shown in Table \ref{tab:em-weak-quantities}.

\subsubsection{Enhancement of PV effects in molecules}
\label{sec:mol_enhance}

The factorized structure of Eq.~(\ref{eq:weak_multipole_master}) shows that the sensitivity of a given experiment to a nuclear PV moment is determined by the electronic factor $\Fn_{\mathcal{S},e}^{(\lambda)}$.  In atoms, the PV-induced mixing of opposite-parity states is suppressed by the large energy gap $\Delta E \sim 1$~eV between electronic states of opposite parity. In molecules, this suppression is dramatically reduced.  Opposite-parity states exist with much smaller energy gaps, such as \cite{safronova2018,arrowsmith2023opportunities,brown2003rotational,hutzler2020polyatomic}:
\begin{itemize}
\item \textbf{Rotational doublets:} $\Delta E_\text{rot} \sim B_\text{rot} \sim 10^{-5}$~eV, set by the rotational constant.
\item \textbf{$\Omega$-doublets:} $\Delta E_\Omega \lesssim 10^{-6}$~eV, arising in electronic states with the total electron angular momentum projection along the molecular axis $\Omega > 0$.
\item \textbf{$\ell$-doublets:} $\Delta E_l \lesssim 10^{-7}$~eV, closely spaced levels of opposite parity present, for example, in linear polyatomic molecules.
\end{itemize}
Moreover, molecular levels can be brought even closer together by applying an external field of the right amplitude \cite{altuntacs2018demonstration,altuntacs2018measuring,demille2008using,karthein2023electroweak}. Since the effective PV mixing scales as $\Fn_{\mathcal{S},e}^{(\lambda)} \propto 1/\Delta E$, the reduction in $\Delta E$ by factors of $10^5$--$10^7$ translates directly into a corresponding enhancement of the electronic factor, and thus of the observable PV effects \cite{safronova2018}.  

\subsection{CP violating nucleus--electron interactions: multipole expansion}
\label{sec:CPV_expansion}

The final class of electron-nucleon (eN) interactions in the Hamiltonian (\ref{eq:H_total}) is those that violate both $P-$ and $T-$symmetry (equivalently, CP). While the Schiff moment or the MQM are manifestations of CP violation inside the nucleus, they are interacting electromagnetically with the electron. The interactions considered in this section are CP violating interactions between the nucleus and the electron cloud. Given our experimental sensitivities, any observed such non-zero effect in an atom or molecule would be a clear sign of a new fundamental CP violating eN interaction \cite{safronova2018,arrowsmith2023opportunities,chupp2019electric}. The derivations required to obtain the terms relevant for experimental searches closely mirror the ones for the $Z_0$-boson interactions above, and we will only present the main results below.

\subsubsection{CP violating eN scalar interaction}
\label{sub:cp_e_N}

This interaction is the CP violating equivalent of the first term in Eq. \ref{eq:HPV_relativistic} \cite{khriplovich1980parity,khriplovich1991}. In the limit of non-relativistic nucleons, the relevant Hamiltonian becomes:

\begin{equation}
H_S = \frac{iG_F}{\sqrt{2}}\, C_S\, A\, \rho(r)\, \gamma^0\gamma_5\,,
\label{eq:HS}
\end{equation}
where $C_S$ is the scalar electron-nucleon coupling constant and $A$ is the mass number. By comparison with Eq. \ref{eq:Q_W_formula}, the $C_S A$ factor can be thought of as a CP violating monopole nuclear charge \cite{safronova2018,pospelov2005electric,chupp2019electric}.  This interaction is one of the leading sources of CP violating effects in paramagnetic systems, together with the electron EDM \cite{safronova2018,acme2018improved,roussy2023improved} (see Sec. \ref{sec:schiff_mom}). Therefore, from a single experiment, only the combination $d_e + k\,C_S$ can be constrained, with a system-dependent coefficient
$k$, and disentangling the two sources requires combining measurements across
species with different $k$
\cite{chupp2019electric,ema2022}. Within the SM, the dominant contribution for such searches arises through the nucleus, via CP odd hadronic interactions that generate $C_S$ through two-photon exchange between the electrons and the nucleons \cite{ema2022}.

The PT violating pion--nucleon couplings that generate nuclear Schiff moments (Sec. \ref{sec:schiff_mom})) can also induce $C_S$, through the CP odd nuclear scalar polarizability arising from
two-photon exchange between the electrons and the nucleus
\cite{Mulder2025ThetaParamagnetic}. The relation between the two observables is, however, model-dependent. It can be obtained by assuming a specific hadronic source of CP violation and computing both matrix elements.

\subsubsection{CP violating eN tensor interaction}

This interaction is the CP violating equivalent of the second term in Eq. \ref{eq:HPV_relativistic} \cite{khriplovich1980parity,khriplovich1991}, and in the non-relativistic limit, the relevant Hamiltonian is:

\begin{equation}
H_T = \frac{2iG_F}{\sqrt{2}}\, C_T\, \rho(r)\, \mathbf{I} \cdot \bm{\gamma}\,,
\label{eq:HT}
\end{equation}
where $C_T$ is the tensor coupling constant, $\mathbf{I}$ is the nuclear spin, and $\bm{\gamma}$ is the vector of Dirac matrices acting on the electron. Once again, this term has an analogous form with the corresponding weak interaction Hamiltonian, Eq. \ref{eq:HW1_direct}, reflecting a nuclear spin-dependent, CP violating, eN interaction. This term is commonly suppressed relative to $H_S$  but can be significant in systems with non-zero nuclear spin \cite{safronova2018,chupp2019electric,graner2016reduced}.

\subsubsection{Molecular amplification of CP violating effects}

The advantage of molecules for CP violation searches can be understood quantitatively through the concept of \emph{molecular polarization}.  Consider the system's two opposite-parity levels $|+\rangle$ and $|-\rangle$, separated by an energy gap $\Delta_\pm$, in the presence of an external electric field $\mathcal{E}$.  The Stark Hamiltonian $H_\text{Stark} = -\mathbf{D}\cdot\bm{\mathcal{E}}$ (where $\mathbf{D}$ is the molecular dipole moment operator) mixes these levels.  If a CP violating interaction with matrix element $\langle +|H_\text{CPV}|-\rangle = \delta_\text{CPV}$ is present, the energy shift of the dressed levels contains a term \cite{sandars1966enhancement,safronova2018,arrowsmith2023opportunities}:
\begin{equation}
\Delta E_\text{CPV} = \frac{\delta_\text{CPV}\, D\mathcal{E}/ \Delta_\pm}{\sqrt{1 + (D\mathcal{E}/\Delta_\pm)^2}} \xrightarrow{D\mathcal{E} \gg \Delta_\pm} \delta_\text{CPV}\,.
\label{eq:mol_enhance_cpv}
\end{equation}
In an atom, where opposite-parity electronic states have $\Delta_\pm \sim 1$~eV and laboratory fields give $D\mathcal{E} \ll \Delta_\pm$, the CP violating shift is suppressed by the small ratio $D\mathcal{E}/\Delta_\pm$.  In a molecule, where rotational or $\Omega$-doublet states have $\Delta_\pm \sim 10^{-5}$--$10^{-6}$~eV, even a modest laboratory field achieves $D\mathcal{E} \gg \Delta_\pm$, saturating the molecular polarization and realizing the complete CP violating shift $\delta_\text{CPV}$ \cite{arrowsmith2023opportunities,safronova2018}.  This is the essential molecular advantage as the molecule acts as a built-in amplifier that brings the internal CP violating matrix element directly to the laboratory observable. This amplification applies both to direct CP violating eN interaction and to CP violation effects induced by nuclear moments.

\subsection{Section summary}
\label{sec:summary}

The development above shows that the entire program of probing nuclear structure and fundamental interactions with atoms and molecules can be understood through a single organizing principle of the multipole expansion of the nucleus--electron interaction.  The full Hamiltonian (Eq.~(\ref{eq:H_total})) decomposes into three parallel expansions of the electromagnetic, weak, and CP violating effects, each generating a sequence of factorized terms as:
\begin{equation}
H_{eN} = \sum_{\lambda,\, \mathcal{S}} \underbrace{\Mn_\mathcal{S}^{(\lambda)}}_{\text{nuclear moment}} \cdot \underbrace{\Fn_{\mathcal{S},e}^{(\lambda)}}_{\text{electronic field operator}}\,,
\label{eq:master_all}
\end{equation}
where $\mathcal{S} \in \{\text{EM}, \text{W}, \text{CPV}\}$ labels the symmetry sector and $\lambda$ labels the multipole rank.  Each term produces a distinct, experimentally identifiable signature in the atomic or molecular spectrum: a level shift (charge radius), a splitting pattern (hyperfine structure), a parity-forbidden transition amplitude (PV), or an energy shift linear in an applied electric field (EDM).

The factorization into nuclear $\times$ electronic parts allows the extraction of a specific nuclear moment $\Mn^{(\lambda)}$, if the electronic factor $\Fn_e^{(\lambda)}$ can be computed or extracted from an independent experiment. Conversely, if the nuclear moment is known (from nuclear theory or independent measurement), the measurement constrains the electronic factor and, through it, the fundamental coupling constants of the Standard Model or its extensions.
Different systems (different atoms, different molecules, different isotopes) provide different electronic factors for the same nuclear moment, enabling cross-checks, redundancy, and the disentangling of multiple contributing effects.

The challenge, and the opportunity, lie in the precision with which both factors
can be determined. Modern laser spectroscopy routinely achieves the resolution
needed to measure energy shifts induced by nuclear and particle-physics
properties, often with relative precision at the $5\%$ level or better. Atomic
and molecular theory has reached percent-level accuracy, or better, for the
electronic factors in many systems. The limiting uncertainty increasingly lies
in the nuclear theory required to compute properties such as charge radii,
electromagnetic moments, Schiff moments, and anapole moments. This is the
subject of the following section, Sec.~\ref{sec:nuclear_theory}.

\section{Nuclear theory for electroweak observables}
\label{sec:nuclear_theory}

The multipole expansion presented in Sec.~\ref{sec:ch2} shows that every atomic and molecular observable discussed in this review factorizes into a nuclear moment $\Mn^{(\lambda)}$ multiplied by an electronic field operator $\Fn_e^{(\lambda)}$.  While the electronic factors are the domain of atomic and molecular theory, the nuclear moments such as charge radii, magnetic dipole and electric quadrupole moments, weak charges, anapole moments, and Schiff moments must be computed within nuclear structure theory. Quantifying and reducing the uncertainties of nuclear calculations are central
challenges for the interpretation of precision measurements.

Hence, the interplay between experimental measurements and theory is central to this program.  On one hand, precise measurements of nuclear electromagnetic observables (Secs.~\ref{sec:charge_radii} and~\ref{sec:EM_moments}) provide benchmarks to validate and improve nuclear theory calculations.  On the other hand, nuclear theory is required to interpret experiments that probe fundamental symmetries---parity violation (Sec.~\ref{sec:PV}) and CP violation (Sec.~\ref{sec:CPV_nuclear})---and to connect the measured atomic and molecular signals to the potential underlying BSM physics through the EFT matching chain of Eq.~(\ref{eq:EFT_chain}).

\begin{figure*}[t]
    \centering
    \includegraphics[width=\textwidth]{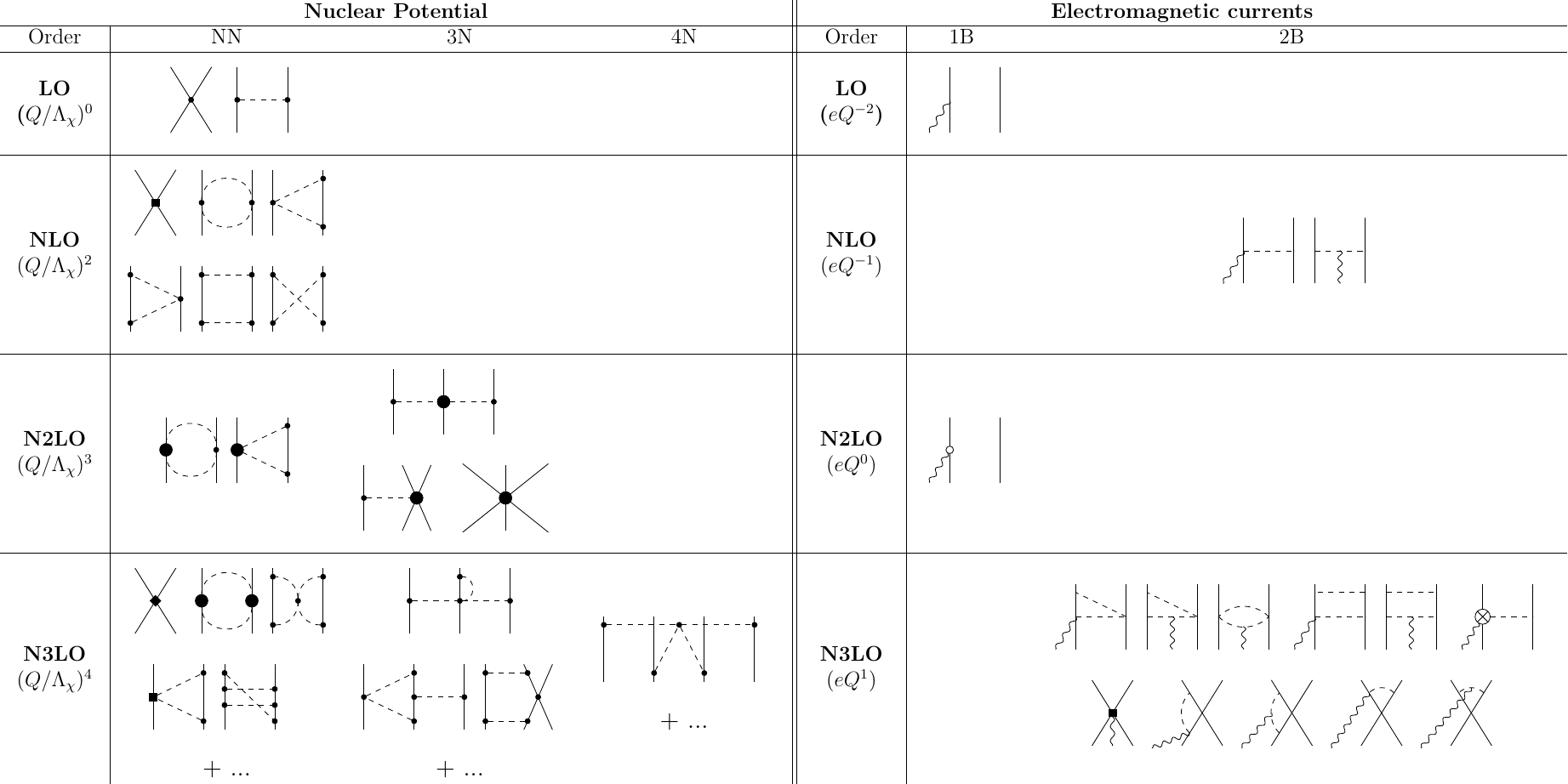}
    \caption{Hierarchy of contributions to the nuclear force in chiral EFT, showing two-, three-, and four-nucleon forces up to N$^3$LO and the corresponding one- and two-body electromagnetic currents.  Solid lines represent nucleons; dashed lines, pions; wavy lines, photons; dots, squares, crossed squares, and diamonds represent vertices proportional to the low-energy constants (LECs) of the theory. The empty circle represents relativistic corrections to the one-body electromagnetic currents. See Refs.~\cite{machleidt2011chiral,pastore2009electromagnetic} for details.}
    \label{fig:2.2.1-xEFT}
\end{figure*}

This section surveys the nuclear theory methods used to compute these quantities.  While a comprehensive treatment of modern approaches in nuclear structure theory is beyond the scope of this review, we refer the reader to Refs.~\cite{bender2003self,caurier2005theshellmodel,epelbaum2009modern,schunck2019energy,hergert2020guided,ekstrom2023ab} for recent in-depth discussions. Here we focus on the aspects most directly relevant to the electroweak observables introduced in Sec.~\ref{sec:ch2}.

\subsection{Nuclear forces from chiral effective field theory}
\label{sec:chiEFT}

At the most fundamental level, the forces between nucleons inside the nucleus originate from quantum chromodynamics (QCD).  However, the strong coupling constant $\alpha_s$ becomes large at the energy scales relevant for nuclear structure ($Q \lesssim 1$~GeV), making perturbative QCD inapplicable~\cite{politzer1973reliable,weinberg1991effective}.  The relevant degrees of freedom at these scales are not quarks and gluons but rather nucleons and pions---the Goldstone bosons of the spontaneously broken chiral symmetry of QCD.

Chiral effective field theory (chiral EFT) provides an appropriate systematic framework for constructing nuclear forces at low-energy scales in terms of these hadronic degrees of freedom~\cite{weinberg1991effective,weinberg1992threebody,epelbaum2009modern,machleidt2011chiral,hammer2020nuclear}.  Following Weinberg~\cite{weinberg1991effective,weinberg1996quantum}, one constructs the most general effective Lagrangian consistent with all symmetries of QCD---in particular chiral symmetry, parity, and time reversal---organized as an expansion in powers of $Q/\Lambda_\chi$, where $Q \sim m_\pi \sim 130$~MeV is the typical momentum scale and $\Lambda_\chi \sim 700$--$1000$~MeV is the breakdown scale of the theory, where heavier mesons or quark and gluon degrees of freedom would become relevant.  Physics at energies above $\Lambda_\chi$ is not explicitly resolved but is instead absorbed into the coefficients of so-called contact interactions.

This expansion generates a hierarchy of contributions to the nuclear force (Fig.~\ref{fig:2.2.1-xEFT}): \textbf{i. Leading order (LO):} One-pion exchange plus two short-range contact interactions, reproducing the long-range part of the nucleon-nucleon potential and the $^1S_0$ and $^3S_1$ scattering lengths. \textbf{ii.~Next-to-leading order (NLO):} Two-pion exchange with subleading vertices, plus additional contact terms. Seven additional low-energy constants (LECs) appear, encoding short-distance physics not resolved by the effective theory. \textbf{iii.~N$^2$LO:} The first three-nucleon forces (3NFs) appear, with two new LECs ($c_D$, $c_E$) that must be fitted to some data (typically few-body systems such as the $^3$H binding energy and the $^4$He charge radius or $n$-$d$ scattering length). \textbf{iv.~N$^3$LO and beyond:} Further refinements of the two- and three-nucleon forces, plus the first four-nucleon forces (which have been shown to be negligible in infinite nuclear matter calculations).

The LECs of the theory encode the short-distance physics that is not explicitly resolved in the effective framework.  In principle, they could be calculated from, e.g., lattice QCD through matching~\cite{eliyahu2020extrapolating,detmold2023constraint}, though in practice, the LECs are fit to experimental data: typically nucleon-nucleon scattering phase shifts and properties of light nuclei~\cite{epelbaum2020high,ekstrom2019global,piarulli2020local,arriola2020nn}.  Several families of chiral interactions have been developed, differing in their choices of regularization scheme, cutoff scale, fitting to properties of heavier nuclei, and chiral order~\cite{entem2003accurate,ekstrom2015accurate,jiang2020accurate,soma2020novel,arthuis2024neutron}.  The existence of multiple well-calibrated interactions is a feature rather than a limitation: it provides a systematic way to estimate the theoretical uncertainty associated with the nuclear force, as discussed in Sec.~\ref{sec:UQ}.

A distinctive advantage of chiral EFT for the program of this review is that it provides not only the nuclear forces but also the {electroweak current operators} (see Fig.~\ref{fig:2.2.1-xEFT}) and {parity-violating nuclear potentials} within the same consistent framework.  Since electroweak currents are obtained by gauging derivatives in the chiral Lagrangian, they depend on the same LECs as the nuclear forces~\cite{krebs2020nuclear}, ensuring consistency between the Hamiltonian used to compute the nuclear wave function and the operators used to evaluate the nuclear moments $\Mn^{(\lambda)}$. This is illustrated in Fig.~\ref{fig:2.2.1-xEFT} where the contact LECs appearing at NLO in the nuclear potential (square vertex) also appear in the N$^3$LO electroweak currents.  Similarly, parity-violating time-conserving (PVTC) and parity-violating time-violating (PVTV) nuclear forces~\cite{devries2020parity} can be derived within the same chiral expansion, as further discussed in Secs.~\ref{sec:anapole_theory} and~\ref{sec:schiff_theory}.

\subsection{Nuclear many-body methods}
\label{sec:many_body}

Given a nuclear Hamiltonian, the next step is to solve the many-body Schr\"odinger equation for any nucleus of interest.  This is a long-standing, formidable computational problem, as the dimension of the many-body Hilbert space grows combinatorially with nucleon number and the size of the single-particle basis. 
A variety of methods have been developed, which can be broadly divided by whether or not they can connect directly to the underlying nuclear forces and QCD via chiral EFT.

Phenomenological methods bypass the challenge of solving the nuclear many-body problem by adapting the problem to particular reduced-space models. As such, they cannot use nuclear interactions or currents constructed from chiral EFT, but rather represent the effects of nuclear forces in a limited region of the nuclear chart by fitting model parameters to experimental data. Thus, they are applicable very broadly across the nuclear chart but face significant difficulties in assessing their uncertainties, since they lack an explicit connection to the underlying QCD and electroweak sectors. 

To date, most calculations of electroweak matrix elements in heavy nuclei have been carried out using two such approaches: the nuclear shell model (NSM) and nuclear density functional theory (DFT). However, \textit{ab initio} methods, based on systematically improvable approximations to the nuclear many-body problem, have opened the way towards calculations directly connected to the underlying forces derived from chiral EFT.

\subsubsection{The nuclear shell model.}
\label{sec:NSM}
The NSM~\cite{caurier2005theshellmodel} exploits the existence of nuclear shell closures to reduce the many-body problem to a tractable size, where the Hilbert space is restricted to a set of {valence} orbitals---the partially filled shells above the chosen closed core. An effective interaction within this valence space is then constructed by fitting interaction matrix elements to experimental data, typically excitation energy levels and ground state energies in the given region.
The Hamiltonian is then diagonalized exactly within the valence space, while further phenomenological parameters are typically introduced in order to reproduce observables other than energies. 

The NSM has been applied extensively to compute electromagnetic moments, Gamow-Teller matrix elements, Schiff moments (in $^{199}$Hg and $^{129}$Xe~\cite{yanase2020largescale}), and anapole moments~\cite{haxton2001atomic}. While it has been a leading theoretical tool for decades, its main limitations are: (i)~the effective Hamiltonian is fitted, not derived, providing no systematic way to assess uncertainties; (ii)~the method relies on an inert core approximation, which does not allow for predictions of absolute quantities such as ground-state energies and radii; (iii)~the shell-model picture breaks down for strongly deformed nuclei, where many shells participate in the collective motion; and (iv)~for electroweak observables such as magnetic dipole ($M1$), electric quadrupole ($E2$), or Gamow-Teller transitions, additional parameters such as effective charges or quenching factors must be introduced to compensate for the missing contributions from excitations outside the valence space or operator deficiencies (e.g., neglected two-body currents), which results in additional inconsistencies between these observables and energies. 

\subsubsection{Nuclear density functional theory.}
\label{sec:dft}
Nuclear DFT~\cite{bender2003self,schunck2019energy} is based on the Hohenberg--Kohn theorem~\cite{hohenberg1964inhomogeneous}: the ground-state energy and density of a many-body system can, in principle, be obtained by minimizing a universal functional of the one-body density alone.  In practice, the energy density functional (EDF) is parameterized with approximately 10--20 adjustable parameters fitted to nuclear masses, radii, and other bulk properties. While, in theory, one ``true'' unique exact functional exists, in practice, many widely used approximations of this functional have been developed (Skyrme, Gogny, relativistic mean-field/covariant DFT), each with different parameterizations that have been optimized for different nuclear properties. Potentially connecting density functionals to underlying nuclear Hamiltonians is still an open challenge in the field, as it would allow for an \textit{ab initio} DFT method that could be connected to the underlying QCD~\cite{drut2010toward, colo2025nuclear}.

The computational cost of DFT scales gently with mass number, making it applicable throughout the nuclear chart, including the heavy and deformed systems of interest for CP violation searches. DFT has been applied to Schiff moments in octupole-deformed nuclei~\cite{dobaczewski2005nuclear,engel2003time}, neutron skins~\cite{reinhard2021nuclear}, and nuclear deformation properties \cite{karthein2023electroweak,Gus25}.  

\subsubsection{\textit{Ab initio} methods}
\label{sec:abinitio}

\begin{figure}[t]
    \centering
    \includegraphics[width=1\linewidth]{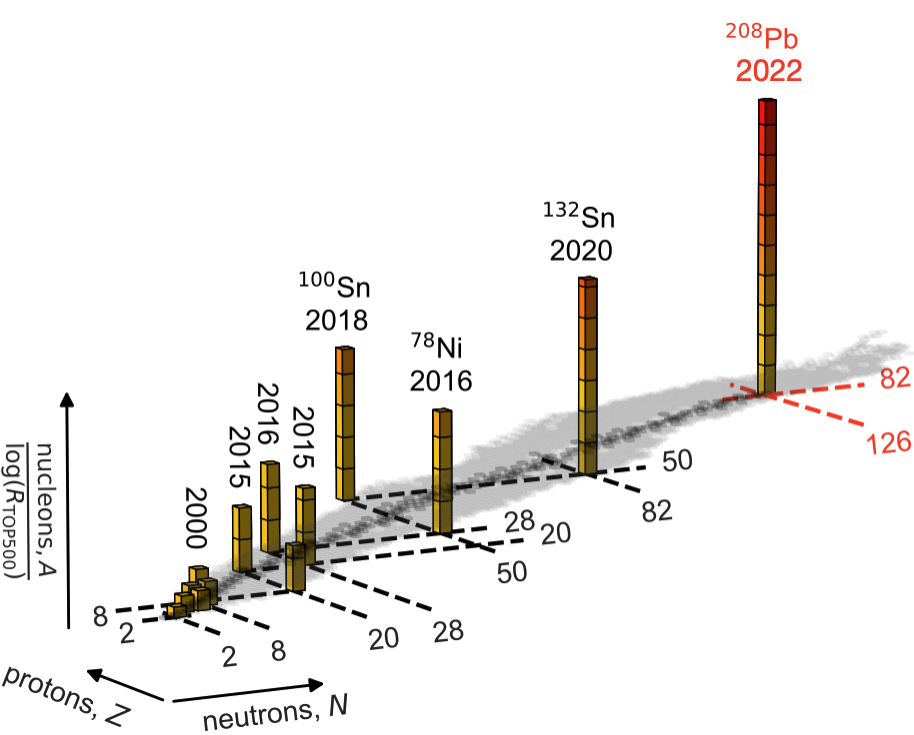}
    \caption{Progress of the reach of \textit{ab initio} theory over the past decade, where bars highlight the years of the first computations of doubly magic nuclei. 
    The height of each bar corresponds to the mass number $A$ divided by the log of the total computing power in flops ($s^{-1}$) in the given year. Figure reproduced from Ref.~\cite{hu2022ab}, under CC BY 4.0.}
    \label{fig:abinitio_theory}
\end{figure}

Over the past two decades, \textit{ab initio} nuclear structure theory has evolved from a framework that was primarily applicable only to light nuclei into a comprehensive, quantitative program capable of describing medium-mass and, increasingly, heavy nuclei to $^{208}$Pb and beyond \cite{hu2022ab}.
These approaches start from some given nuclear Hamiltonian (typically from chiral EFT) and then solve the many-body problem, with the methods discussed below, either quasi-exactly or nonperturbatively via systematically improvable approximations.  
Combined with the systematic expansion of chiral EFT, this paradigm provides a framework for complete uncertainty quantification---a clear qualitative advance over phenomenological methods. 
Particularly significant has been the rapid extension, illustrated in Fig.~\ref{fig:abinitio_theory}, of \textit{ab initio} methods into the tin region and doubly magic heavy nuclei such as $^{132}$Sn and even $^{208}$Pb~\cite{Miyagi2022Heavy,hu2022ab}, allowing access to most of the nuclear regions relevant for new physics searches.

\paragraph{No-Core Shell Model (NCSM).}
The NCSM~\cite{barrett2013abinitio,navratil2016unified} performs a direct diagonalization of the nuclear Hamiltonian in a harmonic-oscillator many-body basis, with all nucleons treated as active (no inert core).  The results converge to the exact solution as the basis is enlarged.  The NCSM has been used to compute charge radii, electromagnetic moments, Gamow-Teller matrix elements, anapole moments~\cite{hao2020nuclear}, and, very recently, the first \textit{ab initio} Schiff moment~\cite{ng2025nuclear}.  Its main limitation, however, is computational cost: the basis dimension grows steeply with mass number, restricting its reach to $A \lesssim 20$. Nevertheless, it often serves as an important benchmark for more approximate methods.

\paragraph{Coupled-cluster theory (CC).}
CC~\cite{hagen2014coupled} starts from a single Slater-determinant reference state and incorporates correlations through an exponential cluster operator: $|\Psi\rangle = e^T |\Phi_0\rangle$, where $T = T_1 + T_2 + \ldots$ creates particle-hole excitations.  Truncating at the singles-and-doubles level (CCSD) with perturbative triples (CCSD(T)) provides an accurate approximation for closed-shell and near-closed-shell nuclei up to $^{208}$Pb~\cite{hu2022ab} and beyond.  Promising extensions to deformed systems~\cite{hu2024abinitio} and electromagnetic observables of open-shell systems are under active development~\cite{bonaiti2024electromagnetic}.

\paragraph{In-Medium Similarity Renormalization Group (IMSRG).}
The IMSRG~\cite{tsukiyama2011inmedium,hergert2016theinmedium} uses a continuous unitary transformation to decouple the ground state (or a target subspace) from the rest of the Hilbert space.  Beyond the standard single-reference approach, two important variants exist:
\begin{itemize}
\item The \emph{valence-space} formulation (VS-IMSRG)~\cite{stroberg2017nucleus,stroberg2019nonempirical} constructs an effective valence-space Hamiltonian derived from the underlying chiral EFT, preserving the connection to QCD.  Unlike the phenomenological shell model, the effective interaction is derived, not fitted, and also allows for the consistent derivation of valence-space operators.
As such, the VS-IMSRG has been applied over most regions of the nuclear chart accessible to the standard shell model, including the $^{208}$Pb region~\cite{hu2022ab}.
In addition to energies, charge radii, electromagnetic moments/transitions, and beta-decay rates across the nuclear chart~\cite{stroberg2021abinitio,gysbers2019discrepancy,Li2026Beta}, it has extensive applications for searches for beyond-standard-model physics such as neutrinoless double beta decay, muon capture, and dark matter direct detection~\cite{Belley2021Ge,belley2024abinitio,Jokiniemi2023Muon,Hu22SDDM} with a recent first extension to parity-violating observables (PV-IMSRG, see Sec.~\ref{sec:anapole_theory}).
\item The \emph{In-Medium Generator Coordinate Method} (IM-GCM)~\cite{yao2020ab,belley2024abinitio} builds the reference state from a deformed mean-field calculation projected onto definite angular momentum.  This approach is well-suited to nuclei exhibiting strong deformation and applicable for including the octupole-deformed physics of interest for calculations of Schiff moments.
\end{itemize}

\paragraph{Other methods.}
We note that many other \textit{ab initio} methods exist, with some examples listed below.
Quantum Monte Carlo methods~\cite{carlson2015quantum} provide highly accurate results for light nuclei and have been used for electromagnetic moments and form factors~\cite{pastore2009electromagnetic, pastore2013quantum}.  
Nuclear lattice effective field theory~\cite{elhatisari2024wavefunction} places nucleons on a spacetime lattice and has recently achieved accurate charge radii with high-fidelity chiral interactions. Self-Consistent Green's Function~\cite{soma2020self} solves the Dyson equation to obtain a propagator from which nuclear properties can be obtained. While not yet explicitly applied to new physics searches, these methods potentially provide valuable new benchmarks for such calculations in medium and heavy systems.

\subsection{Electromagnetic operators and many-body currents}
\label{sec:EM_operators}

The nuclear moments $\Mn^{(\lambda)}$ introduced in Sec.~\ref{sec:ch2} are expectation values of specific operators in the nuclear ground state.  The form of these operators must be consistent with the Hamiltonian used to compute the wave function.  In chiral EFT, both the nuclear forces and the electromagnetic current operators are derived from the same Lagrangian, ensuring this consistency.

\subsubsection{The charge radius operator}

The charge radius operator, at leading order, includes one-body, nucleon finite-size, and relativistic corrections:
\begin{equation}
\hat{r}^2_\text{ch} = \frac{1}{Z}\sum_{i=1}^{A} \frac{1+\tau_3^{(i)}}{2}\, r_i^2 + r^2_{so} + r_p^2 + \frac{N}{Z}r_n^2 + \frac{3}{4M_p^2}\,,
\label{eq:r2_operator}
\end{equation}
where $r_p^2$ and $r_n^2$ are the proton and neutron charge radii, $r^2_{so}$ is the spin-orbit correction~\cite{heinz2025improved} and the Darwin-Foldy term $3/(4M_p^2)$ is a relativistic correction. 
Further corrections coming from two-body charge-density contributions from meson exchange currents have also been shown to be important for accurate charge radii in light nuclei~\cite{king2025formfactors}.

\subsubsection{The magnetic dipole operator}

The magnetic dipole moment (Sec.~\ref{sec:term_mu}, Fig.~\ref{fig:multipoles}a) is given by
\begin{equation}
    \boldsymbol{\mu} = -\frac{i}{2}\lim_{q\rightarrow0}\nabla_q \times \mathbf{j}(\mathbf{q})\,,
\end{equation}
where $\mathbf{j}(\mathbf{q})$ is the spatial electromagnetic current at momentum transfer $\mathbf{q}$.  These vector currents arise from the electroweak interactions inside the nucleus and also enter in beta decays and other rare electroweak processes in chiral EFT~\cite{krebs2020nuclear}, making them central to fundamental symmetry tests.

The dominant contribution is a one-body operator:
\begin{equation}
    \mu_{1B} = \mu_N\sum_i\left(g^l_i\, l_{i,z}+g^s_i\,\sigma_{i,z}\right),
    \label{eq:mu_1B}
\end{equation}
where $\mu_N = e\hbar/(2m_p)$ is the nuclear magneton, $g^l_i$ and $g^s_i$ are the orbital and spin $g$-factors ($g^l_p = 1$, $g^l_n = 0$, $g^s_p = 5.586$, $g^s_n = -3.826$~\cite{tiesinga2021codata}), and $l_{i,z}$ and $\sigma_{i,z}$ are the $z$-projections of the orbital angular momentum and spin of the $i$th nucleon.  This corresponds to the LO one-body operator in chiral EFT.

However, neither phenomenological nor \textit{ab initio} methods can reproduce experimental magnetic moments using only this one-body operator.  Chiral EFT shows that \emph{two-body currents} appear at the next chiral order (NLO) and must be included for consistency~\cite{pastore2009electromagnetic,krebs2020nuclear}.  At NLO, these two-body contributions come from parameter-free pion-exchange diagrams~\cite{seutin2023magnetic} and can be decomposed into intrinsic and Sachs (center-of-mass dependent) contributions:
\begin{equation}
    \mu_{2B} = \sum_{i<j} \left(\mu_{ij}^{\rm intr} + \mu^{\rm  Sachs}_{ij}\right),
\end{equation}
with
\begin{align}
    \mu_{ij}^{\rm intr} &= \mu_N(\boldsymbol{\tau}_i \times \boldsymbol{\tau}_j)_z\, V_{\rm{intr}}(\mathbf{r}_{ij})\,, \\
    \mu_{ij}^{\rm Sachs} &= \mu_N(\boldsymbol{\tau}_i \times \boldsymbol{\tau}_j)_z\, (\mathbf{R}_{ij}\times \mathbf{r}_{ij})_z\, V_{\rm{Sachs}}(r_{ij})\,,
\end{align}
where $\mathbf{r}_{ij} = \mathbf{r}_i - \mathbf{r}_j$ and $\mathbf{R}_{ij} = (\mathbf{r}_i+\mathbf{r}_j)/2$ are the relative and center-of-mass coordinates, and $\boldsymbol{\tau}_i$ is the isospin operator.  The intrinsic potential is
\begin{equation}
\begin{split}
     V_{\rm{intr}}(r_{ij}) = -\frac{g_A^2 m_\pi}{32\pi F^2_\pi}&\bigg\{\left(1+\frac{1}{m_\pi r_{ij}}\right)\left[(\boldsymbol{\sigma}_i\times\boldsymbol{\sigma}_j)\cdot \hat{r}_{ij}\right]\hat{r}_{ij}\\ 
    &-(\boldsymbol{\sigma}_i\times\boldsymbol{\sigma}_j)\bigg\} e^{-m_\pi r_{ij}}
\end{split}
\end{equation}
with $m_\pi = 138$~MeV, $F_\pi = 92.3$~MeV, and $g_A = 1.27$.  The Sachs potential involves the tensor operator $S_{ij} = 3(\hat{r}_{ij}\cdot\boldsymbol{\sigma}_i)(\hat{r}_{ij}\cdot\boldsymbol{\sigma}_j) - \boldsymbol{\sigma}_i\cdot\boldsymbol{\sigma}_j$ (see Ref.~\cite{seutin2023magnetic} for the full expression).

A recent study~\cite{seutin2023magnetic,miyagi2024impact} found that including these two-body currents at NLO systematically improves the agreement with experimental magnetic moments across the nuclear chart, as discussed further in Sec. \ref{sec:EM_properties} and shown in Fig.~\ref{fig:dipolecurrent}.  This result is the magnetic-moment analog of the resolution of the $g_A$ quenching puzzle for Gamow-Teller beta decays~\cite{gysbers2019discrepancy}: in both cases, the inclusion of consistently derived two-body currents from chiral EFT as well as many-body correlations eliminates the need for further phenomenological adjustments to improve agreement with data.

\subsubsection{The electric quadrupole operator and collectivity}
\label{sec:quadrupole_theory}

The electric quadrupole moment (Sec.~\ref{sec:term_Q}, Fig.~\ref{fig:multipoles}b) is given microscopically by
\begin{equation}
    Q_{2\mu} = \sum_{i=1}^Z e\, r_i^2\, Y_{2\mu}(\hat{r}_i)\,,
    \label{eq:Q_operator}
\end{equation}
where $Y_{2\mu}$ is the spherical harmonic of rank 2, and the sum runs over protons.  The expectation value of this operator in the nuclear ground state yields the intrinsic quadrupole moment, which is nonzero only if the nucleon distribution breaks spherical symmetry.

Systematic measurements of quadrupole moments across isotopic chains reveal the onset of deformation away from closed shells~\cite{bohr1975nuclear,casten2001nuclear}.  Near magic numbers, quadrupole moments are typically small, consistent with nearly spherical ground states.  Away from shell closures, enhanced quadrupole moments appear, signaling collective motion where many nucleons coherently contribute to deformation.

The collective nature of the quadrupole moment makes it a challenging observable to describe microscopically.  The phenomenological shell model, starting from a spherical potential, requires the introduction of so-called effective charges to reproduce $E2$ transition strengths, compensating for missing excitations outside the valence space. In contrast, \textit{ab initio} methods that explicitly include multi-particle multi-hole excitations---such as the NCSM---can reproduce $E2$ strengths without effective charges~\cite{henderson2018testing,caprio2022robust}.  The VS-IMSRG and other \textit{ab initio} approaches starting from spherical symmetry consistently underpredict $E2$ strengths due to neglected many-body physics when truncated at the two-body level~\cite{stroberg2022systematics}.  Methods that include deformation explicitly, such as DFT~\cite{sassarini2022nuclear}, the IM-GCM~\cite{belley2024abinitio}, and coupled-cluster with angular momentum projection~\cite{hu2024abinitio,sun2025multiscale}, can also reproduce $E2$ strengths without phenomenological adjustments.

\begin{figure*}[t]
    \centering
    \includegraphics[width=\linewidth]{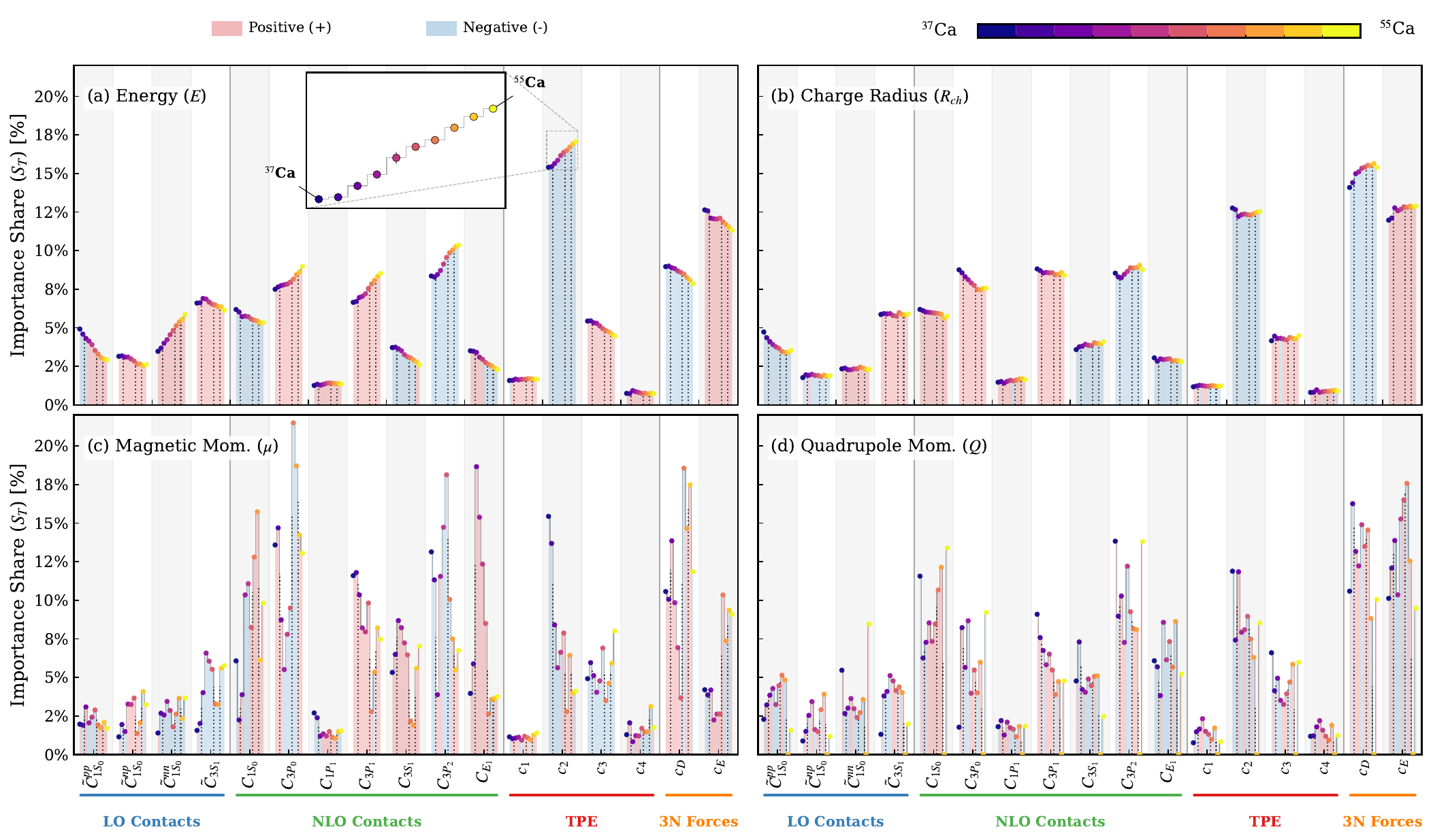}
    \caption{Relative importance of each low-energy constant (LEC) of the chiral EFT interaction for different nuclear observables along the Ca isotopic chain.  Ground-state energies and charge radii probe bulk effects of the nuclear force, while electromagnetic moments (magnetic dipole and electric quadrupole) show a highly nucleus-dependent sensitivity, making them complementary probes of the nuclear interaction.  Figure reproduced from Ref.~\cite{munoz2026linking}, with permission from the authors.}
    \label{fig:SHAPvalues}
\end{figure*}

A recent emulator-based study~\cite{munoz2026linking} investigated how different nuclear observables depend on the LECs of the chiral EFT potential along the Ca isotopic chain.  While ground-state energies and charge radii show consistent sensitivity to the LECs across the chain---probing bulk properties of the nuclear force---electromagnetic moments exhibit a highly nucleus-dependent sensitivity (Fig.~\ref{fig:SHAPvalues}).  This makes electromagnetic moments particularly valuable for probing how nuclear structure is shaped by specific features of the nuclear interaction.

\subsection{Parity-violating nuclear observables: the anapole moment}
\label{sec:anapole_theory}

As discussed in Sec.~\ref{sec:anapole}, the nuclear anapole moment is defined as $\mathbf{a} = -\pi\int r^2\,\mathbf{j}_N(\mathbf{r})\,d^3r$---an expectation value of the ordinary electromagnetic current $\mathbf{j}_N$ in a nuclear state that contains parity-violating admixtures from the weak nucleon-nucleon interaction.  Computing the anapole moment therefore requires: (i)~the parity-violating nuclear potential $V_{NN}^{PV}$; (ii)~the nuclear ground-state wave function in the presence of this potential; and (iii)~the anapole moment operator.

\subsubsection{The parity-violating nuclear potential}

The parity-violating nucleon-nucleon interaction can be parameterized either in the traditional one-meson-exchange model, in terms of six weak meson-nucleon couplings ($h_\pi$, $h_\rho^{0,1,2}$,$h_\omega^{0,1}$)~\cite{desplanques1980unified,haxton2001atomic}, or using parity-violating nuclear forces derived in chiral EFT~\cite{zhu2005nuclear,devries2020parity}, which provides a more systematic parameterization in terms of the parity-breaking pion-nucleon couplings and short-range LECs. The leading-order potential in chiral EFT corresponds to the pion-exchange term in the meson-exchange models, and the contributions of heavier mesons can be matched to the five short-range LECs $C_{0,...,4}$ appearing at NLO in chiral EFT~\cite{deVries2014study}. The ``best values'' and ``reasonable ranges'' estimated by Desplanques, Donoghue, and Holstein (known as the DDH potential) ~\cite{desplanques1980unified} for the meson-nucleon couplings carry uncertainties of order 100\%, highlighting the large degree of theoretical uncertainty in these parameters.  One of the main goals of anapole moment measurements is to constrain these experimentally.

In the single-particle approximation, the parity-violating potential reduces to an interaction between one unpaired nucleon and the nuclear core~\cite{flambaum1997anapole}:
\begin{equation}
    V_{NN,\text{SP}}^{PV} = \frac{G_F}{\sqrt{2}}\frac{g_i}{2m}\left[\boldsymbol{\sigma}\cdot\mathbf{p}\,\rho(r) + \rho(r)\,\mathbf{p}\cdot\boldsymbol{\sigma}\right],
\end{equation}
where $g_i$ ($i = p,n$) parameterizes the nucleon-nucleus weak coupling.

\subsubsection{The anapole moment operator}

The anapole moment operator is given by~\cite{flambaum1997anapole,hao2020nuclear}:
\begin{equation}
    \hat{\mathbf{a}} = \sum_i^A\frac{\pi e}{m}\left[\mu_i\left(\mathbf{r}_i\times\boldsymbol{\sigma}_i\right)-\frac{q_i}{2}\left(\mathbf{p}_i r_i^2 + r_i^2\mathbf{p}_i\right)\right],
    \label{eq:anapole_operator}
\end{equation}
where $\mu_i$, $\mathbf{r}_i$, $\mathbf{p}_i$, $\boldsymbol{\sigma}_i$ are the magnetic moment, position, momentum, and spin of the $i$th nucleon, and $q_i = 1$ (0) for a proton (neutron).  The first (spin) term typically dominates over the second (convection) term.  Two-body currents can also be added to this operator~\cite{haxton2002nuclear,hao2020nuclear}; given their importance for magnetic dipole moments (Sec.~\ref{sec:EM_operators}), they are expected to contribute at a similar level for anapole moments.

\subsubsection{Single-particle estimates}

In the single-particle approximation, the anapole contribution to the nuclear-spin-dependent PV parameter $\eta_\text{anapole}$ can be estimated as~\cite{flambaum1984nuclear,flambaum1985enhancement}:
\begin{equation}
    \begin{split}
     \eta_{\mathrm{anapole}} &= \frac{9}{10}\frac{\alpha\mu_i}{m r_0}g_i A^{2/3}\frac{K}{I(I+1)}\\
     &\approx  1.15\times 10^{-3}\, \mu_i g_i A^{2/3}\frac{K}{I(I+1)}\,,
    \end{split}
    \label{eq:eta_anapole_SP}
\end{equation}
where $\alpha$ is the fine-structure constant, $r_0 \approx 1.2$~fm, $A$ is the mass number, $\mu_i$ is the unpaired nucleon magnetic moment, and $K = (I+1/2)(-1)^{I+l_i+1/2}$. The dimensionless constants $g_i$ are derived from the mean nucleon-nucleus weak potential for the unpaired nucleon. 
These constants are defined only roughly; here we use the values from Ref.~\cite{fadeev2019time} of $g_p = 3.4\pm0.8$ and $g_n = 0.9\pm0.6$.

Similarly, the direct $C_2$ NSD contribution (Sec.~\ref{sec:W_expansion}) in this approximation is~\cite{flambaum1980p}:
\begin{equation}
    \eta_{\text{axial}} = C_2\,\frac{1/2-K}{I(I+1)}\,,
\end{equation}
where $C_2 = C_{2,p}$ or $C_{2,n}$ for an unpaired proton or neutron, and the combined spin-flip contribution from the weak charge is~\cite{flambaum1985enhancement}:
\begin{equation}
    \eta_{\mathrm{hf}} \approx -\frac{1}{3}Q_W\frac{\alpha\mu}{m_p r_0 A^{1/3}} \approx 2.5\times 10^{-4}\, A^{2/3}\, \mu\,.
\end{equation}
The last effect is typically much smaller than the anapole contribution due to its numerical prefactor.

These single-particle estimates are useful for gauging the scaling of the effect with atomic number, but they can deviate from more detailed calculations by more than 100\%, even for light nuclei~\cite{hao2020nuclear}.  Reliable extraction of the weak coupling constants from experiments requires \textit{ab initio} nuclear structure calculations.

\subsubsection{\textit{Ab initio} calculations}

To compute \textit{ab initio} anapole moments, one evaluates the ground-state wave function in the presence of $V_{NN}^{PV}$ using first-order perturbation theory~\cite{hao2020nuclear}:
\begin{equation}
    |\Psi_\text{gs}\rangle = |\Psi_0, P\rangle + \sum_k \frac{\langle \Psi_k, -P|V_{NN}^{PV}|\Psi_0, P\rangle}{E_0-E_k}\,|\Psi_k, -P\rangle\,,
    \label{eq:PV_admixture_anapole}
\end{equation}
where $|\Psi_0, P\rangle$ is the unperturbed ground state of definite parity $P$, and the sum runs over all states of opposite parity.  The anapole moment is then $\mathbf{a} = \langle\Psi_\text{gs}|\hat{\mathbf{a}}|\Psi_\text{gs}\rangle$, and $\eta_\text{anapole}$ is extracted from the relation $\mathbf{a} = (G_F/\sqrt{2}|e|)\,\eta_\text{anapole}\,\mathbf{I}$.

To date, two \textit{ab initio} frameworks have been applied to the anapole moment.  The NCSM has evaluated the polarization contribution for light isotopes up to $^{25}$Mg~\cite{hao2020nuclear}, representing the computational limit of this quasi-exact method due to its combinatorial scaling with nucleon number.  Recently, the Parity-Violating IMSRG (PV-IMSRG)~\cite{belley2026abinitio} has been developed to compute parity-violating observables using a unitary transformation that induces a parity-conserving component to the anapole moment, allowing it to be evaluated as a ground-state expectation value.  The PV-IMSRG has been benchmarked against the NCSM, where it was found to be in good agreement: while many-body truncation effects are present, results are within a factor of two of the exact calculations---a much smaller discrepancy than what is found between different phenomenological methods.  Furthermore, the method is systematically improvable and can, in principle, reach much heavier nuclei than the NCSM.

\begin{figure}[t]
    \centering
    \includegraphics[width=1\linewidth]{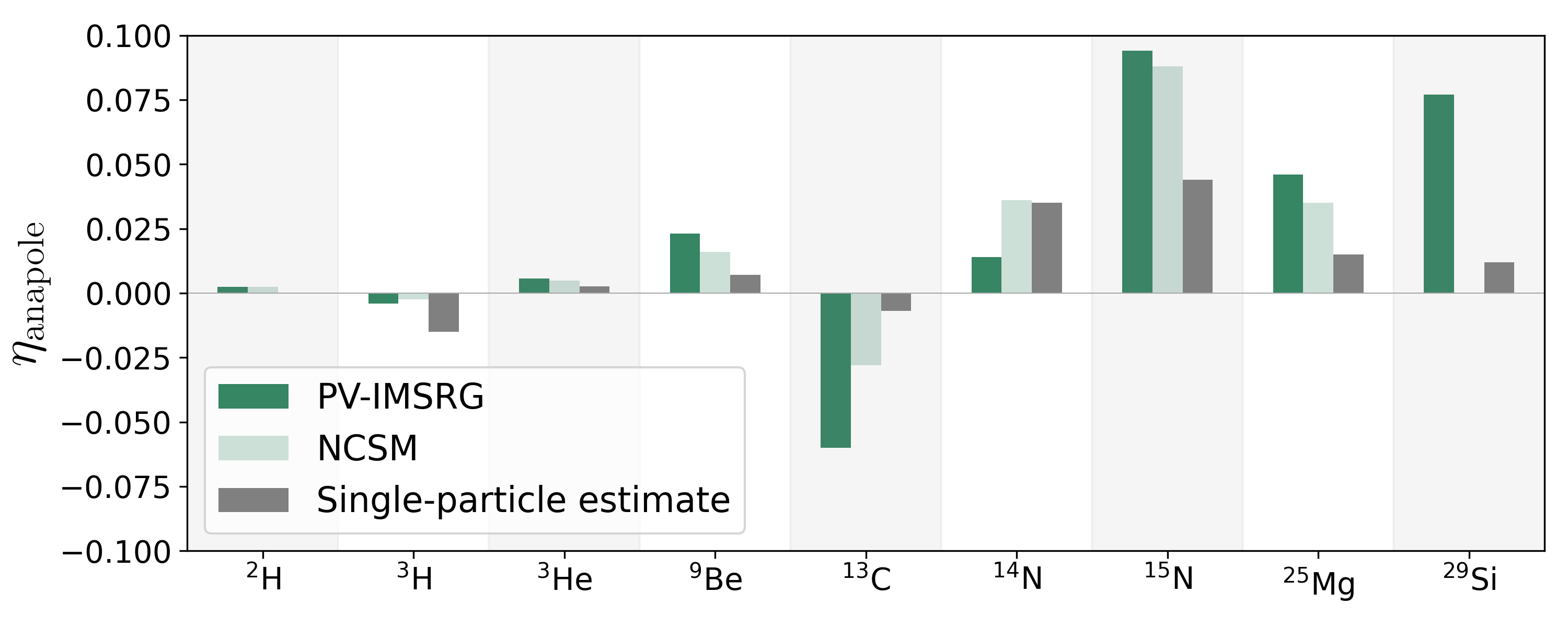}
    \caption{Calculation of the nuclear-spin dependent contribution of the anapole $\eta_{\rm anapole}$ moment using the PV-IMSRG, compared to the exact results from the NCSM and single-particle estimates. Figure adapted from~\cite{belley2026abinitio}, with permission from the authors}
    \label{fig:annapolemoment_theory}
\end{figure}

Notably, as shown in Fig.~\ref{fig:annapolemoment_theory}, the \textit{ab initio} results differ significantly from single-particle estimates.  For example, the anapole moment of $^{29}$Si computed with the PV-IMSRG (benchmarked against the NCSM) is more than six times larger than the single-particle estimate of Eq.~(\ref{eq:eta_anapole_SP}), highlighting the importance of many-body correlations and the limitations of single-particle approximations for this observable.

\subsection{CP violating nuclear observables: Schiff moments}
\label{sec:schiff_theory}

As discussed in Sec.~\ref{sec:ch2}, the nuclear Schiff moment $\mathbf{S}$ is the leading CP violating nuclear observable in diamagnetic systems, arising as the residual effect of the nuclear charge distribution after Schiff screening.  Its computation requires the CP violating (PVTV) nuclear potential $V_{PT}$ and a detailed treatment of the nuclear many-body problem.

\subsubsection{Connection to CP violating parameters}

Using $V_{PT}$ derived at leading order in chiral EFT~\cite{devries2020parity}, contributions from the potential to the nuclear Schiff moment can be decomposed linearly as~\cite{engel2025nuclear}:
\begin{equation}
    S = a_0\, g\, \bar{g}_\pi^{(0)} + a_1\, g\, \bar{g}_\pi^{(1)} + a_2\, g\, \bar{g}_\pi^{(2)} + A_1 \tilde{C}_1+ A_2\tilde{C}_2+a_n d_n+a_p d_p\,,
    \label{eq:SchiffCoefficient}
\end{equation}
where $\bar{g}_\pi^{(0,1,2)}$ are the isoscalar, isovector, and isotensor pion-nucleon CP violating couplings (the same parameters that appear at the hadronic level of the EFT matching chain, Eq.~(\ref{eq:EFT_chain})), $\tilde{C}_1$ and $\tilde{C}_2$ are contact terms in the PV-potential, $g \approx 13.3$ is the strong pion-nucleon coupling constant, and the $a$ and $A$ coefficients contain all of the nuclear structure information. The nucleon's dipole moments, $d_p$ and $d_n$, can be extracted more precisely from experiments in free nucleons where there is no Schiff screening. As such, the focus for nuclear theory lies in computing the parameters associated with the PV-potential, in particular the $a_{0,1,2}$ associated with the parity-violating pion-exchanges.

The Schiff moment is evaluated using perturbation theory in $V_{PT}$, which is justified by the extremely small size of the CP violating potential compared to the nuclear Hamiltonian:
\begin{equation}
    S = \sum_{k \neq 0} \frac{\langle \Psi_0\, J^\pi | \hat{S}_z | \Psi_k\, J^{-\pi} \rangle\, \langle \Psi_k\, J^{-\pi} | V_{\text{PT}} | \Psi_0\, J^\pi \rangle}{E_0 - E_k} + \text{c.c.}\,,
    \label{eq:Schiff_pert}
\end{equation}
where $|\Psi_0\, J^\pi\rangle$ is the ground state with angular momentum $J$ and parity $\pi$, the sum runs over all states of opposite parity, and $\hat{S}_z$ is the $z$-projection of the Schiff operator (Eq.~(\ref{eq:Schiff})) with $J = M$.

\subsubsection{Phenomenological calculations}

Other than single-particle estimates~\cite{flambaum1986time,dmitriev2005effects}, two main methods have been used to compute Schiff moments in heavy nuclei: the nuclear shell model and DFT.

\paragraph{Shell-model calculations.}
The NSM has been applied to $^{129}$Xe and $^{199}$Hg---isotopes with moderate deformation that are accessible to the shell-model framework.  Several variants have been used: the pair-truncated shell model~\cite{yoshinaga2013nuclear,teruya2017effects,yoshinaga2020schiff}, which restricts the model space to nucleon pairs with total angular momentum 0 or 2, and the large-scale shell model~\cite{yanase2020largescale}, which works in an extended model space to better capture deformation effects.

\begin{figure*}[t]
    \centering
    \includegraphics[width=1\linewidth]{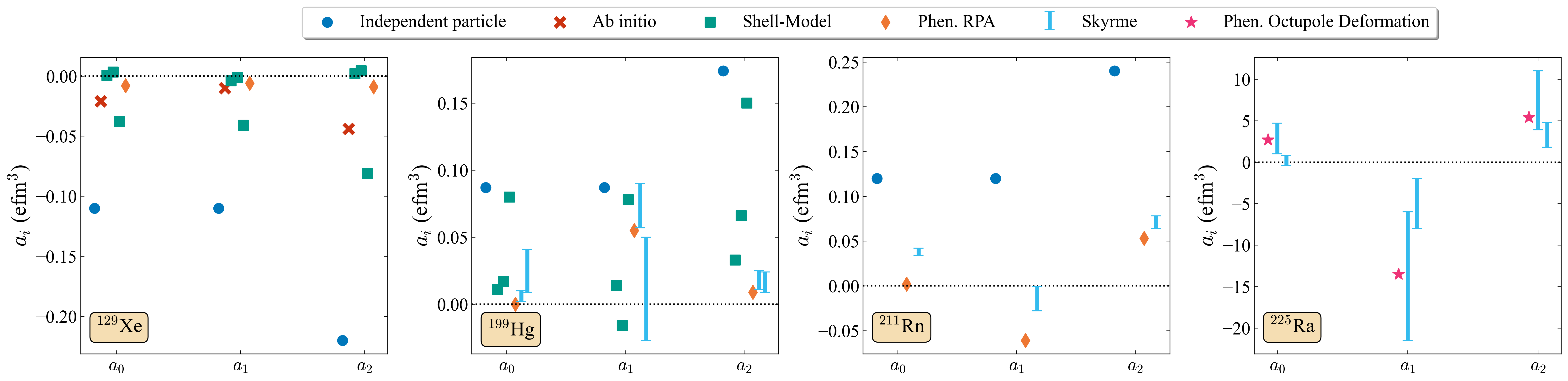}
    \caption{Calculations of the isoscalar ($a_0$), isovector ($a_1$), and isotensor ($a_2$) coefficients of the Schiff moment (Eq.~(\ref{eq:SchiffCoefficient})) using different nuclear structure methods.  The spread between models highlights the need for \textit{ab initio} calculations with controlled uncertainties.  See text for details.}
    \label{fig:SchiffMoment_theory}
\end{figure*}

\paragraph{DFT calculations.}
For $^{129}$Xe and $^{199}$Hg, DFT calculations have used either Skyrme functionals with the random-phase approximation (RPA) for excited states~\cite{ban2010fully}, or simpler mean-field approaches based on Woods-Saxon potentials~\cite{dmitriev2003pandt,dmitriev2005effects}.  For $^{225}$Ra, where octupole deformation plays a central role in enhancing the Schiff moment (as discussed in Sec.~\ref{sec:schiff_mom}), both Skyrme-based DFT~\cite{dobaczewski2005nuclear,dobaczewski2018correlating} and rigid-rotor models with octupole-deformed Woods-Saxon potentials~\cite{spevak1997} have been employed.

The results from these different calculations are summarized in Fig.~\ref{fig:SchiffMoment_theory}.  A large spread is found between the predictions of different models, with discrepancies of factors of 2--5 for some coefficients.  This spread is the primary source of theoretical uncertainty in the interpretation of EDM experiments in diamagnetic systems, and it highlights the need for \textit{ab initio} calculations with controlled uncertainties.

\subsubsection{\textit{Ab initio} calculations}

The first \textit{ab initio} calculation of a nuclear Schiff moment was performed for $^{19}$F using the NCSM~\cite{ng2025nuclear}.  Combined with relativistic molecular calculations for HfF$^+$, an experimental bound on a nuclear Schiff moment from the JILA electron EDM measurement was extracted~\cite{roussy2023improved}.  Although the resulting bounds on the pion-nucleon couplings $\bar{g}_\pi^{(i)}$ are not yet the most stringent, this work establishes the foundation for constraining these couplings using \textit{ab initio} nuclear methods.

The PV-IMSRG~\cite{belley2026abinitio} framework introduced for the anapole moment calculations (Sec.~\ref{sec:anapole_theory}) can similarly be used for the Schiff moment and has, again, been benchmarked against NCSM results.  A preliminary evaluation of the Schiff moment of $^{129}$Xe has been performed, yielding results in agreement with previous shell-model calculations (Fig.~\ref{fig:SchiffMoment_theory}), and a careful assessment of theoretical uncertainties is underway.

\subsection{Uncertainty quantification}
\label{sec:UQ}

A defining advantage of \textit{ab initio} approaches is the possibility of systematic uncertainty quantification.  The theoretical uncertainty of a nuclear-structure calculation has several sources, which have so far been estimated independently.

\subsubsection{Interaction uncertainties.}
The LECs of the chiral EFT Hamiltonian are fitted to data with finite precision, and their values are correlated. The resulting uncertainty can be propagated to nuclear observables by sampling over the LECs within their allowed range.  The main challenge to doing so is that it requires many evaluations of the same quantity under variation of the LECs, and due to the high computational cost of \textit{ab initio} methods, this is not feasible directly. However, this approach can be implemented using \emph{many-body emulators}---fast surrogate models that approximate the output of the full many-body calculation as a function of the LECs~\cite{ekstrom2019global,belley2025global,munoz2026linking}.  Such emulators enable the exploration of a large parameter space at a fraction of the computational cost, and the resulting samples can be weighted in a Bayesian framework to obtain posterior distributions for nuclear observables~\cite{hu2022ab,belley2024abinitio}.  These emulators have recently been improved to perform global emulation over multiple isotopes simultaneously~\cite{belley2025global,munoz2026linking}, and have enabled sensitivity analyses revealing which LECs most strongly influence each observable (Fig.~\ref{fig:SHAPvalues}).

\subsubsection{Chiral truncation uncertainties.}
The chiral EFT expansion is truncated at a finite order, and the omitted higher-order terms contribute an uncertainty that can be estimated from the observed convergence pattern~\cite{furnstahl2015quantifying,melendez2017bayesian,wesolowski2021rigorous}.

\subsubsection{Many-body truncation uncertainties.}
The many-body methods involve truncations (e.g., IMSRG(2) vs.\ IMSRG(3)) whose errors can be assessed by comparing results at different truncation levels.  Recent IMSRG(3) and many-body perturbation theory developments~\cite{stroberg2024imsrg3,he2024imsrg3, svenson2026Bayesian} have provided the first systematic assessments of these errors for medium-mass nuclei.

\subsubsection{Comprehensive uncertainty budgets.}
The first comprehensive uncertainty quantification for a nuclear electroweak observable was the \textit{ab initio} prediction of the $^{208}$Pb neutron skin~\cite{hu2022ab}.  A subsequent study quantified the uncertainty of the nuclear matrix element for neutrinoless double-beta decay in $^{76}$Ge~\cite{belley2024abinitio}.  Extending this program to the full set of electroweak observables in Table~\ref{tab:moments}---including Schiff moments, anapole moments, and magnetic quadrupole moments in heavy nuclei---is a major goal for the coming decade.

\subsection{Challenges and outlook}
\label{sec:outlook_theory}

Despite the rapid progress described above, significant challenges remain before \textit{ab initio} methods can provide the nuclear matrix elements needed for the full program of electroweak searches with atoms and molecules.

\subsubsection{Heavy and deformed nuclei.}
Many of the nuclei of greatest interest---$^{199}$Hg, $^{225}$Ra, $^{229}$Pa, and the actinides used in radioactive molecule experiments---are both heavy ($A \gtrsim 200$) and strongly deformed.  Current \textit{ab initio} methods can reach these masses for bulk properties but face challenges for delicate quantities like Schiff moments that require accurate treatment of collective correlations.  Methods specialized for deformed systems---the IM-GCM~\cite{yao2020ab} and deformed coupled-cluster~\cite{hagen2022angular}---offer promising routes, but have not yet been applied beyond the medium-mass region.

\subsubsection{Consistent symmetry-violating operators.}
The PVTC and PVTV operators needed for anapole and Schiff moment calculations must be consistently derived and transformed through the same many-body procedure as the Hamiltonian. While development has been done with the PV-IMSRG~\cite{belley2026abinitio}, the consistent treatment of these operators in the IMSRG and CC frameworks is still computationally demanding and is an active area of development.

\subsubsection{Higher-body corrections.}
Assessing the convergence of the many-body expansion scheme---and extending to higher-body operators where necessary~\cite{svensson2025bayesian}---differs for different observables and will be crucial for reaching the precision required by next-generation experiments.

\subsubsection{Connection to experiment.}
The electromagnetic observables discussed in Secs.~\ref{sec:charge_radii} and~\ref{sec:EM_moments}---charge radii and electromagnetic moments across isotopic chains---provide benchmarks for validating nuclear calculations.  Good agreement with these ``known'' quantities builds confidence in predictions for the ``unknown'' electroweak and CP violating moments.  This interplay, mediated by the factorized structure of the atomic/molecular Hamiltonian (Eq.~(\ref{eq:master_all})), is a recurring theme of this review.

\section{Electromagnetic properties of exotic nuclei}
\label{sec:EM_properties}

The electromagnetic multipole expansion developed in Sec.~\ref{sec:EM_expansion} shows that the energy levels of atoms and molecules encode
the charge radius $\langle r^2\rangle$ (Sec.~\ref{sec:term_FS}), the magnetic dipole moment $\mu$ (Sec.~\ref{sec:term_mu}), and the electric quadrupole moment $Q$ (Sec.~\ref{sec:term_Q}) of the nucleus.  These PT conserving moments (Fig.~\ref{fig:multipoles}a,b and Table~\ref{tab:moments}) are the $known$ sector of the multipole expansion:~they can be measured with high precision using established techniques, and they serve simultaneously as probes of nuclear structure and as benchmarks for the nuclear theory methods (Sec.~\ref{sec:nuclear_theory}) that are needed to compute the $unknown$ symmetry-violating moments discussed in subsequent chapters.

\begin{figure*}[t]
    \centering
    \includegraphics[width=0.49 \linewidth]{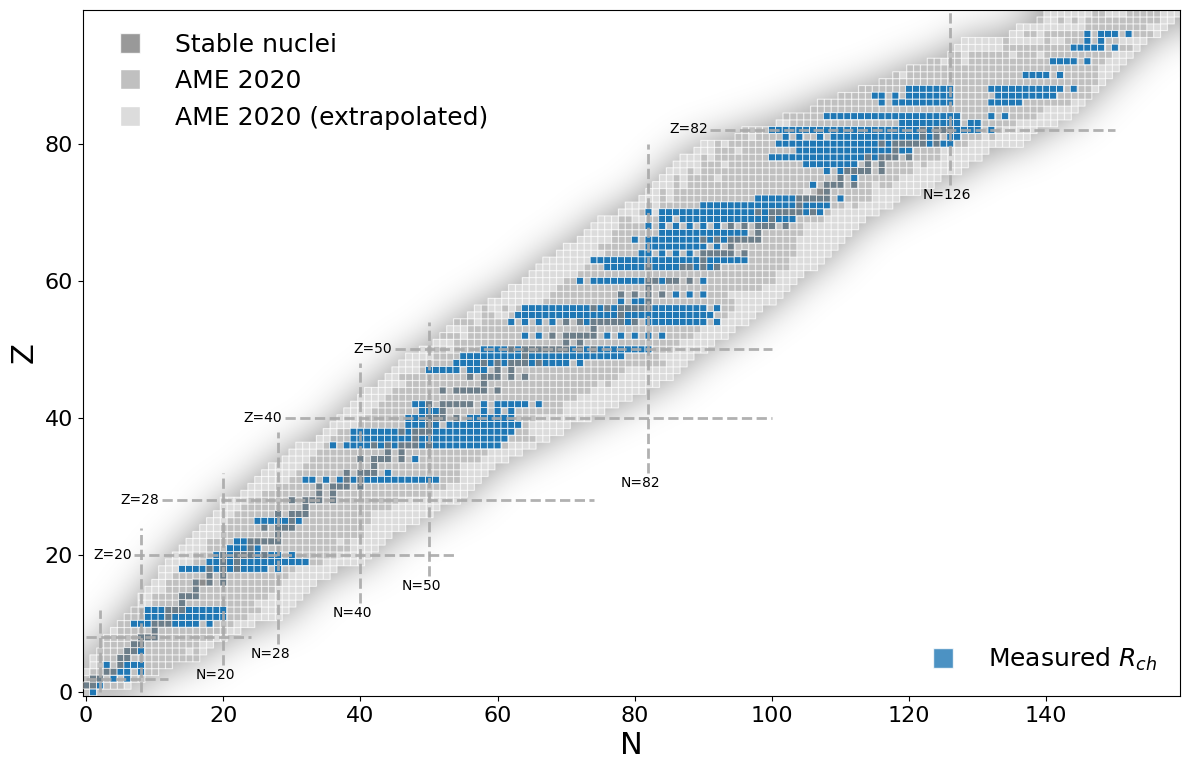}
    \includegraphics[width=0.49 \linewidth]{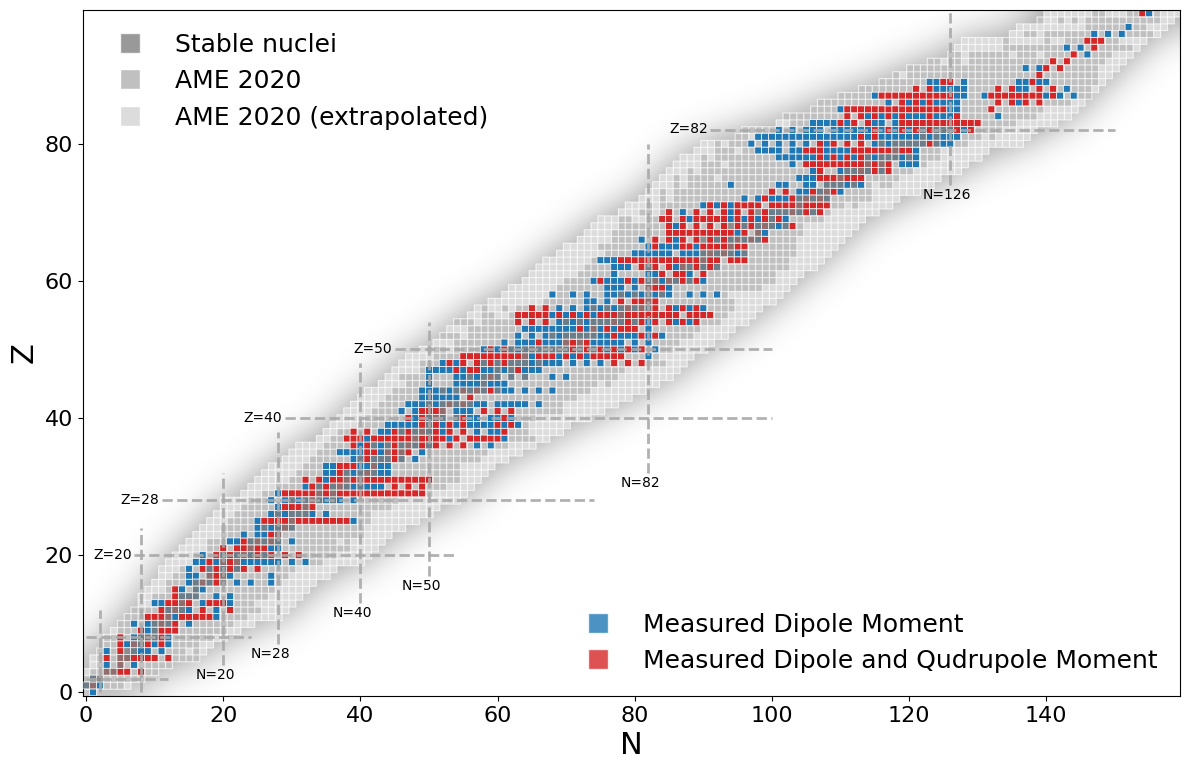}
    \caption{Isotopes for which nuclear charge radii (left) and nuclear electromagnetic moments (right) have been measured.  The reach of laser spectroscopy now spans nearly the entire nuclear chart, from helium to the actinides.  Stable isotopes are shown in black; radioactive isotopes studied by different techniques are indicated by color.  See Ref.~\cite{angeli2013table} for details.}
    \label{fig:radii_chart}
\end{figure*}

Figure~\ref{fig:radii_chart} provides an overview of the nuclei for which
ground-state charge radii and electromagnetic moments have been
measured. To date, laser spectroscopy has
been used to study more than a thousand ground and isomeric states of unstable
nuclei, spanning elements from helium ($Z=2$) to californium ($Z=98$). As
illustrated in Fig.~\ref{fig:radii_chart}, the reach of these measurements now
extends across nearly the entire nuclear chart, including many isotopes far from
stability.

In this section, we review how these electromagnetic properties are measured in exotic (short-lived, far-from-stability) nuclei, how the results have reshaped our understanding of nuclear forces and nuclear structure, and where the field is heading.  The section focuses on symmetry-conserving observables. The parity-violating and CP violating sectors are treated in Sec. \ref{sec:symm_violation}.

\subsection{Experimental techniques}
\label{sec:exp_techniques}

The electromagnetic moments $\Mn^{(\lambda)}$ entering the multipole expansion can be extracted from the measured isotope shifts and hyperfine structure splittings using the factorization derived in Sec.~\ref{sec:ch2}: the isotope shift gives access to $\delta\langle r^2\rangle$ through Eq.~(\ref{eq:IS_total}), the magnetic hyperfine constant $A_\text{hfs}$ gives $\mu$ through Eq.~(\ref{eq:Ahfs}), and the quadrupole hyperfine constant $B_\text{hfs}$ gives $Q$ through Eq.~(\ref{eq:EF_quad}).  In each case, the required electronic form factors ($F$, $K$, $B_e$, $V_{zz}$) must be computed from atomic or molecular theory, or calibrated against independent measurements (e.g., electron scattering or muonic atom data).  
Laser spectroscopy has been established as the workhorse technique for measuring electromagnetic properties of exotic nuclei~\cite{yang2023laser,campbell2016laser}.  By probing the hyperfine structure and isotope shifts of atomic or ionic transitions, one can simultaneously determine the nuclear spin $I$, the magnetic dipole moment $\mu$, the electric quadrupole moment $Q$, and the change in mean-square charge radius $\delta\langle r^2\rangle$ of ground and isomeric states of a given nucleus.  
Several complementary methods have been developed, each optimized for different experimental conditions:
\subsubsection{Fluorescence collinear laser spectroscopy} In collinear laser spectroscopy (CLS)~\cite{neugart2017collinear,koszorus2024collinear}, the ions (or atoms) travel at keV energies along a beamline, collinear with a laser beam.  The kinematic compression of the velocity distribution in the fast beam provides Doppler-limited resolution of $\sim$10--100~MHz, sufficient to resolve the hyperfine structure and isotope shifts of most elements.  CLS is the standard technique at ISOL facilities and has been applied to hundreds of isotopes from helium to uranium.  The use of gas-filled radiofrequency ion traps to produce bunched beams has increased the sensitivity by several orders of magnitude, enabling measurements on species produced at rates as low as $\sim$100 ions/s~\cite{garcia2016unexpectedly}.
\subsubsection{Collinear resonance ionization spectroscopy} The Collinear Resonance Ionization Spectroscopy (CRIS) technique~\cite{Flan13,cris2020} combines the high resolution of collinear spectroscopy with the high efficiency of resonance ionization.  Atoms in the fast beam are stepwise excited and ionized by pulsed lasers, with the resulting ions being detected with near-unity efficiency.  CRIS can achieve sensitivities down to $\sim$10 ion/s while maintaining MHz-level resolution \cite{de2020measurement}.
\subsubsection{In-source resonance ionization spectroscopy} Laser ionization can also be performed directly inside the hot cavity of the ion source, where the atoms are produced ~\cite{fedosseev2017ion}.  This can provide complementary access to exotic species with the shortest half-lives ($\lesssim$10~ms), at the cost of reduced spectral resolution ($\sim$1--5~GHz) due to Doppler and pressure broadening \cite{Mar18,reponen2021evidence}.  
In-source methods such as in-gas-jet spectroscopy~\cite{kudryavtsev2013insource,Ferrer2017} offer a promising route to combine high efficiency with high resolution by performing the ionization in a supersonic gas jet.

\subsubsection{Ion-trap spectroscopy}

Ions confined in radiofrequency (Paul) or Penning traps can be interrogated by laser and/or radiofrequency fields for extended periods, enabling interrogation times that are several orders of magnitude longer than in beam-based spectroscopy \cite{Savard2000,Werth2007}. Consequently, the attainable spectral resolution approaches the Fourier limit, while repeated interrogation of the same trapped ion(s) allows efficient state preparation and readout and greatly enhances the signal-to-noise ratio, making precision measurements feasible even with only a few stored ions.

Recent developments on quantum logic \cite{Schmidt2005,Rosenband2007} and correlation spectroscopy \cite{Mano19} have enabled precision measurements with single trapped ions at unprecedented levels of accuracy, opening new opportunities for nuclear physics studies based on trapped radioactive ions.

Compared with conventional collinear laser spectroscopy, trapped-ion techniques can improve the precision of hyperfine constants and nuclear $g$-factor measurements by more than an order of magnitude, with sub-Hz or part-per-billion precision demonstrated in favorable cases \cite{counts2020evidence}. Although these methods have so far been applied mainly to stable isotopes, their combination of long interrogation times, laser or sympathetic cooling, and high detection efficiency makes them particularly promising for future measurements of electromagnetic moments, hyperfine anomalies, and higher-order multipole moments in nuclei far from stability \cite{deGroote2024TrappedAtoms}.

The techniques mentioned above are available at different radioactive ion beam (RIB) facilities worldwide: i. \textbf{ISOLDE} (CERN, Switzerland). An ISOL facility, using 1.4-GeV proton beams on thick targets to produce isotopes across the nuclear chart.  Home to COLLAPS, CRIS, and RILIS~\cite{catherall2017isolde}; ii. \textbf{FRIB} (Michigan State University, USA). A projectile-fragmentation facility producing the most neutron-rich isotopes.  It has a dedicated beamline for fluorescence collinear laser spectroscopy, BECOLA~\cite{minamisono2013becola}, and the recent Resonance Ionization Spectroscopy Experiment, RISE, is in operation \cite{Brin26}; iii.
\textbf{TRIUMF-ISAC} (Vancouver, Canada). An ISOL facility with CFBS and TITAN for spectroscopy and mass measurements; iv. \textbf{IGISOL} (Jyv\"askyl\"a, Finland).  It uses ion-guide techniques to produce refractory elements inaccessible at conventional ISOL facilities~\cite{moore2014igisol}, and is the house of multiple ion trap and laser spectroscopy experiments such as RIS\cite{reponen2021evidence}, CLS \cite{groote24}, and RAPTOR \cite{raptor23}; v. \textbf{BRIF, China} The Beijing Radioactive Ion-beam Facility hosts the newly developed Precision LAser Spectroscopy for Exotic Nuclei (PLASEN) \cite{Mei26} system for collinear resonance ionization spectroscopy; vi.  \textbf{RIBF} (RIKEN, Japan). A projectile-fragmentation facility
specialized for the most neutron-rich nuclei, with SLOWRI
\cite{Wada2011SLOWRI} for laser spectroscopy, and the KISS facility,
which combines multinucleon transfer reactions with in-gas-cell
resonance ionization spectroscopy ~\cite{hirayama2024ingas,Hir22}.
Upcoming facilities such as FAIR (Darmstadt), SPIRAL2 (GANIL), RAON (South Korea), and HIAF (China), are expected to expand the reach of laser spectroscopy to more exotic species.

\subsection{Nuclear charge radii}
\label{sec:charge_radii}

The mean-square charge radius $\langle r^2\rangle$ is the monopole term in the electromagnetic expansion (Sec.~\ref{sec:term_FS}).  Its change between isotopes, $\delta\langle r^2\rangle^{A,A'}$, is extracted from the measured isotope shift using the factorized expression of Eq.~(\ref{eq:IS_total}).  Systematic measurements of charge radii across isotopic chains are powerful tools for probing nuclear structure, nuclear forces, and the properties of nuclear matter.

\subsubsection{Bulk properties and the nuclear force}

The approximate scaling of the root-mean-square charge radius with mass number, $\sqrt{\langle r^2\rangle} \sim r_0 A^{1/3}$ with $r_0 \approx 1.2$~fm, reflects the saturation of nuclear matter. The short-range, saturating character of the nuclear force produces nuclei with approximately constant interior density.  However, systematic measurements reveal deviations from this global scaling that encode detailed information about the nuclear force~\cite{degroote2021,miyagi2025nuclear}.

\begin{figure*}[t]
    \centering
    \includegraphics[width=\linewidth]{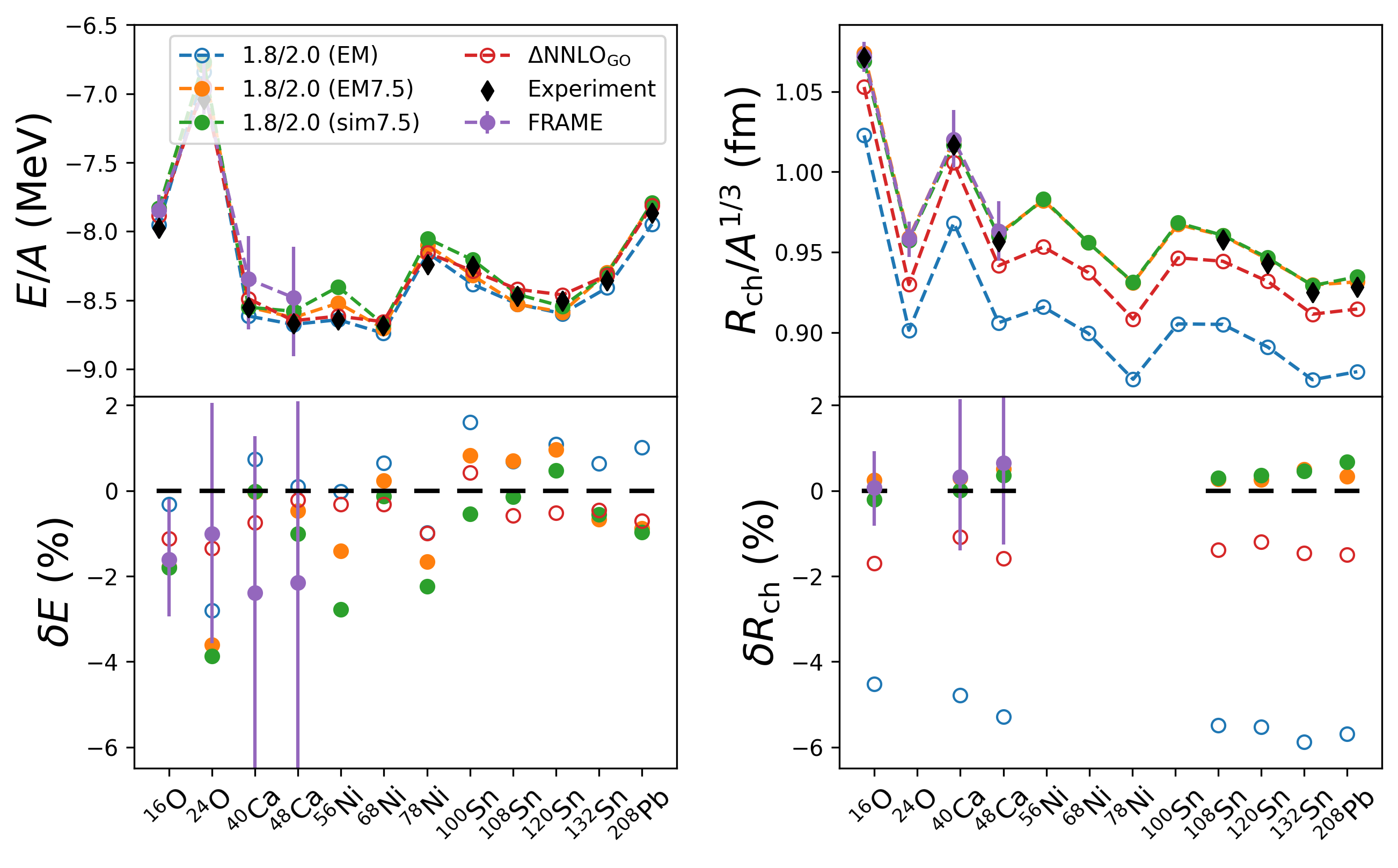}
    \caption{Nuclear binding energies (left) and charge radii (right) computed using multiple chiral EFT interactions using the VS-IMSRG~\cite{arthuis2024neutron}, with relative errors shown below.  While binding energies are well reproduced by several interactions, charge radii reveal systematic discrepancies that constrain the three-body sector of the nuclear force.  Results from the FRAME emulator framework~\cite{belley2025global} illustrate the interaction-parameter uncertainty.  Experimental data are shown where available.}
    \label{fig:radiivsenergies}
\end{figure*}

Charge radii have become a key benchmark for chiral EFT interactions (Sec.~\ref{sec:chiEFT}).  A commonly used interaction, labeled as  EM1.8/2.0 ~\cite{entem2003accurate,hebeler2011improved},  reproduces binding energies across the nuclear chart  but systematically underestimates charge radii~\cite{simonis2017saturation,garcia2016unexpectedly,koszorus2021potassium}.  This discrepancy points to deficiencies in the available inter-nucleon interactions, which produce saturation densities that are too high compared with empirical estimates~\cite{drischler2021chiral}.  The tension has motivated the construction of new interaction families that include radii in their calibration protocols~\cite{ekstrom2015accurate,jiang2020accurate,arthuis2024neutron}, suggesting that charge radii are essential for refining the balance between two- and three-nucleon forces.  The constraints provided by binding energies and charge radii are illustrated in Fig.~\ref{fig:radiivsenergies}. Interactions that reproduce binding energies well can still fail for charge radii, and vice versa, highlighting the need to inform nuclear forces from both observables simultaneously.

An important success of \textit{ab initio} many-body methods is that different many-body methods yield consistent charge radii and binding energies when using the same nuclear interaction, even with different many-body truncation schemes (See for example Fig. \ref{fig:abinitio_radii}). This indicates that the remaining discrepancies with experiment are dominated by the nuclear interaction rather than the many-body method, making charge radii a particularly important probe of the nuclear force.

Recent developments in nuclear emulators have made it possible to quantify the uncertainties associated with the low-energy coupling constants that define the nuclear force, revealing that charge-radius predictions from \textit{ab initio} methods are consistent with experimental values within the estimated uncertainties \cite{munoz2026linking}.

\subsubsection{Shell effects and magic numbers}

Near magic numbers (the traditional shell closures at $N$ or $Z = 2, 8, 20, 28, 50, 82, 126$), charge radii exhibit characteristic signatures: kinks in the isotope shift trends, reflecting the abrupt change in the filling of nuclear orbitals at a shell closure.  These kinks, which are visible as changes in the slope of $\delta\langle r^2\rangle$ vs.\ $N$, have been used to confirm or question the existence of magic numbers \cite{Gar20emer}.

\begin{figure}[t]
    \centering
    \includegraphics[width=\linewidth]{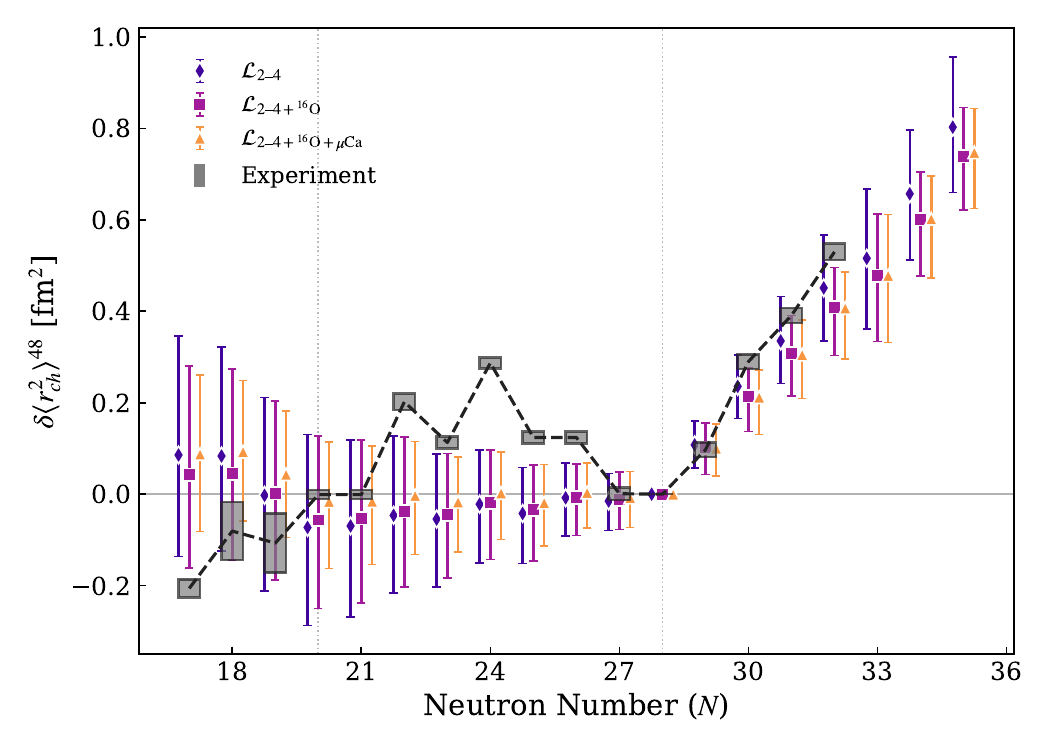}
    \caption{Comparison of experimental differential mean-squared charge radii (with respect to $^{48}$Ca) and \textit{ab initio} predictions under the variation of the LECs of chiral forces. LECs are constrained using 3 different likelihoods: only using few-body data (purple), adding observables in $^{16}$O (magenta) and finally adding magnetic moments in the calcium chain (orange).  Figure adapted from Ref.~\cite{munoz2026linking}, with permission from the authors.}
    \label{fig:frame_ca_radii}
\end{figure}

The calcium isotopic chain offers a key benchmark for the development of theoretical models as it possesses a proton magic number ($Z=20$) and multiple neutron magic numbers at $N=20$ and $28$, as well as suggested for $N=32,34$ \cite{52Camagic,54Camagic}. It has remained a challenge for several decades to explain the marked odd-even staggering effect and parabolic behavior observed between $N=20$ and $N=28$ \cite{garcia2016unexpectedly}. A recent study using emulators to sample the LECs of chiral forces found that they could not reproduce these trends even when using different observables to calibrate the LECs~\cite{munoz2026linking} as shown in Fig.~\ref{fig:frame_ca_radii}.

A landmark result was the measurement of charge radii of neutron-rich calcium isotopes up to $^{52}$Ca at ISOLDE~\cite{garcia2016unexpectedly}. The results revealed a large increase in charge radius beyond $N = 28$, later observed as well for the matter radii \cite{Tan20}.  This was in contrast with the doubly magic nature suggested for $^{52}$Ca and posed an exacting test for DFT and \textit{ab initio} nuclear theory.  Subsequent  VS-IMSRG calculations reproduced the qualitative trend but not the magnitude, suggesting that the nuclear interaction needs further refinement in this region~\cite{garcia2016unexpectedly}. Laser spectroscopy measurements of potassium isotopes ($Z=19$) towards and beyond $N=32$~\cite{kreim2014nuclear,koszorus2021potassium}, show a similar large increase of the nuclear radii. Improved DFT calculations have provided a relatively good description of the nuclear radii in this region \cite{koszorus2021potassium}.

\begin{figure}[t]
    \centering
    \includegraphics[width=\linewidth]{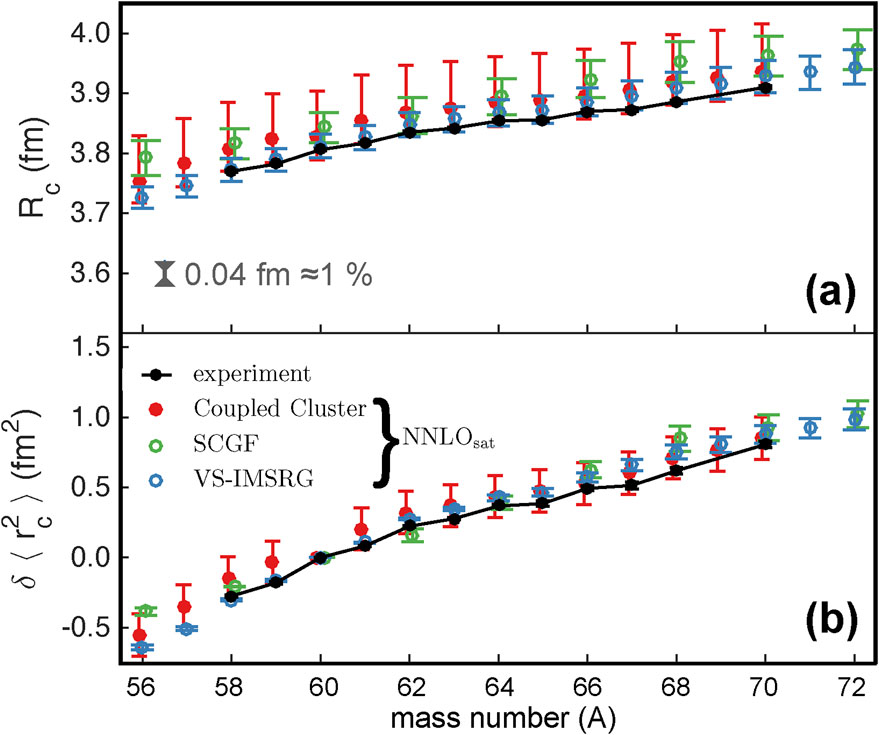}
    \caption{\textit{Ab initio} charge radii compared with experiment in the Nickel isotopic chain.  The comparison illustrates both the successes and remaining challenges of chiral EFT interactions combined with different many-body methods.  Figure adapted from Ref.~\cite{malbrunot2022nickel}, under CC BY 4.0.}
    \label{fig:abinitio_radii}
\end{figure}

For the nickel isotopes, with the proton magic number $Z=28$, the evolution of their charge radii appears to be rather smooth, exhibiting relatively small odd--even staggering, as shown in Figure~\ref{fig:abinitio_radii}. The figure compares the experimental results with predictions from several \textit{ab initio} calculations, illustrating both the successes and the remaining challenges of current chiral EFT interactions when combined with different many-body methods.

\subsubsection{Deformation and shape coexistence}

\begin{figure*}[t]
    \centering
    \includegraphics[width=\linewidth]{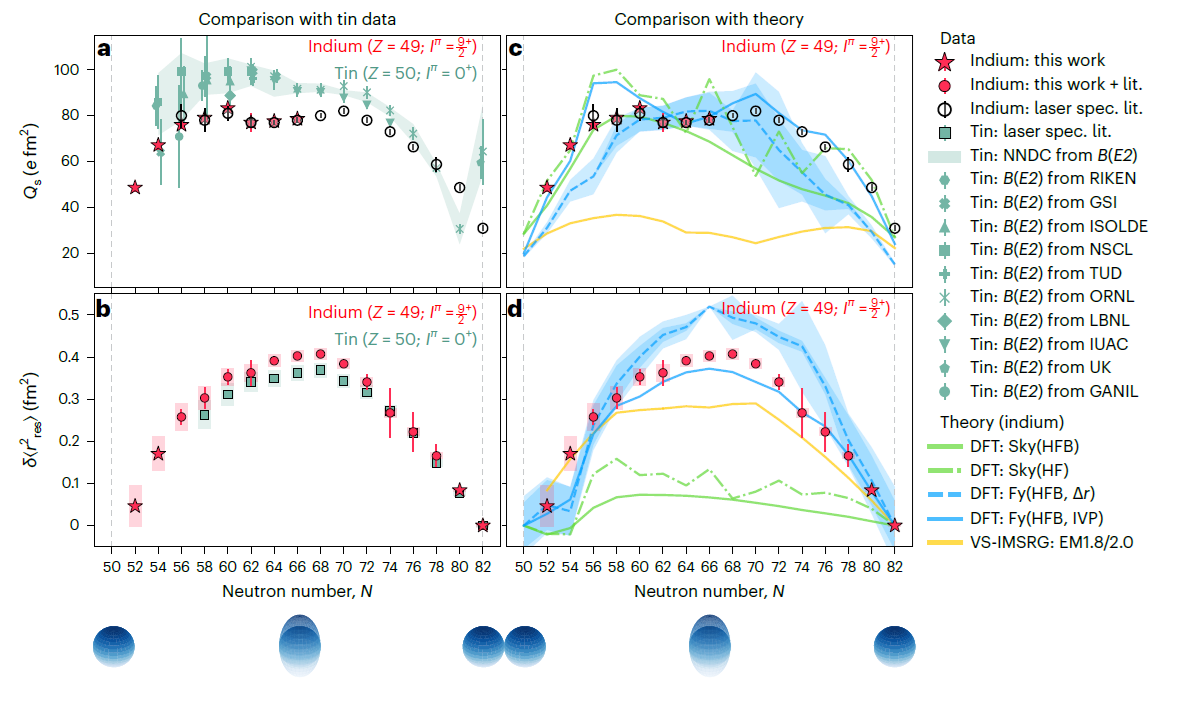}
    \vspace{-1cm}
\caption{{Evolution of nuclear deformation between the neutron magic numbers $N=50$ and $82$ for tin and indium.}
(a,c) Spectroscopic quadrupole moments $Q_s$ with experimental uncertainties for even-$N$ isotopes of indium ($Z=49$) and tin ($Z=50$), compared with nuclear theory. The quadrupole moments of indium isotopes are obtained from laser spectroscopy measurements, while tin values are derived from $B(E2)$ measurements.
(b,d) Residual differential charge radii $\delta\langle r^2\rangle_{\mathrm{res}}$, compared between isotopes and with theoretical predictions.
 Theoretical results include selected DFT and \textit{ab initio} calculations. The blue spheres illustrate the evolution of deformation between the closed shells. Figure reproduced from \cite{Kar24}, under CC BY 4.0.}
    \label{fig:Sn_In}
\end{figure*}

The quadrupole deformation of a nucleus affects its charge radius \cite{Gel22}. In a semi-classical view, a deformed nucleus can have a larger root-mean-square radius than a spherical one of the same volume, due to the anisotropy of the charge distribution.  A simple relationship between the charge radius and the deformation parameter $\beta_2$ can be expressed to leading order \cite{bohr1975nuclear}:
\begin{equation}
\langle r^2\rangle \approx \langle r^2\rangle_\text{sph}\left(1 + \frac{5}{4\pi}\beta_2^2\right),
\label{eq:r2_deformation}
\end{equation}
where $\langle r^2\rangle_\text{sph}$ is the spherical contribution.  Sudden changes in the charge radius along an isotopic chain, therefore, signal the onset or disappearance of deformation. Figure~\ref{fig:Sn_In} shows the relative changes in the nuclear charge radii for both Sn and In isotopes between the neutron closed shells at $N=50$ and $N=82$. A clear parabolic behavior appears, reflecting the evolution of deformation ($\beta_2$), with a maximum at midshell and minima at the neutron closed shells. Further discussion of the Figure in the context of the quadrupole moment is given in Sec.~\ref{sec:quadrupole}.

The presence of nearly degenerate nuclear states with different shapes, known as shape coexistence, has been evidenced through the simultaneous measurement of ground-state and isomeric charge radii in regions such as the mercury and lead isotopes \cite{Mar18}, where intruder configurations with different deformations coexist at low excitation energy.

\subsubsection{Nuclear size far from stability}

Moving towards the drip lines, charge radii become a probe of the interplay between nuclear attraction, Coulomb repulsion, and weak binding.  Neutron-rich isotopes with large neutron-to-proton asymmetries exhibit departures from the $A^{1/3}$ scaling, including halo structures (where one or two loosely bound neutrons extend far beyond the nuclear core) and neutron skins.

In light nuclei, the halo structure of $^{6,8}$He~\cite{wang2004nuclear}, $^{11}$Li~\cite{sanchez2006nuclear}, and $^{11}$Be~\cite{nortershauser2009nuclear} has been confirmed through their anomalously large charge radii.  In medium-mass nuclei, the sudden increase in charge radii of neutron-rich potassium and calcium isotopes beyond $N = 28$~\cite{garcia2016unexpectedly,koszorus2019precision} discussed above highlights the role of weakly bound orbitals and correlations absent in stable nuclei.  Improvement of many-body methods to include higher-order correlations yielded no improvement in reproducing these large radii~\cite{heinz2025improved}, pointing to the nuclear interaction as the primary source of the remaining discrepancy.

\begin{figure*}[t]
    \centering
    \includegraphics[width=\linewidth]{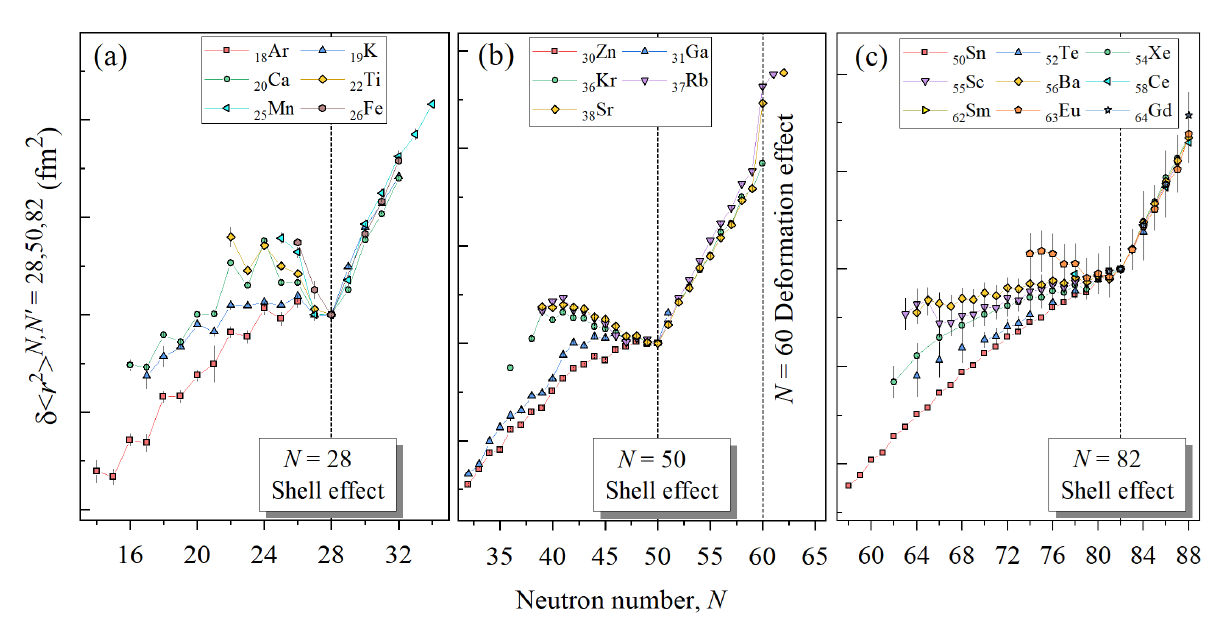}
 \caption{Changes in mean-square charge radii in the calcium ($Z=20$), nickel ($Z=28$), and tin ($Z=50$) isotopic chains as a function of neutron number $N$, relative to a reference isotope with a magic neutron number. A pronounced kink is observed at the neutron shell closures. Beyond these magic numbers, neutron-rich isotopes exhibit a similar slope in the increase of charge radii across all three elements, whereas neutron-deficient isotopes show a stronger dependence on proton number. Figure reproduced from \cite{yang2023laser}.}
    \label{fig:Ca_Ni_Sn}
\end{figure*}

Fig.~\ref{fig:Ca_Ni_Sn} shows the evolution of the charge radii in different regions of the nuclear chart, relative to the values at the neutron magic numbers $N=28, 50,$ and $82$. The trends exhibit a clear kink at these shell closures, followed by a nearly universal slope for neutron-rich isotopes. In contrast, before reaching the neutron magic numbers, the evolution of the charge radii shows a marked dependence on the atomic number $Z$.

\subsubsection{Odd-even staggering}

The odd-even staggering (OES) of charge radii---the systematic difference between radii of even-$N$ and odd-$N$ isotopes---provides insight into nuclear pairing correlations.  The three-point indicator, defined as
\begin{equation}
\Delta^{[3]}(N) = \frac{1}{2}\left[\langle r^2\rangle(N+1) - 2\langle r^2\rangle(N) + \langle r^2\rangle(N-1)\right],
\label{eq:OES}
\end{equation}
is used to isolate the effects from the relative changes due to the addition of an unpaired nucleon.  The magnitude of the OES correlates with pairing gaps and shell closures, linking charge radii to pairing properties of the nuclear force.

\begin{figure}[t]
    \centering
    \includegraphics[width=\linewidth]{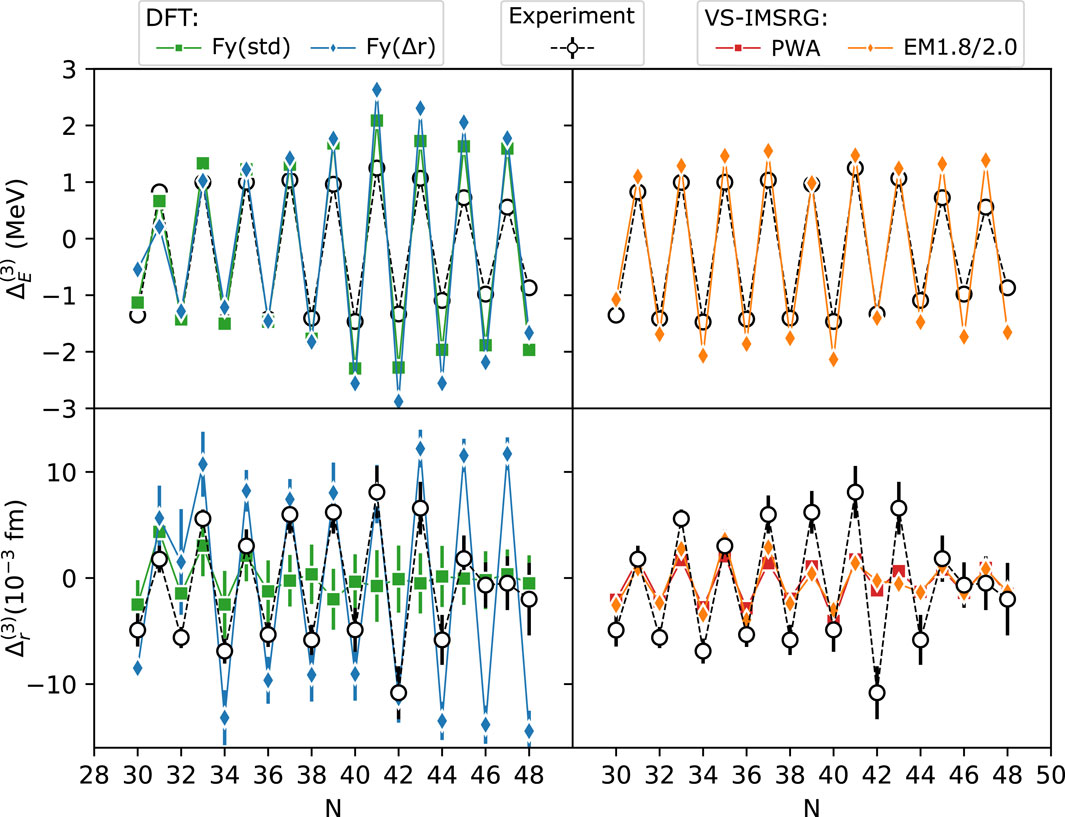}
    \caption{Odd-even staggering of nuclear charge radii in selected isotopic chains, compared with \textit{ab initio} calculations.  The staggering amplitude is sensitive to pairing correlations and shell structure, providing a strict test of nuclear forces that is largely independent of the global charge radius trend.  Figure reproduced from Ref.~\cite{de2020measurement}, under CC BY 4.0.}
    \label{fig:OES}
\end{figure}

OES offer a compelling test of nuclear theory because it probes relative changes between adjacent isotopes, filtering out global trends.  \textit{Ab initio} methods have been found to reproduce OES as well as or better than DFT in copper and tin isotopes~\cite{de2020measurement,Gus25}, even when using interactions that consistently predict smaller absolute radii, as shown in Fig.~\ref{fig:OES}.  This indicates that OES probes complementary details of nuclear structure such as pairing and single-particle occupation, which are largely independent of the overall nuclear saturation properties.

\subsubsection{Mirror nuclei and the nuclear equation of state}

The difference in charge radii between mirror nuclei---pairs where the numbers of protons and neutrons are interchanged ($Z \leftrightarrow N$) provides a probe of isospin-asymmetric components of the nuclear force.  Since mirror pairs are nearly identical in structure, any difference in their radii is driven primarily by isospin-breaking effects, including the Coulomb interaction and charge-symmetry-breaking nuclear forces.

It has been shown that mirror charge radius differences are proportional to the slope parameter $L$ of the nuclear symmetry energy at saturation density~\cite{brown2017mirror,yang2018mirror}, a quantity of central importance for the physics of neutron stars.  Experimental measurements of mirror radii in the silicon~\cite{konig2024silicon} and calcium chains have been used to constrain $L$, providing independent information to neutron-skin measurements (PREX/CREX \cite{adhikari2021accurate,adhikari2022precision}) and gravitational-wave observations.

\subsubsection{Electron scattering}

Elastic electron-nucleus scattering measures the nuclear charge form factor $F_\text{ch}(q^2)$ (Sec.~\ref{sec:FS_formfactor}) over a range of momentum transfers, from which the charge density distribution $\rho_c(r)$ and its moments ($\langle r^2\rangle$, $\langle r^4\rangle$, \ldots) can be extracted.  Electron scattering provides \emph{absolute} charge radii (not just differences between isotopes, as in laser spectroscopy) and can map out the full shape of the charge distribution.  However, it requires relatively large target quantities and is therefore limited to stable or long-lived isotopes.  The SCRIT facility \cite{Tsu17} at RIKEN and the proposed ELISe experiment at FAIR \cite{Antonov2011ELISe} aim to extend electron scattering to unstable nuclei using stored radioactive beams.

\subsubsection{Muonic atom spectroscopy}

When a muon replaces an electron in an atom, its larger mass ($m_\mu \approx 207\, m_e$) brings its wave function much closer to the nucleus, increasing the overlap with the nuclear charge distribution by a factor of $\sim 207^3 \sim 10^7$.  The resulting muonic X-ray energies are highly sensitive to the nuclear charge radius and higher radial moments~\cite{fricke2004nuclear,Bey26}.  Muonic atom spectroscopy can provide precise {absolute} charge radii measurements for stable nuclei and is being extended to some radioactive species at PSI (Switzerland)\cite{Bey26}.  The ``proton radius puzzle'', a previous discrepancy between the proton charge radius measured in muonic hydrogen and in electronic systems,  stimulated a broad program of muonic atom measurements across the periodic table~\cite{antognini2013proton,pohl2016laser}.

\subsubsection{M\"ossbauer spectroscopy}
\label{sec:mossbauer}

An alternative, solid-state route to the nuclear charge radius exploits the recoilless emission and resonant absorption of nuclear $\gamma$ rays, the M\"ossbauer effect~\cite{mossbauer1958}.  When the resonant nuclei are bound in a crystal lattice, a finite fraction of transitions proceed without phonon recoil, so that the emitted and absorbed photons retain the natural linewidth ($\Gamma/E_\gamma \sim 10^{-12}$ for the $14.4$~keV transition of $^{57}$Fe), and a small Doppler velocity ($\delta E = E_\gamma v/c$, typically mm/s) tunes source and absorber through resonance.

The centroid of the resonance is displaced by the {isomer shift}, which originates from the same finite-size interaction as the field shift of Sec.~\ref{sec:term_FS} (Eq.~(\ref{eq:FS_shift})), but with the roles of the two factors interchanged.  Whereas the optical isotope shift compares one electronic transition across two isotopes and so probes $\delta\langle r^2\rangle^{A,A'}$ between the ground states of different nuclei (Eq.~(\ref{eq:IS_FS})), the M\"ossbauer isomer shift compares one nuclear transition in two electronic environments, labeled as source $S$ and absorber $A$. They are therefore sensitive to the change in mean-square charge radius between the {excited and ground states of the same nucleus},
\begin{equation}
\delta_\text{IS} = \underbrace{\frac{Ze^2}{6\epsilon_0}\,\Delta|\psi_e(0)|^2}_\text{electronic} \times \underbrace{\delta\langle r^2\rangle_{eg}}_\text{nuclear}\,,
\label{eq:isomer_shift}
\end{equation}
with 
\begin{equation}
\delta\langle r^2\rangle_{eg} \equiv \langle r^2\rangle_e - \langle r^2\rangle_g\,,
\label{eq:isomer_shift2}
\end{equation}
and $\Delta|\psi_e(0)|^2 = |\psi_e(0)|^2_A - |\psi_e(0)|^2_S$ is the difference in electronic contact density between the two environments.  The factorization is similar in structure to Eqs.~(\ref{eq:FS_shift})--(\ref{eq:F_def}); the electronic prefactor is the M\"ossbauer analog of the field-shift factor $F$ and is obtained either from relativistic electronic-structure calculations or, more robustly, by forming ratios of isomer shifts in isoelectronic compounds, which cancel it.  In this way, $\delta\langle r^2\rangle_{eg}$ has been extracted for the $\gtrsim 40$ nuclides with a transition suitable for the technique~\cite{kalvius1974,shenoy1978}.

 M\"ossbauer measurements have been applied to several radioactive species such as $^{129}$I (populated from $^{129m}$Te) and $^{237}$Np ~\cite{spijkervet1979}, and \emph{emission} M\"ossbauer spectroscopy implants short-lived parents produced online at ISOLDE-CERN (e.g.\ $^{57}$Mn, $t_{1/2}=1.5$~min) whose decay feeds the resonant level~\cite{mokhlesgerami2016}.  Two requirements nonetheless limit the reach of the method: the transition energy must lie in the recoilless window ($\sim 10$--$100$~keV), and the intermediate state must have a lifetime of $\sim 1$--$100$~ns, while the resonant nuclei must be held in a lattice in sufficient quantity.  The technique, therefore, accesses a fixed roster of M\"ossbauer-active isotopes and yields the ground-to-excited radius difference rather than the isotope-chain trends $\delta\langle r^2\rangle^{A,A'}$ that laser spectroscopy maps far from stability.  Synchrotron-based nuclear resonant scattering relaxes the source requirement by exciting the M\"ossbauer level directly with pulsed X-rays, extending the approach to resonances such as $^{193}$Ir that are impractical with decay sources~\cite{alexeev2019}.

The natural meeting point between this technique and the present review is the $^{229}$Th isomer (Sec.~\ref{sec:Th229_structure}).  Its $^{229m}\text{Th}\to{}^{229}\text{Th}$ transition is effectively the lowest-energy M\"ossbauer-like transition known, and the isomer shift measured in the electronic spectrum yields $\delta\langle r^2\rangle_{eg} = 0.0105(13)$~fm$^2$~\cite{safronova2018thoriumradii,Thiel18}---the observable of Eq.~(\ref{eq:isomer_shift}) obtained optically rather than through a $\gamma$ resonance.

\subsubsection{Higher radial moments and King-non-linearity}

\label{sec:IS_newphysics}

The isotope shift expansion (Eq.~(\ref{eq:FS_higher})) contains contributions from $\langle r^4\rangle$, $\langle r^6\rangle$, and higher radial moments of the charge distribution, which probe the nuclear surface diffuseness and the detailed shape of $\rho_c(r)$.  In heavy elements ($Z \gtrsim 50$), the $\langle r^4\rangle$ contribution can reach the percent level of the total field shift and must be accounted for in precision extractions of $\delta\langle r^2\rangle$.  Recent \textit{ab initio} calculations have begun to provide predictions for fourth-order charge density moments~\cite{companysfranzke2025r4}, connecting these quantities to neutron radii and the nuclear equation of state.

Precision atomic spectroscopy provides a powerful probe of nuclear structure and possible new physics by exploiting the extreme sensitivity of electronic energy levels to small nuclear effects. Since nuclei are finite-size objects, their charge distributions induce tiny corrections to atomic spectra, typically suppressed by the ratio of nuclear to atomic scales. At leading order, these effects are governed by the ratio of the root-mean-square charge radius $\sqrt{\langle r^2 \rangle}$, being of the order of the Fermi scale to that of the Bohr one, but higher moments such as $\langle r^4 \rangle$ encode more detailed information about the nuclear charge distribution \cite{reinhard2020beyond}. Exploring this higher-moment ``frontier'' opens a new avenue for studying nuclear structure, while also setting the stage for precision tests of physics beyond the Standard Model \cite{counts2020evidence}.

A key observable in this context is the so-called King linearity \cite{king2013isotope}. To leading order, isotope shifts measured in different atomic transitions factorize into electronic and nuclear contributions, leading to a linear relation between appropriately normalized shifts. Deviations from this relation, known as King non-linearity (KNL), provide a sensitive diagnostic of subleading effects. These can arise either from higher-order Standard Model contributions---such as terms involving $\langle r^4 \rangle$, quadratic field shifts, or nuclear polarization---or from new forces. 

Measurement of King non-linearity could be related to such 5th force searches as follows. A new boson, mediating an electron-neutron interaction, would produce an isotope-dependent shift with a different nuclear scaling than the SM contributions \cite{delaunay2017probing,berengut2018probing,frugiuele2017constraining}. The nuclear dependence of these contributions is trivial due to Gauss law \cite{counts2020evidence,hur2022evidence}, and thus it only relies on the knowledge of the electronic structure, assuming that the other SM contributions are under control. The sensitivity of King plot searches for new forces improves with the number of isotopes and transitions measured.  Extending the measurements to long isotopic chains of exotic nuclei---accessible at RIB facilities using the laser spectroscopy techniques of Sec.~\ref{sec:exp_techniques}---would increase the lever arm and help separate nuclear-structure from new-physics contributions.  Proposals to use optical clocks on different isotopes of the same element~\cite{flambaum2018isotope}, or molecular isotopologue comparisons, would push the sensitivity to new bosons into unexplored territory.

Recent high-precision measurements have begun to observe such non-linearities, pushing experimental sensitivity to unprecedented levels \cite{hur2022evidence}. We are now in a position where sub-kHz spectroscopy has allowed us to establish King NL in several systems, most notably
Yb\textsuperscript{+} \cite{counts2020evidence,hur2022evidence,Ono:2021ogd,Door:2024qqz} and Ca\textsuperscript{+} ions \cite{Rehbehn:2021zlr, Wilzewski:2024wap}.
There are many more systems where King NL could be measured in the future (see, e.g.,~\cite{solaro2020improved}). 
This remarkable experimental progress raises the question of whether such high-precision data can also be leveraged to probe nuclear structure with unprecedented accuracy.
The central difficulty in describing and interpreting King NL is the limited theoretical control over non-perturbative SM contributions (see, for instance, ref.~\cite{Berengut:2025nxp} for a recent review). 
Understanding the SM contributions remains an active area of research, and it is not yet fully settled whether all relevant sources have been correctly identified. 

In the pioneering work of ref.~\cite{counts2020evidence}, which first established the observation of King NL, the $\langle r^4 \rangle$ and $\langle r^2 \rangle^2$ contributions were, for instance, treated as related effects. Subsequent studies, however, recognized them as distinct contributions \cite{hur2022evidence}, which received some support from the \textit{ab initio} nuclear-structure calculations presented in~\cite{Door:2024qqz}. The situation is further complicated by the fact that not all SM corrections generate independent sources of King NL. Some higher-order shifts in the electronic transition energy can be absorbed into common prefactors in the King linearity relation, thereby obscuring which effects genuinely contribute to the King NL.

One way to interpret the measurements and the different contributions in a controlled manner is to use an effective field theory framework based on non-relativistic QED (NRQED) \cite{Assi:2025nim}. The problem exhibits a clear hierarchy of scales, from the nuclear size and mass down to the electronic momentum and binding energy, allowing for a systematic expansion in small parameters. Within this framework, nuclear structure effects are encoded in a finite set of Wilson coefficients, including charge-radius moments and polarizabilities, which can be matched onto observable energy shifts.

\subsubsection{Searching for ultralight DM with oscillating charge radius}

An oscillating charge radius can also be used to search for new physics, such as ultralight DM. As explained in section~\ref{sec:nuclear_clock}, it is particularly challenging to probe ultralight dark matter that couples to the nuclear sector. In~\cite{Banerjee:2023bjc}, it was pointed out that this class of DM models leads to oscillation of the charge radius.
As atomic energy levels are sensitive to finite nuclear size effects, this induces corresponding oscillations in atomic transition frequencies. The effect is enhanced in heavy nuclei, where sensitivity to nuclear structure is larger, making precision spectroscopy a particularly powerful probe.

The DM's induced frequency modulation can be written schematically as
\begin{equation}
\delta \nu_i(t) \;=\; F_i \,\delta \langle r^2 \rangle(t)
\;\simeq\; F_i \, d_\phi \,\phi_0 \cos(m_\phi t)\,,
\end{equation}
where $F_i$ is the field-shift coefficient for transition $i$, $d_\phi$ parametrizes the coupling of the dark-matter field $\phi$ to nuclear matter, and $m_\phi$ is the dark-matter mass. Comparing transitions with different $F_i$ allows one to isolate this effect, analogously to King-plot constructions, and to distinguish it from standard nuclear contributions.
Using high-precision optical clock systems, such as $^{171}\mathrm{Yb}^+$, the authors show that existing and near-future measurements can probe these oscillatory signals with high sensitivity. By analyzing transitions with different sensitivities to nuclear size, one can either search for time-dependent frequency modulations or place bounds on the ultralight DM couplings to nucleons. This establishes a close conceptual link between King non-linearity searches and ultralight dark-matter detection: both rely on the factorization of electronic and nuclear effects and on identifying controlled violations of this structure, whether static (as in KNL) or time-dependent (as induced by UDM).

\subsection{Nuclear electromagnetic moments}
\label{sec:EM_moments}

\subsubsection{Magnetic dipole moments}

The magnetic dipole moment $\mu$ (Sec.~\ref{sec:term_mu}, Fig.~\ref{fig:multipoles}a) can be extracted from the magnetic hyperfine constant $A_\text{hfs}$ through the factorized relation of Eq.~(\ref{eq:Ahfs}).  Systematic measurements of magnetic moments across isotopic chains, enabled by laser spectroscopy at RIB facilities, probe the evolution of nuclear structure, particularly the single-particle and collective contributions to the nuclear current distribution \cite{Neyens2005,dobaczewski2025electromagnetic}.

\begin{figure}[t]
    \centering
    \includegraphics[width=0.8\linewidth]{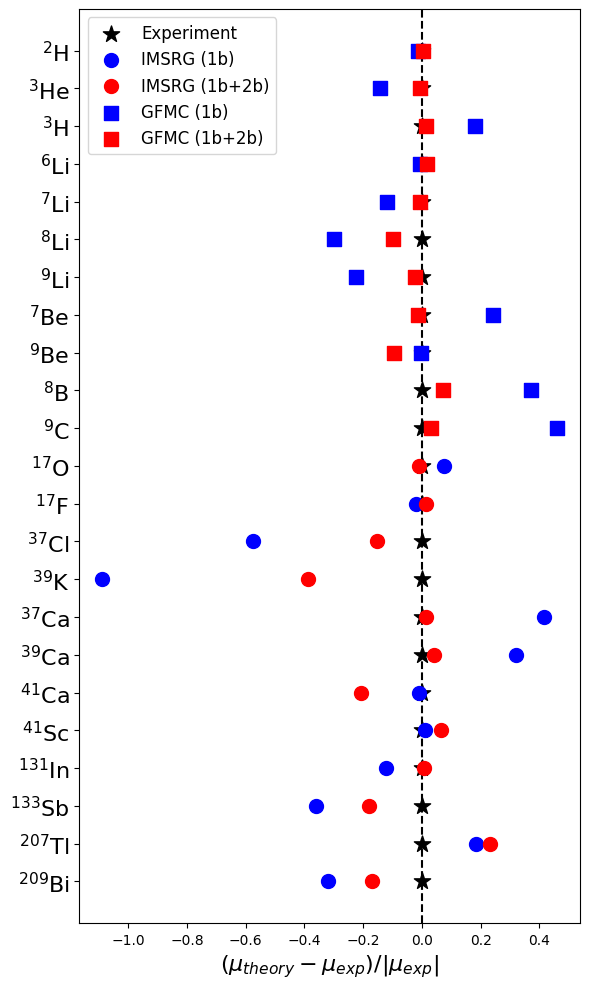}
    \caption{Magnetic dipole moments computed using Green's Function Monte Carlo (squares)~\cite{pastore2013quantum} and the VS-IMSRG~\cite{miyagi2024impact} (circles) with one-body currents only (blue) or one-body plus two-body currents (red), compared with experiment (black).  The inclusion of consistently derived two-body currents from chiral EFT removes the need for phenomenological quenching factors and improves agreement across a wide range of nuclear masses.}
    \label{fig:dipolecurrent}
\end{figure}

In the extreme single-particle (Schmidt) limit, the magnetic moment is determined by the orbital and spin angular momenta of the unpaired nucleon.  Deviations from the Schmidt values can arise from configuration mixing, core polarization, and meson-exchange currents \cite{Papu13, Gar15,Groote17, papuga2014shell,vernon2022nuclear}.  As discussed in Sec.~\ref{sec:EM_operators}, the inclusion of two-body currents from chiral EFT~\cite{miyagi2024impact} has been shown to systematically improve the agreement between \textit{ab initio} calculations and experiment, resolving a long-standing discrepancy that had previously required phenomenological quenching factors.  This is illustrated in Fig.~\ref{fig:dipolecurrent}. The blue points, computed with one-body currents only, systematically deviate from experiment, while the red points, which include two-body currents, fall close to the experimental values across a wide range of masses from light to heavy-mass nuclei.

Magnetic moments are particularly sensitive to the specific single-particle orbital occupied by the valence nucleon, making them powerful probes of shell evolution \cite{Neyens2005,vernon2022nuclear}.  Changes in the ground-state spin and magnetic moment along an isotopic chain can signal the crossing or inversion of single-particle levels \cite{Papu13}, a phenomenon that has been observed in different regions, including the ``island of inversion'' around $N = 20$ in the sodium and magnesium chains, and the $N = 28$ region in the potassium chain \cite{Papu13}.

\subsubsection{Electric quadrupole moments}
\label{sec:quadrupole}
The nuclear electric quadrupole moment $Q$ (Sec.~\ref{sec:term_Q}, Fig.~\ref{fig:multipoles}b) can be extracted from the quadrupole hyperfine constant $B_\text{hfs}$ using the relation shown in Eq. \ref{eq:EF_quad}.  The quadrupole moment provides a measurement of the anisotropy of the nuclear charge distribution and encodes information on nuclear deformation.

As discussed in Sec.~\ref{sec:EM_operators}, the quadrupole moment is inherently tied with shell closures, being small near magic numbers indicating a spherical ground state~\cite{Kar24,yang2023laser}), and growing away from them, interpreted as enhanced collectivity.

The spectroscopic quadrupole moment $Q_s$ and the reduced transition probability $B(E2)$ provide important information on collective nuclear phenomena. While $Q_s$ probes the static quadrupole moment of a given state, $B(E2)$ reflects the collectivity of quadrupole transitions between states. In the rotational model, both observables can be related through the intrinsic quadrupole moment $Q_0$,
\begin{equation}
B(E2; J_i \rightarrow J_f) = \frac{5}{16\pi} Q_0^2 \, \langle J_i K\, 2\, 0 | J_f K \rangle^2,
\end{equation}
\begin{equation}
Q_s(J) = \frac{3K^2 - J(J+1)}{(J+1)(2J+3)} \, Q_0,
\end{equation}
such that $B(E2)$ determines the magnitude of the deformation ($Q_0^2$), whereas $Q_s$ provides both its magnitude and sign. In Fig. \ref{fig:Sn_In}, the $Q_s$ values for indium isotopes and those derived from $B(E2)$ measurements in tin thus offer a consistent view of the evolution of nuclear collectivity between the neutron shell closures at $N=50$ and $N=82$. 

As seen in Fig.~\ref{fig:Sn_In}, \textit{ab initio} calculations (IMSRG) describe well the relative changes in the nuclear charge radii, but fall short in reproducing the spectroscopic quadrupole moments. DFT calculations, however, perform fairly well for both observables. As discussed in Sec.~\ref{sec:EM_operators}, the collectivity of the quadrupole moment makes it a challenging test of microscopic theory.  As discussed in Section ~\ref{sec:quadrupole_theory}, the phenomenological shell model requires effective charges to compensate for missing excitations, while \textit{ab initio} methods that explicitly include deformation (NCSM, IM-GCM) can reproduce $E2$ observables without such adjustments~\cite{caprio2022robust,henderson2018testing}. 

\subsubsection{Higher-order moments}

\paragraph{Magnetic octupole moment.}
The magnetic octupole moment~\cite{butler1996intrinsic, butler2016octupole, cao2020landscape} ($\Omega$, requiring $I \geq 3/2$) has been measured in a few stable atoms (e.g., $^{133}$Cs \cite{gerginov2003observation}, $^{209}$Bi \cite{li2022re}, $^{173}$Yb \cite{degroote2021}) from the octupole hyperfine interaction.  It probes the next term in the magnetization distribution beyond the dipole and is sensitive to nuclear deformation and configuration mixing.  Extension to unstable isotopes and molecular systems remains an open frontier. 

\begin{figure}[t]
    \centering
    \includegraphics[width=1\linewidth]{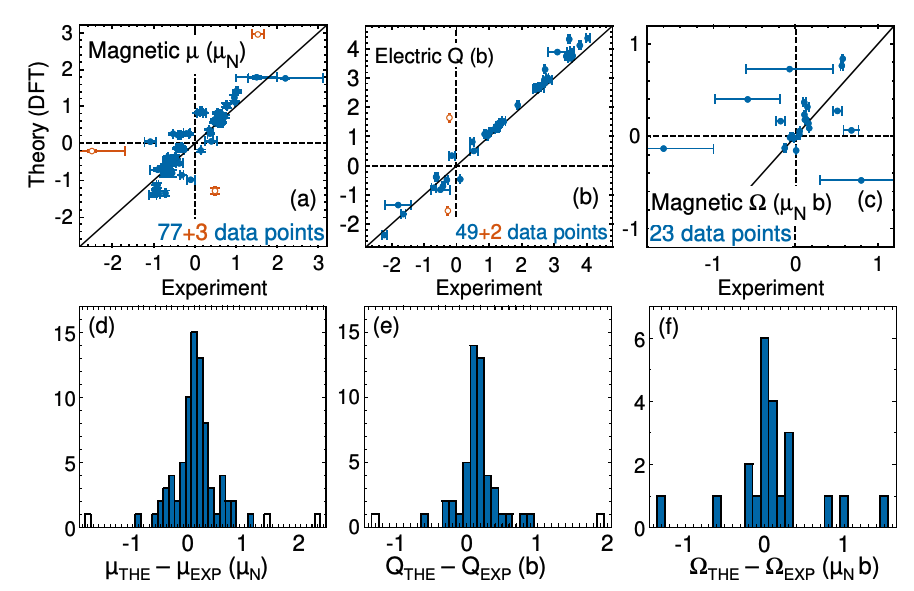}
    \caption{Magnetic dipole moments (a), quadrupole moments (b) between the gadolinium and osmium isotopic chains, and magnetic octupole moments (c) across the nuclear chart computed using nuclear DFT. A generally good agreement is found, except in the case of a few outliers indicated by open symbols. Distribution of the residuals between theory and experiments is shown in panels d, e, and f, respectively. Figure reproduced from Ref. \cite{dobaczewski2026electromagneic}, under CC BY 4.0 and Ref. \cite{dobaczewski2025electromagnetic}, with permission from the authors.}
    \label{fig:dftmoments}
\end{figure}

From the theory side, this moment has been evaluated using nuclear DFT~\cite{dobaczewski2025electromagnetic, dobaczewski2026electromagneic} alongside the magnetic moment and quadrupole moment. Results are presented in Fig.~\ref{fig:dftmoments} for isotopes between gadolinium and osmium for the magnetic dipole and electric quadrupole moments and across the nuclear chart for the octupole moment. The magnetic dipole moment is evaluated using uniquely the leading-order one-body operator.

\paragraph{Electric hexadecapole moment.}
The electric hexadecapole ($I \geq 2$) is even more challenging to access experimentally.  Proposals exist to use molecular systems, where the higher-order electric field gradients are enhanced by the directed molecular bond, to access these higher multipoles \cite{Hammerle1973}.

\subsection{Recent highlights across the nuclear chart}
\label{sec:highlights}

The combination of improved RIB production, advanced laser spectroscopy techniques, and \textit{ab initio} nuclear theory has produced a series of notable results in recent years.  We highlight a selection that illustrates the breadth of the program.

\subsubsection{Light nuclei (\texorpdfstring{$Z\le 16$}{Z<16}): halos, clustering, and shell evolution}
Nuclei with $Z\le 16$ are accessible to quasi-exact \textit{ab initio} calculations such as the no-core shell model and quantum Monte Carlo methods of Sec.~\ref{sec:nuclear_theory}. Thus, measurements of their nuclear properties provide important guidance for the development of nuclear theory. Laser spectroscopy has provided measurements of nuclear electromagnetic properties of short-lived isotopes in this region, but accurate extraction of nuclear charge radii are challenging as the nuclear-volume contribution to the isotope shift is smaller than the mass shift by three to four orders of magnitude, so extracting $\delta\langle r^2\rangle$ demands atomic mass-shift and QED calculations of comparable accuracy (Sec.~\ref{sec:exp_techniques}).

The lightest halo nuclei remain the defining benchmarks.  The neutron-halo isotopes $^{6}$He~\cite{wang2004nuclear}, $^{8}$He~\cite{mueller2007he8}, $^{11}$Li~\cite{sanchez2006nuclear}, and the one-neutron halo $^{11}$Be~\cite{nortershauser2009nuclear} were measured by trapping and collinear techniques on systems with at most three electrons, where the mass shift is calculable to the required precision; their radii encode how the loosely bound halo neutrons polarize and recoil against the core, producing nonmonotonic trends that only realistic treatments of nucleon correlations reproduce.  Proton halos appear on the other side of stability: the two-proton-halo candidate $^{17}$Ne was characterized through its charge radius and moments~\cite{geithner2008ne}, and the one-proton-halo candidate $^{8}$B is the target of an ongoing laser-spectroscopy program at ATLAS/ANL \cite{maass2019boron}.

Clustering leaves an additional important imprint. The charge radius of $^{12}$Be, measured at TRIUMF, exceeds the trend of the lighter beryllium isotopes and signals the breakdown of the $N=8$ shell closure and the onset of cluster correlations~\cite{krieger2012be12}.  Pushing precise mass-shift calculations to a five-electron system, laser spectroscopy of $^{10,11}$B enabled their first charge radii measurements~\cite{maass2019boron}.  The extracted charge radii difference is consistent with a cluster picture and benchmarks no-core shell model and Green's function Monte Carlo predictions.  Most recently, the charge radius of $^{13}$C was determined by laser spectroscopy of the helium-like ion $^{13}$C$^{4+}$, whose two-electron structure restores the atomic-theory precision available in the lightest systems. The measurement improved the electronic charge radius of $^{13}$C, revealed a $3\sigma$ tension with the muonic-atom value, and benchmarked chiral effective field theory calculations including two-body charge operators~\cite{muller2025c13}.

Higher in mass, charge radii trace the evolution of nuclear shape and the disappearance of the $N=20$ shell closure in the ``island of inversion.''  Collinear laser spectroscopy mapped the complete chain of magnesium isotopes $^{21-32}$Mg, exposing shape changes correlated with the neutron shell structure from the $N=8$ cluster region to the deformed $N=20$ region~\cite{yordanov2012mg}.  The first charge radii of radioactive aluminium ($^{27-32}$Al) followed, where a reduced radius near $N=19$ hints at a residual $N=20$ shell effect visible in this chain but not in its neighbours~\cite{heylen2021al}. On the neutron-deficient side, Resonance Ionization Spectroscopy Experiment (RISE) at FRIB extended the chain in the opposite direction, measuring the neutron-deficient isotopes from $^{25}$Al down to the proton-drip-line nucleus $^{22}$Al~\cite{Brinson2026AlChargeRadii}.  These proton-rich data reveal an increase of the charge radius toward the drip line, with $^{22}$Al and $^{23}$Al of nearly equal size, and a comparison with their mirror partners that tracks the calculated proton skins and follows the systematics of well-bound nuclei.  

Extending the reach to silicon, the charge radius of $^{32}$Si was measured and confronted with nuclear lattice effective field theory, the in-medium similarity renormalization group, and density functional theory; completing the mirror pair $^{32}$Ar--$^{32}$Si, its difference constrains the slope $L$ of the symmetry energy in the nuclear equation of state to $L\lesssim 60$~MeV~\cite{konig2024silicon,konig2024siliconerratum}, linking the properties of light nuclei to the physics of neutron-rich matter. 

The proton-drip-line measurement of $^{22}$Al~\cite{Brinson2026AlChargeRadii} shows that laser spectroscopy, combined with the unique capabilities of FRIB, can now reach the limits of existence on the proton-rich side. On the neutron-rich side, the radii of the most exotic helium, lithium, and beryllium isotopes, and of the one-proton-halo candidate $^{8}$B, remain prime targets. Toward small $Z$, the uncertainty for the charge radii measurements is limited by atomic theory, as the extraction of nuclear properties requires mass-shift and QED calculations at a precision currently reached only for systems with a few electrons. At present, significant gaps remain in our knowledge of nuclear charge radii for
key elements such as nitrogen, oxygen, fluorine, phosphorus, and sulfur. These
gaps are largely due to the difficulty of producing the relevant isotopes and to
the complexity of their atomic structures, for which many of the accessible
ground-state transitions lie in the deep-ultraviolet spectral region.

\subsubsection{The calcium region: a challenge for nuclear theory}

As already briefly touched upon in Sec.~\ref{sec:charge_radii}, the calcium isotopic chain ($Z = 20$) is considered a major benchmark for nuclear theory, anchored by the doubly magic $^{40}$Ca and $^{48}$Ca and extending in both directions to short-lived isotopes.  On the neutron-rich side, an observed rapid increase in the charge radii beyond $N = 28$, towards the proposed $N = 32$ and $N = 34$ subshell closures, was revealed by laser spectroscopy measurements of $^{49,51,52}$Ca ~\cite{garcia2016unexpectedly}.  On the proton-rich side, the charge radii of $^{36,37,38}$Ca were found to decrease with a pronounced odd-even staggering across the chain. The results can be reproduced within DFT with the so-called Fayans functional, with a dedicated pairing term, pointing to the role of nucleonic superfluidity in weakly bound systems~\cite{miller2019calcium}. While improved nuclear density functional theory has successfully described the
observed systematics of the calcium charge radii \cite{miller2019calcium,koszorus2021potassium}, a
comparable description within \textit{ab initio} nuclear theory remains an
open challenge.

Adding or removing a single proton provides complementary tests.  In potassium ($Z = 19$), collinear resonance ionization spectroscopy extended the chain to $^{52}$K and found gradual increase of the nuclear raii towards $N~=~32$, in contrast to expectations commonly observed at magic numbers~\cite{kreim2014nuclear,koszorus2019precision,koszorus2021potassium}.  In scandium ($Z = 21$), the charge radii of $^{47-49}$Sc display a markedly reduced odd-even staggering relative to calcium despite the single-proton difference. The models that reproduce the calcium radii fail for scandium, while the trend across the $N = 28$ isotones follows the seniority symmetry of the $0f_{7/2}$ nuclear shell~\cite{bai2025scandium}.

Deeper into the $fp$ shell, the charge radii trace the erosion of the $N = 28$ closure and the onset of collectivity toward $N = 40$.  The manganese chain ($Z = 25$) was mapped across this region, its radii revealing increased proton and neutron excitations as $N = 40$ is approached~\cite{heylen2016manganese,Bab15,Bab16}.  The remaining $fp$-shell elements, such as titanium, vanadium, chromium, iron, and cobalt, are only beginning to be reached by laser spectroscopy, limited by unfavorable atomic transitions and low production yields \cite{Karadimas2026}.

\subsubsection{The nickel region, shape coexistence, and shell evolution}

Laser spectroscopy measurements are now spanning the nickel, copper, and zinc chains across the $N = 40$ subshell and toward the doubly magic $^{78}$Ni at $N = 50$, where shape coexistence has been suggested \cite{Yang16a,Nowacki2016Ni78,Taniuchi2019Ni78}.  For Ni, collinear laser spectroscopy of $^{58-68,70}$Ni provided another test of \textit{ab initio} methods ~\cite{malbrunot2022nickel,kaufmann2020nickel} (see Fig.  \ref{fig:abinitio_radii}).  On the proton-rich side, the radii of $^{55,56}$Ni reveal a behavior at $N = 28$ strikingly similar to that of the calcium isotones, while neutron-deficient $^{54}$Ni anchors the mirror-pair determination of the symmetry energy discussed above~\cite{sommer2022nickel,pineda2021charge}.

Adding a single proton, the copper isotopes ($Z = 29$) exhibit a pronounced odd-even staggering in their charge radii that persists across $N = 40$. The magnitude and trend of this staggering are well reproduced by VS-IMSRG calculations, lending confidence to the predictive power of these \textit{ab initio} methods\cite{de2020measurement} (See Figure \ref{fig:OES}).  Two protons above the shell gap, the zinc isotopes ($Z = 30$) were studied from $^{62}$Zn to $^{80}$Zn, with the variation of their charge radii, the inversion of the odd-even staggering near $N = 40$, and the ground-to-isomer radius differences are all accounted for by proton excitations across the $Z = 28$ gap in large-scale shell-model calculations, reflecting the cumulative influence of the $^{68}$Ni and $^{78}$Ni cores~\cite{xie2019zinc}.  Extending toward $N = 40$ in the next element, the germanium isotopes ($Z = 32$, $^{68-74}$Ge) display a large odd-even staggering reproduced only when ground-state quadrupole correlations and deformation are included, underscoring the growing role of collectivity in this region~\cite{wang2024germanium,Kane20}.

Together, these chains map how nuclear size responds to the filling of the $fp$ and $g_{9/2}$ orbitals around $Z = 28$.  The interplay of the $Z = 28$ shell gap, the $N = 40$ subshell, and the approach to the doubly magic $^{78}$Ni provides a complete data set to understand the evolution of collective nuclear properties, which can be described by \textit{ab initio} methods and DFT calculations.

\subsubsection{The tin region}
The tin isotopic chain ($Z = 50$) has been studied by different laser spectroscopy experiments, providing access to two doubly magic nuclei, $^{100}$Sn ($N = 50$) and $^{132}$Sn ($N = 82$). It offers a comprehensive test of nuclear models across these two major nuclear shells.  Charge radii are now established from $^{104}$Sn to $^{134}$Sn, combining the neutron-rich chain with a recent extension to the neutron-deficient side using two collinear techniques at ISOLDE~\cite{gorges2019laser,Yordanov2020Sn,Rod20,Gus25}.  The radii follow the archlike, approximately parabolic trend expected from the evolution of quadrupole deformation between closed shells (Fig.~\ref{fig:Sn_In}), with an odd-even staggering that grows more pronounced toward $N = 50$ and $N = 82$ and a discontinuity, or kink, at the doubly magic $^{132}$Sn.  Standard density functionals struggle to reproduce these kinks consistently, whereas a Fayans functional captures the discontinuities at both $^{132}$Sn and $^{208}$Pb~\cite{gorges2019laser}. The local trends along the chain are also described by density functional theory and VS-IMSRG calculations, which point to appreciable beyond-mean-field correlations in the neutron-deficient isotopes~\cite{Gus25}.

With a single proton hole in the $Z = 50$ shell, the indium isotopes ($Z = 49$) trace a similar valence space, exhibiting marked differences.  Collinear resonance ionization spectroscopy has been performed from $^{101}$In up to $^{131}$In, nearly the complete range between $N = 50$ and $N = 82$, yielding charge radii, magnetic dipole and electric quadrupole moments, and spins~\cite{vernon2022nuclear,Kar24,Ver25}.  The nuclear magnetic moments revealed an abrupt change at $N = 82$~\cite{vernon2022nuclear}, while the parabolic trends in both the charge radii and quadrupole moments toward $N = 50$ support the doubly magic character of $^{100}$Sn~\cite{Kar24}.  The indium quadrupole moments and the tin $B(E2)$ values together provide the consistent view of deformation shown in Fig.~\ref{fig:Sn_In}. VS-IMSRG calculations reproduce the relative charge radii well but underestimate the spectroscopic quadrupole moments, whereas density functional theory performs reasonably well for both.  Laser spectroscopy of the high-spin isomers $^{127,129}$In added a new test, showing that the charge radius and intrinsic quadrupole moment both shrink when protons and neutrons are fully aligned in $^{129}$In ($N = 80$), an effect absent in $^{127}$In and not consistently reproduced by current theory~\cite{Ver25b}.

The neighboring antimony ($Z=51$) and cadmium ($Z=48$) isotopic chains provide additional probes on the robustness of the $Z=50$ shell. As a single proton outside the closed shell, the odd-$A$ Sb isotopes closely follow single-particle behavior, making them an ideal testing ground for shell-model and \textit{ab initio} descriptions of electromagnetic moments. Recent collinear laser spectroscopy measurements from $^{113}$Sb to $^{133}$Sb established a consistent set of spins, magnetic dipole moments, and electric quadrupole moments with substantially improved precision, demonstrating that modern shell-model calculations reproduce the experimental moments across the entire chain, while VS-IMSRG calculations reveal deficiencies primarily in the renormalization of the electromagnetic operators~\cite{Lechner2023Sb}. These results emphasize the sensitivity of electromagnetic moments to missing meson-exchange-current contributions and provide valuable benchmarks for extending \textit{ab initio} methods beyond the doubly magic $^{132}$Sn region. 

On the proton-deficient side of the shell closure, laser spectroscopy of cadmium isotopes has uncovered signatures of enhanced collectivity approaching midshell and demonstrated the persistence of shell effects around $N=82$. Measurements of charge radii, spins, and electromagnetic moments across long Cd isotopic chains revealed a surprisingly smooth evolution of the ground-state properties while exposing subtle structural changes associated with the filling of the neutron $h_{11/2}$ orbital and the onset of deformation \cite{Yordanov2013Cd}. Together with the Sn, In, and Sb results, these measurements provide a comprehensive experimental picture of shell evolution in the vicinity of the doubly magic nucleus $^{132}$Sn and constitute clear benchmarks for modern shell-model, density-functional, and \textit{ab initio} approaches.

\subsubsection{Heavy and superheavy elements}
In the region around the doubly magic $^{208}$Pb, charge radii and moments exhibit pronounced shape effects.  Neutron-deficient mercury isotopes present an odd-even shape staggering, in which alternating isotopes switch between near-spherical and strongly deformed configurations~\cite{Mar18}. An analogous large staggering was later found in the bismuth isotopes ($Z = 83$) at the same neutron number $N = 105$, while the intervening semi-magic lead chain remains nearly spherical~\cite{Bar21}.  These shape-coexistence phenomena, indicated by measurements derived from in-source resonance-ionization spectroscopy at ISOLDE, remain stringent benchmarks for density functional theory and beyond-mean-field calculations.

Laser spectroscopy of heavier and superheavy elements poses unique challenges, owing to the complexity of the open-shell atomic structure and the difficulty of producing and isolating these species at rates as low as a few atoms per second.  Nonetheless, charge radii and moments have been measured across the heavy actinides and their neighbors. Collinear resonance ionization spectroscopy of $^{222-233}$Ra ($Z = 88$) and resonance-ionization studies of actinium ($Z = 89$) map the onset of octupole (reflection-asymmetric) deformation in this region~\cite{lynch2018radium,verstraelen2019actinium}, with related measurements extending to francium ($Z = 87$)~\cite{Voss13,Flan13,lynch2014francium,wil17}.  
The charge radii and
moments of these nuclei are of direct relevance for CP violation searches (Sec.~\ref{sec:CPV_nuclear}), as several candidate nuclei
for EDM experiments, such as $^{225}$Ra and $^{229}$Pa, lie in this region.

Laser spectroscopy of super-heavy nuclei poses additional challenges because production rates are often limited to only a few atoms per second, while the isotopes of interest typically have short half-lives. The development of Radiation Detected Resonance Ionization Spectroscopy (RADRIS) at GSI overcame these challenges by combining efficient stopping and neutralization of fusion products with resonance ionization and $\alpha$-decay tagging, enabling optical spectroscopy at the atom-at-a-time level~\cite{Backe2007,Lautenschlager2016}.

This technique led to the first laser spectroscopy of a transfermium element through resonance ionization measurements of nobelium ($Z=102$), establishing nobelium as the heaviest element studied by optical spectroscopy to date~\cite{Laatiaoui2016}. Subsequent isotope-shift measurements on $^{252-254}$No provided the first determination of changes in the mean-square charge radii in the transfermium region. Combined with relativistic atomic calculations and nuclear density functional theory, these measurements indicated a central depression of the proton density in the deformed nobelium isotopes and provided new tests of both atomic and nuclear structure models~\cite{Raeder2018}.

Recent advances have considerably extended the reach of laser spectroscopy in this region. By combining online RADRIS measurements with off-line resonance-ionization spectroscopy on reactor-produced samples, isotope shifts were measured for the fermium isotopic chain ($^{245-257}$Fm), while the nobelium data were extended beyond the deformed $N=152$ shell gap~\cite{Warbinek2024}. The extracted charge radii exhibit a strikingly smooth evolution across the shell gap, indicating that, unlike the pronounced kinks observed at major spherical shell closures, the weak deformed $N=152$ shell has only a modest influence on nuclear sizes. Instead, the measured radii are largely governed by bulk properties of the nuclear density, providing stringent benchmarks for modern energy-density-functional calculations and demonstrating the growing capability of laser spectroscopy to probe the structure of the heaviest nuclei.


\subsubsection{Highly charged ions with exotic isotopes}
\label{sec:HCI}

Highly charged ions (HCI)---atoms stripped of most or all of their electrons---provide a qualitatively different atomic environment for probing nuclear properties and fundamental physics.  By removing the screening electrons, HCI amplifies relativistic, QED, and nuclear-size effects, making them sensitive to physics that is masked in neutral or single-charged ions ~\cite{kozlov2018hci,safronova2018}.

\paragraph{QED tests with nuclear structure input.}
In few-electron HCI (H-like, He-like, Li-like), the binding energies and transition frequencies can be computed to high precision using bound-state QED, with the nuclear charge radius as the dominant nuclear input.  Measurements of the $1s$ Lamb shift in H-like uranium~\cite{gassner2005lamb} and the $g$-factor of the bound electron in H-like ions~\cite{sturm2014gfactor} test QED in the strong-field regime ($Z\alpha \sim 1$).  Extending these measurements to radioactive isotopes---using storage rings (ESR at GSI, CR at FAIR) or Penning traps (HITRAP, ARTEMIS)---would access nuclei far from stability, where nuclear charge radii from laser spectroscopy (Sec.~\ref{sec:charge_radii}) provide the required input and where new nuclear effects may emerge.

\paragraph{Nuclear charge radii from HCI.}
The $2p_{3/2} \to 2p_{1/2}$ fine-structure transition in Li-like ions scales as $\sim (Z\alpha)^4$ and depends on the mean-square charge radius $\langle r^2\rangle$ (Eq.~(\ref{eq:r2_def})) with a sensitivity that increases steeply with $Z$.  This provides an alternative route to absolute charge radii for heavy elements~\cite{yerokhin2020hci,Staiger2025EUV}, in addition to the muonic atom and electron-scattering methods discussed in Sec.~\ref{sec:exp_techniques}.

\paragraph{Nuclear moments in HCI.}
The hyperfine structure of H-like and Li-like HCI probes the nuclear magnetic dipole moment $\bm{\mu} = g_I\muN\mathbf{I}$ (Sec.~\ref{sec:term_mu}) and its spatial distribution through the Bohr--Weisskopf effect (Eq.~(\ref{eq:BW})), with QED corrections that are calculable to high precision.  The specific difference of hyperfine splittings between H-like and Li-like ions of the same isotope cancels the leading nuclear-structure uncertainty, enabling a clean test of bound-state QED at the $10^{-4}$ level~\cite{shabaev2001hci}.  Extending these measurements to radioactive isotopes at FAIR's CRYRING and HITRAP facilities would probe nuclear magnetization distributions far from stability.

\paragraph{Highly charged actinide ions as optical clocks.}
Level crossings of valence orbitals along isoelectronic sequences can
bring transitions in HCI into the optical range while endowing them
with exceptional sensitivity to a possible variation of the fine
structure constant $\alpha$~\cite{kozlov2018hci}. Californium ions
near the $5f$--$6p$ crossing exhibit the largest $\alpha$ sensitivity
identified in an atomic system~\cite{berengut2012optical}. Detailed
clock proposals exist for Cf$^{15+}$ (618~nm, enhancement factor
$K_{\alpha}=+47$) and Cf$^{17+}$ (485~nm, $K_{\alpha}=-43.5$), with
systematic shifts analyzed at the level required for fractional
frequency uncertainties of order
$10^{-19}$~\cite{porsev2020optical,barontini2022qsnet}. Co-trapping
the two charge states is particularly attractive, since the opposite
signs of their large $K_{\alpha}$ coefficients allow common systematic
effects to cancel in a differential
measurement~\cite{barontini2022qsnet}. Similar level crossing enhancements have
been suggested in even more exotic Es
ions~\cite{dzuba2015highly}.


\subsubsection{Spectroscopy of radioactive molecules}

\begin{figure}[t]
    \centering
    \includegraphics[width=\linewidth]{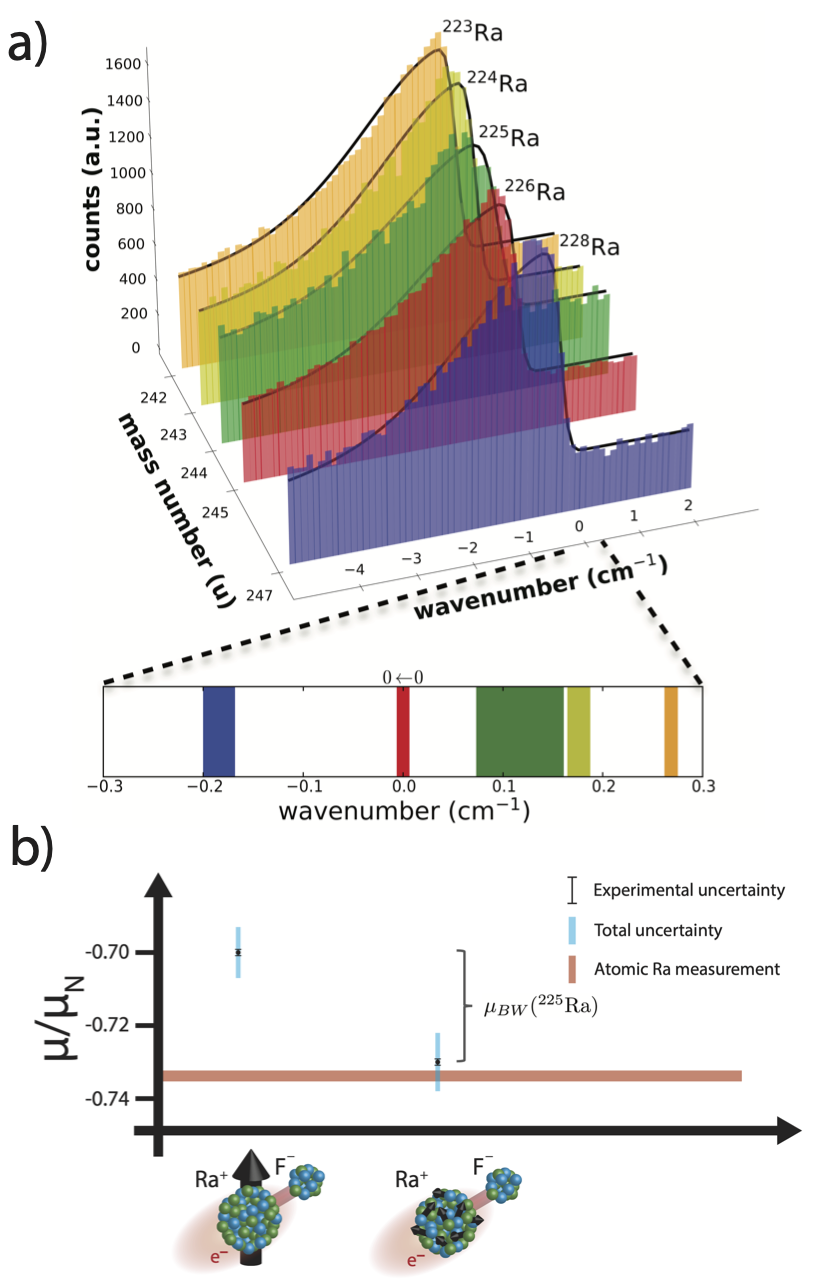}
    \caption{a) First observation of the isotope shift effects in a radioactive molecule, RaF, for five Ra isotopes: $^{223-226,228}$Ra. Figure adapted from Ref. \cite{udrescu2021isotope}, under CC BY 4.0. b) First observation of the distribution of the nuclear magnetization in a molecule, $^{225}$RaF. Figure adapted from Ref. \cite{wilkins2023observation}.}
    \label{fig:raf_measurements}
\end{figure}

A new development is the extension of laser spectroscopy from atoms to molecules containing short-lived nuclei \cite{udrescu2021isotope,garcia2020spectroscopy,wilkins2023observation,udrescu2024precision,conn2025production,wilkins2026ionization}.  A program focused on laser spectroscopy measurements of radioactive molecules (Fig. \ref{fig:raf_measurements} a)), with initial results for radium monofluoride (RaF), have been growing at ISOLDE-CERN~\cite{udrescu2021isotope,garcia2020spectroscopy,athanasakis2025electron,Atha24}. The initial results demonstrated the feasibility of extracting nuclear charge radii from molecular spectra, using the molecular isotope shift formalism of Eq.~(\ref{eq:mol_IS}).  The molecular measurements provide access to the $R$-dependence of the electron density at the nucleus (Sec.~\ref{sec:term_FS}), yielding additional observables, vibrational and rotational isotope shifts, that are absent in atoms.  

Recently, the first observation of the effect of the distribution of the nuclear magnetization on the molecular energy levels has been observed in $^{225}$RaF \cite{wilkins2023observation} (see Fig. \ref{fig:raf_measurements} b)). This work proved the high sensitivity of this radioactive species to minuscule nuclear effects, opening the way for precision nuclear structure studies in molecular systems, with potential applications to elements that are difficult to study in atomic form, as well as for future nuclear symmetry violation searches using radioactive molecules.(Secs.~\ref{sec:W_expansion} and~\ref{sec:CPV_expansion}). Concurrently, this result proved the reliability of \textit{ab initio} quantum chemistry, at the sub-percent level, for calculations of electronic form factors relevant for future P- and/or T-violation phenomena.

Another exciting direction of research involving radioactive molecules is the prospect of them being laser-coolable. A highly effective laser cooling scheme has been recently proposed for RaF \cite{udrescu2024precision}. Its implementation is currently actively pursued, towards future symmetry violation searches combining the high sensitivity of molecules containing octupole-deformed nuclei with the high precision and quantum control offered by state-of-the-art ultracold molecular techniques \cite{langen2024quantum, demille2024quantum}. Work towards effective offline production of radioactive species, as well as spectroscopic studies of polyatomic radioactive molecules, such as RaOH \cite{conn2025production}, is an ongoing direction of research.

Laser spectroscopy of radioactive molecules has recently been extended
beyond radium to the actinides. Chemically and isotopically pure beams
of $^{227}$Ac$^{19}$F$^{+}$were produced at ISOLDE by injecting CF$_4$ into the target and
ionizing the resulting fluorides in an arc discharge source, enabling
the first spectroscopic study of a gas phase actinium
molecule~\cite{AthanasakisKaklamanakis2025AcF}. The measurements 
were accompanied by combined electronic and nuclear structure
calculations of the sensitivity of $^{227}$AcF to CP violating
properties of the octupole deformed $^{227}$Ac nucleus.

\subsection{Open challenges and future directions}
\label{sec:open_EM}

Despite the remarkable progress of the last two decades, several challenges remain:

\subsubsection{Reaching the most exotic nuclei.}
Many isotopes of interest---particularly those near the neutron drip line and in the heavy actinide region---are produced at rates too low for current laser spectroscopy techniques.  Next-generation facilities (FRIB, FAIR, SPIRAL2) and new techniques \cite{miracls2020,udrescu2024doppler} aim to extend measurements to species produced at rates below 1 ion/s, with half-lives below 1~ms.

\subsubsection{Absolute charge radii.}
Laser spectroscopy measures differences in charge radii between isotopes.  Extracting absolute radii requires calibration to a reference isotope whose radius is known from electron scattering or muonic atom data.  For many elements, such reference data are unavailable or insufficiently precise, introducing systematic uncertainties that propagate through the entire isotopic chain.  New electron-scattering facilities for radioactive beams (SCRIT, ELISe) and muonic atom measurements at PSI aim to address this gap.

A complementary approach exploits extreme-ultraviolet spectroscopy of highly charged ions confined in electron-beam ion traps \cite{Staiger2025EUV}. Instead of measuring isotope shifts within a single element, this method determines transition-energy differences between different elements, which are directly sensitive to their charge-radius differences through precision relativistic many-body calculations. Recently, this technique was used to determine the charge-radius difference between natural ytterbium and lutetium, reducing the uncertainty in the absolute radius of $^{175}$Lu by a factor of three and resolving a long-standing anomaly in the odd--even staggering of charge radii in the rare-earth region \cite{Staiger2025EUV}.

\subsubsection{Atomic theory for electronic factors.}
The extraction of nuclear properties from measured isotope shifts and
hyperfine structure requires the electronic factors $F$, $K$, $B_e$, and
$V_{zz}$ from atomic theory.  How critical
these calculations are depends strongly on the element.  When three or
more stable isotopes exist whose absolute radii are known from muonic
atom and electron scattering data~\cite{fricke2004nuclear,angeli2013table},
the factors $F$ and $K_\text{SMS}$ can be calibrated semi-empirically
through King plot analyses~\cite{king2013isotope}, and the
atomic theory uncertainty is largely eliminated.  This option
is not possible for most of the elements with odd proton number that
possess only one or two stable isotopes \cite{Sahoo_2020,ohayon2022sodium}.
For these isotopic chains the extracted $\delta\langle r^2\rangle$ rests
entirely on computed electronic factors, whose uncertainty is then
frequently the dominant systematic on the charge radii extraction, exceeding the
experimental precision by an order of magnitude or
more~\cite{sahoo2025advancements}.  For simple systems
(alkali-like ions, few-electron systems), relativistic many-body
methods reach sub-percent precision, and quantum electrodynamics
corrections to the isotope shift factors have now become relevant at
the precision frontier~\cite{skripnikov2024isotope}.  For complex
open-shell atoms, uncertainties of 5--10\% or more remain, and their
reduction through relativistic coupled-cluster, configuration
interaction, and multiconfiguration methods is an active field of
development, recently reviewed in Ref.~\cite{sahoo2025advancements}.
Work towards atomic structure calculations with the
VS-IMSRG is ongoing~\cite{tenkila2022ab}. 

For the
lightest elements the nuclear volume effect is reduced by three to four
orders of magnitude below the mass shift, and only few-body QED
calculations make the extraction possible at all.  Because several of these
theory-limited chains carry electroweak physics, improved
electronic structure calculations translate directly into
fundamental physics reach.

\subsubsection{Molecular spectroscopy of exotic nuclei.}
The first molecular spectroscopy results on RaF~\cite{udrescu2021isotope,udrescu2024precision} have demonstrated the technique, but systematic measurements across isotopic chains and in different rovibrational states, which is needed to fully exploit the molecular isotope shift formalism of Eq.~(\ref{eq:mol_IS}), are still in their infancy.  Extending molecular spectroscopy to other species (RaOH \cite{conn2025production}, RaOH$^+$, RaOCH$_3^+$\cite{fan2021optical,Yu21}) and to heavier radioactive nuclei is a challenge for the coming decade.

\subsubsection{Confrontation with \textit{ab initio} theory.}
As highlighted in Sec.~\ref{sec:nuclear_theory}, the electromagnetic observables reviewed in this section---charge radii, magnetic moments, quadrupole moments---provide the essential validation for the nuclear theory methods that are needed to compute the symmetry-violating moments (anapole, Schiff, MQM) targeted by the experiments of Sec. \ref{sec:symm_violation}.  The electromagnetic sector is where theory is tested against known quantities; the symmetry-violating sector is where theory makes predictions for unknown quantities.  Ensuring that the same nuclear Hamiltonian and the same many-body method accurately reproduce the electromagnetic observables is a prerequisite for trusting the predictions for symmetry violation.  This interplay---experiment informing theory, theory enabling experiment---is the driving force behind the continued investment in precision spectroscopy of exotic nuclei.

\section{Probing symmetry-violating nuclear properties}
\label{sec:symm_violation}

The preceding section focused on $P$ and $T$ conserving interactions between the electrons and nuclei (Sec.~\ref{sec:EM_expansion}), which probe the distribution of charge and magnetization inside the nucleus.  In this section we will explore $P$ violating (Sec.~\ref{sec:PV} and \ref{sec:NSD_PV}) and the CP violating phenomena (Sec.~\ref{sec:CPV_nuclear}) inside atoms and molecules.  These observables encode information about the weak interaction at the quark level, the hadronic parity-violating force between nucleons, or possible new sources of CP violation beyond the Standard Model \cite{safronova2018,arrowsmith2023opportunities}.  Their measurement in atoms and molecules---using the factorized framework of Sec.~\ref{sec:ch2} and the nuclear theory of Sec.~\ref{sec:nuclear_theory}---is one of the central goals of the field.

A unifying theme of this section is the qualitative advantage of molecules over atoms for symmetry-violation searches.  As shown in Sec.~\ref{sec:mol_enhance}, the small opposite-parity splittings in molecules can amplify symmetry-violating effects by factors of $10^5$--$10^7$ relative to atoms. Similarly, in polar molecules, a modest external electric field can fully polarize the system, converting the internal symmetry-violating matrix element directly into a laboratory-frame observable.
Combining this molecular enhancement with nuclei possessing large symmetry-violating moments---such as octupole-deformed nuclei in the case of Schiff moments, quadrupole-deformed nuclei for MQM searches, or heavy nuclei for anapole moments---yields total sensitivity gains of up to five orders of magnitude compared to other atomic and molecular systems, motivating the use of radioactive molecules at RIB facilities \cite{arrowsmith2023opportunities}.

Such AMO-based techniques provide a powerful avenue for probing physics BSM, complementary to searches performed at particle colliders \cite{navas2024review}. There are two regimes to be considered: when the new particles or the relevant dynamics scale of the BSM is heavier than the typical energy of the experiment, say the nuclear scale or size, and when it is lighter. Both cases are relevant and discussed in this review. When the new particles are too heavy to be produced directly, their virtual effects can generate small deviations in low-energy observables that can be described within an effective field theory (EFT) framework \cite{chupp2019electric}. The exceptional agreement between many precision measurements and SM predictions, therefore, translates into sensitivity to mass scales far above those directly accessible at colliders \cite{arrowsmith2023opportunities,safronova2018}. Classic examples include flavor-changing processes, electric dipole moments, and parity-violating observables, all of which constrain new interactions through their contributions to higher-dimensional operators (see, e.g., Refs.~\cite{Cirigliano:2013lpa,erler2013weak,Langacker:2008yv} for earlier reviews).

Nuclear, atomic, and molecular systems are particularly valuable in this program because they combine high experimental precision with enhanced sensitivity to fundamental interactions \cite{arrowsmith2023opportunities,safronova2018}. A prominent example is atomic parity violation (APV) \cite{wood1997}, where the exchange of the $Z$ boson induces tiny parity-violating effects in atomic transitions. Measurements of the weak nuclear charge, most notably in cesium \cite{wood1997}, provide precision tests of the electroweak sector and place strong constraints on various BSM physics models \cite{safronova2018}. The interpretation of APV experiments relies crucially on advances in atomic and nuclear many-body theory, illustrating the close interplay between precision nuclear calculations and searches for new fundamental interactions. Future improvements in atomic structure calculations, isotope-ratio measurements, and parity-violating observables in nuclei and molecules, as further discussed below, are expected to extend the sensitivity of these probes to multi-TeV and even higher scales of new physics (see Refs.~\cite{safronova2018, ginges2004violations} for earlier reviews). Similar logic applies to searches for dark matter (DM), which can be produced thermally, typically above the MeV scale, where its interactions with the nuclei can be described using EFT (see~\cite{Cirelli:2024ssz} for a recent review).

The situation changes qualitatively when the new degrees of freedom are light, say, compared to the characteristic momentum transfer of the experiment. In this case, the mediator must be treated explicitly, and its effects are no longer captured by local effective operators. Light bosons arise naturally in many extensions of the SM, including axions, dilatons, relaxions, dark photons, Higgs-portal scalars, and other pseudo-Nambu--Goldstone bosons associated with approximate symmetries~\cite{Graham:2015ouw,Irastorza:2018dyq,safronova2018}. Their exchange can generate new long-range interactions, often referred to as fifth forces, which can induce composition-dependent forces, spin-dependent interactions, and apparent violations of the equivalence principle (EP). Precision measurements in atomic, molecular, and nuclear systems therefore provide some of the most sensitive probes of light weakly-coupled particles through EP tests, inverse-square-law measurements, and spin-dependent force searches~\cite{Adelberger:2009zz,Fischbach:1999bc}.

An especially compelling possibility is that a light bosonic field constitutes some or all of the dark matter abundance. Ultralight dark matter (ULDM) can be produced through the misalignment mechanism~\cite{Preskill:1982cy,Abbott:1982af,Dine:1982ah}, a non-thermal production mechanism requiring only gravitational interactions and effective for masses below the eV scale. Theoretically, ultralight scalars and pseudoscalars are ubiquitous in extensions of the SM, including the QCD axion~\cite{Preskill:1982cy,Abbott:1982af,Dine:1982ah}, dilatons~\cite{Arvanitaki:2014faa} (see however~\cite{Hubisz:2024hyz}), relaxions~\cite{Graham:2015ifn,Banerjee:2018xmn}, Higgs-portal scalars~\cite{Piazza:2010ye}, and other axion-like particles~\cite{Dine:2024bxv}. Owing to their enormous phase-space occupancy, such fields behave as coherent classical waves. For a scalar field $\phi$ of mass $m_\phi$, the local dark matter density implies an oscillation amplitude $\phi_0 \simeq \sqrt{2\rho_{\rm DM}}/m_\phi$ and oscillation frequency $m_\phi$. Couplings of $\phi$ to SM operators therefore induce oscillatory variations of masses, couplings, and other observables, transforming precision experiments into detectors of time-dependent signals \cite{safronova2018,arrowsmith2023opportunities}.

\subsection{Parity violation and the nuclear weak charge}
\label{sec:PV}

\subsubsection{The nuclear weak charge as a low-energy test of the Standard Model}

At low momentum transfer, the parity violating interaction between
electrons and quarks is described by the effective Lagrangian ~\cite{musolf1994intermediate}
\begin{equation}
\mathcal{L}^{eq}_{\mathrm{PV}}
=\frac{\GF}{\sqrt{2}}\sum_{q=u,d}
\left[\,C_{1q}\,\bar e\gamma^{\mu}\gamma_{5}e\,\bar q\gamma_{\mu}q
      +C_{2q}\,\bar e\gamma^{\mu}e\,\bar q\gamma_{\mu}\gamma_{5}q\,\right],
\label{eq:pv_eq_lagrangian}
\end{equation}
where $\GF{}$ is the Fermi constant, $C_{1q}=2g_{A}^{e}g_{V}^{q}$ couples
the axial electron current to the quark vector current, and
$C_{2q}=2g_{V}^{e}g_{A}^{q}$ is its chirality counterpart, with
$g_{V}^{f}~=~T_{3}^{f}~-~2Q_{f}\sin^{2}\thetaW$ and $g_{A}^{f}~=~T_{3}^{f}$
the vector and axial couplings of the $Z^{0}$ boson to fermion $f$ of
weak isospin $T_{3}^{f}$ and electric charge $Q_{f}$ (in the modern
notation of the Particle Data Group, $C_{1q}=g_{AV}^{eq}$ and
$C_{2q}=g_{VA}^{eq}$).

\begin{figure}[t]
    \centering
    \includegraphics[width=\linewidth]{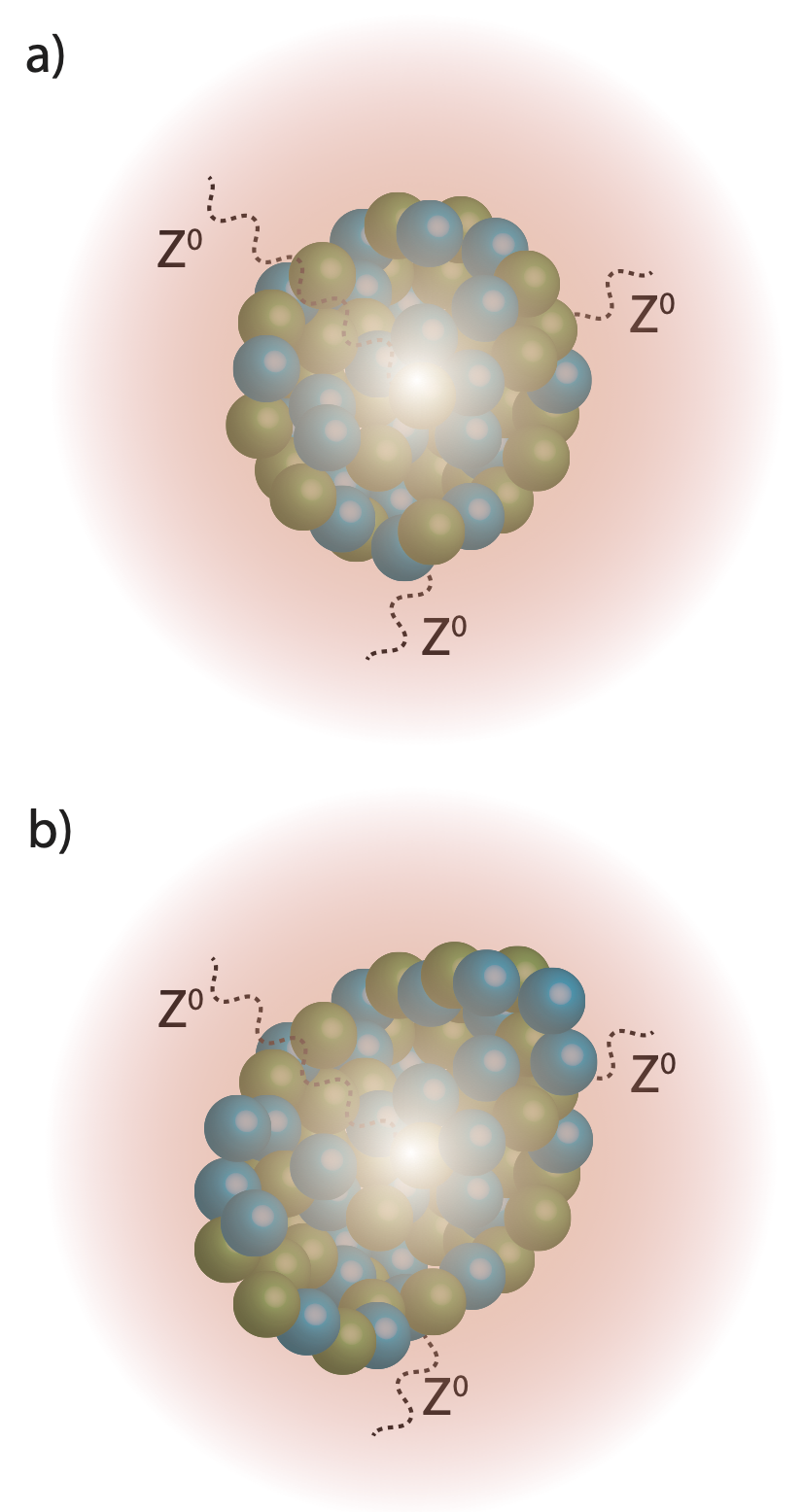}
    \caption{Dominant nuclear spin independent parity-violating effects from $Z^0$-boson exchange.  (a)~The spherically symmetric proton (blue) and neutron (green) distributions give rise to the nuclear weak charge $Q_W$.  (b)~Quadrupole deformation of the nucleon distributions leads to the nuclear weak quadrupole moment.}
    \label{fig:weak_charge_interaction}
\end{figure}

The nuclear-spin-independent (NSI) parity-violating interaction (Sec.~\ref{sec:W_expansion}) is governed by the nuclear weak charge $Q_W$ (Eq.~(\ref{eq:QW_def})), which, including one-loop and leading two-loop electroweak radiative corrections~\cite{marciano1983radiative,erler2013weak}, is given by:
\begin{equation}
\begin{split}
    Q_W(Z,N) = &2\left[Z\left(C_{1,p}+0.00005\right)+N\left(C_{1,n}+0.00006\right)\right] \\
& \times \left(1-\frac{\alpha}{2\pi}\right).
\end{split}
\label{eq:QW_full}
\end{equation}
At tree level, $Q_W \approx Z(1-4\sin^2\theta_W) - N \approx -N$ (Fig.~\ref{fig:weak_charge_interaction}a).  A precision measurement of $Q_W$, combined with accurate atomic theory for the electronic factor $k_\text{PV}$ (Eq.~(\ref{eq:MPV})), allows the extraction of $\sin^2\theta_W$ at low momentum transfer, supplementing high-energy measurements performed at colliders and providing a strong test of the Standard Model's electroweak sector \cite{wood1997,safronova2018,davoudiasl2014muon}.

\subsubsection{Experimental milestones}

The experimental program for measuring atomic parity violation (APV) has developed over nearly five decades, using two main techniques:

\paragraph{\textbf{Optical rotation.}}

The first observations of PV effects in atoms were made using the optical rotation technique \cite{barkov1978,barkov1979parity}, which can be explained as follows. Linearly polarized light is an equal superposition of right- and left-handed circularly polarized light. For two atomic energy levels, with an allowed magnetic dipole transition of amplitude $A_{M_1}$, PV can induce an additional allowed electric dipole transition, proportional to the PV effect, $A_{PV}$. This leads to different indices of refraction for the left- and right-circularly polarized light components and thus to a rotation of an initially linearly polarized light, upon passing through an atomic vapor. The angle of this rotation is proportional to $\mathrm{Im}(A_{PV}/A_{M_1})$, where $\mathrm{Im}$ denotes the imaginary part, and its measurement allows the extraction of the underlying PV effects \cite{khriplovich1991} (see Sec. \ref{sec:appendix_optical_rot}). Measurements in $^{209}$Bi~\cite{macpherson1991,warrington1993}, $^{208}$Pb~\cite{meekhof1993,phipp1996}, and $^{205}$Tl~\cite{edwards1995,vetter1995} proved the existence of PV effects at the percent level, but did not reach sufficient precision to facilitate tests of the SM.

\paragraph{\textbf{Stark interference.}}

The most precise atomic PV measurement to date was performed in $^{133}$Cs by the Boulder group~\cite{wood1997}, using the Stark interference technique. In this approach, the signal comes from the measured transition rate between two atomic energy levels of the same parity, in the presence of the PV effects. The PV signal is amplified by an externally applied electric field, $\mathbf{E}$, which mixes states of opposite parity in the atom through the Stark effect. This leads to an additionally allowed $E_1$ transition amplitude, $A_{E_1} = \beta E$, ($\beta$ is the vector transition polarizability), which interferes with $A_{PV}$. The quantity obtained in such a measurement is $\mathrm{Im}\left(A_{PV}/\beta\right)$, from which $Q_W$ can be extracted using the electronic factor $k_\text{PV}$ from atomic theory.

The Boulder experiment \cite{wood1997} achieved a relative accuracy of $0.35\%$:
$\text{Im}(E_\text{PV}/\beta)_\text{NSI} = -1.5935(56)$~mV/cm.  Extracting $Q_W$ required atomic theory at comparable precision, which was achieved through two decades of relativistic many-body calculations~\cite{dzuba1985relativistic,derevianko2000reconciliation,porsev2009precision,sahoo2021new}, including QED corrections, the Breit interaction, and the neutron skin effect.  The final result is consistent with the SM prediction, with a theoretical uncertainty of $0.3\%$ \cite{sahoo2021new,roberts2022comment}, comparable to the experimental one \cite{wood1997}.

The Stark interference method has also been applied to Yb isotopes~\cite{antypas2019isotopic}, where the PV amplitude is $\sim 100$ times larger than in Cs~\cite{demille1995parity}.  First results for $^{170,172,174,176}$Yb reached $\sim 0.5\%$ single-isotope accuracy, providing the first experimental observation of the $N$-dependence of atomic PV and providing new bounds on BSM bosons mediated electron-nucleon interactions.  The measurement of PV ratios between isotopes partially cancels atomic theory uncertainties and provides sensitivity to the neutron skin (Sec.~\ref{sec:neutron_skin_PV}).  Upgrades aim to reach sub-$0.1\%$ accuracy and to measure NSD PV in $^{171,173}$Yb, which have a non-zero nuclear spin.

Several additional experimental programs are underway with the goal of measuring parity-violating effects in various atomic and molecular systems using the Stark interference technique:

\begin{itemize}
\item \textbf{Francium.} at TRIUMF: Fr ($Z=87$) is expected to display PV effects $\sim 18$ times larger than Cs, with similar atomic theory precision \cite{dzuba1995calculation,safronova2000high,dzuba2001calculations,gomez2008nuclear,sahoo2010ab,ginges2018testing,roberts2020nuclear,gwinner2022studies}.  The FrPNC collaboration aims to trap $\sim 10^6$ Fr atoms in a Magneto Optical Trap (MOT), for which a PV measurement at the $1\%$ level is predicted to be achieved in a few hours of measurement time, with an order of magnitude improvement expected in the future \cite{gwinner2022studies}.

\item \textbf{Ra$^+$ in ion traps.}:  Having a large atomic number and a relatively simple electronic structure, similar to that of Cs, Ra$^+$ is a promising candidate for precision tests of the SM at low energy \cite{versolato2011atomic,nunez2013towards,nunez2014ra}. Using this species is further motivated by the long interaction times and reduced systematics offered by trapped ions platforms. Recently, progress has been achieved towards trapping and laser cooling  Ra$^+$ for future precision measurements \cite{ready2026laser}. This technique is also pursued using Ba$^+$ ions \cite{williams2013method,fortson1993possibility,koerber2003radio,kleczewski2012coherent} with a projected relative uncertainty in the PV signal of $0.1\%$ using entanglement-based techniques \cite{craik2026entanglement}. A similar approach has also been proposed for neutral alkali atoms trapped in a 2D optical lattice \cite{kastberg2019optical}.

\item \textbf{Chiral molecules.}: PV effects manifest as an energy difference between left- and right-handed enantiomers \cite{quack1989structure,quack2008high}. This could potentially be related to the homochirality of biomolecules~\cite{yamagata1966hypothesis,kondepudi1985weak}.  Over the past decades, several attempts have been made to measure an energy difference between enantiomers using high-resolution rotational, rovibrational, electronic or NMR spectroscopy  \cite{daussy1999limit,ziskind2002improved,darquie2010progress}, but no conclusive non-zero signal has been observed to date. However, progress is being made both in improving the resolution of the techniques used and in finding chiral molecules with a higher PV signal \cite{patterson2013enantiomer,janssen2014detecting,beaulieu2017attosecond,stickler2021enantiomer}. Chiral molecules containing a heavy nuclei, e.g. At($Z=85$) and U($Z=92$), are particularly interesting in this context, as the PV signal scales strongly with the atomic number, $Z$, of the heaviest nucleus in the molecule, roughly as $Z^5$ \cite{berger2007electroweak,wormit2014strong,gaul2020chiral}.
\end{itemize}

\subsubsection{New physics from weak charge measurements}

\begin{figure}[t]
    \centering
    \includegraphics[width=\linewidth]{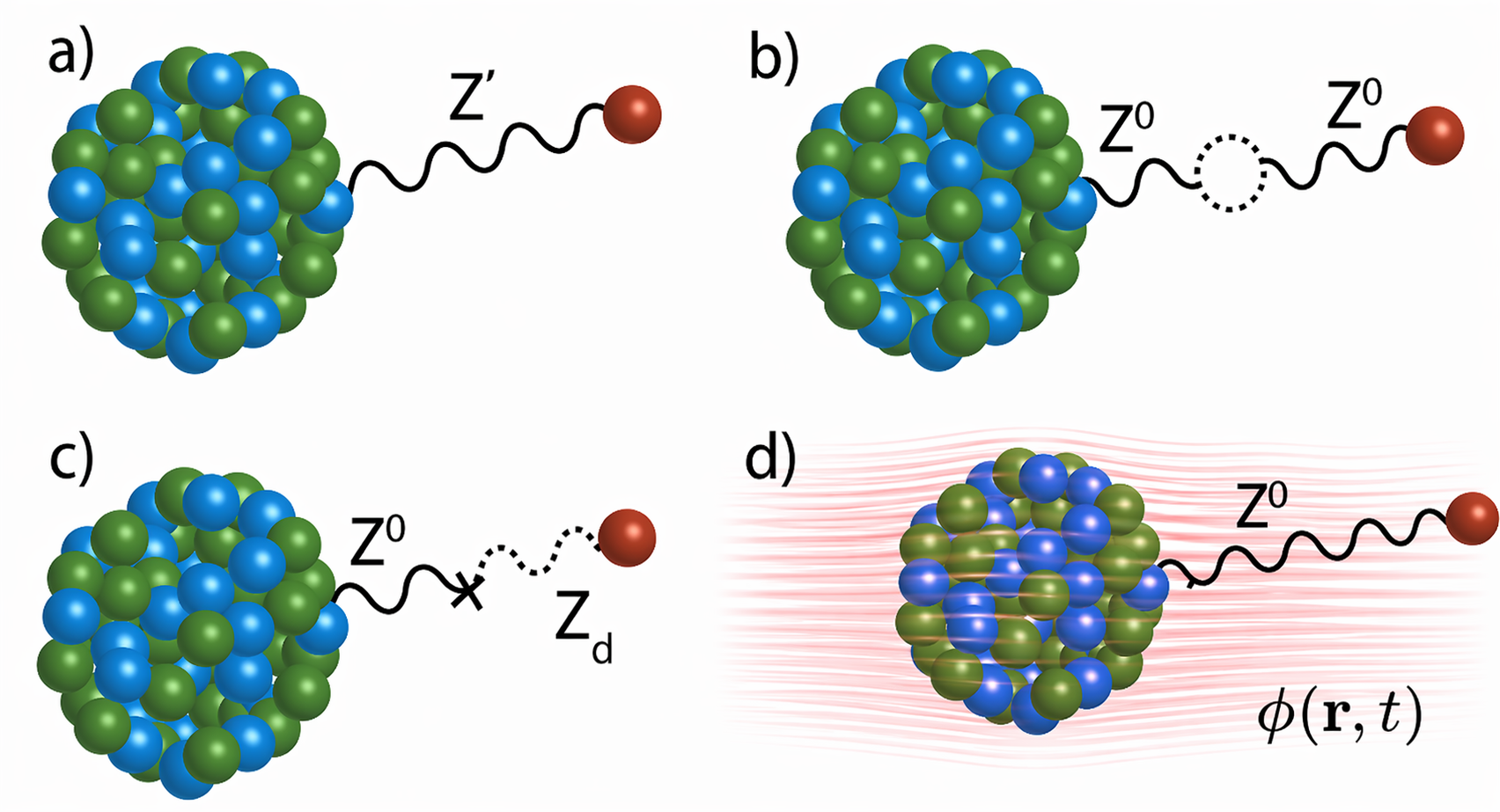}
    \caption{New physics probed by weak charge measurements.  (a)~Exchange of a new massive $Z'$ boson.  (b)~Vacuum polarization corrections to the $Z^0$ self-energy from new particles.  (c)~Kinetic mixing of a dark boson $Z_d$ with the SM $Z^0$.  (d)~Interaction with a cosmic pseudoscalar field, $\phi(\mathbf{r},t)$ (e.g., the axion).}
    \label{fig:NP_weak}
\end{figure}

Precision measurements of $Q_W$ can constrain BSM physics through several channels \cite{davoudiasl2012muon,davoudiasl2012dark,andreas2012update,derevianko2018detecting} (Fig.~\ref{fig:NP_weak}):

\paragraph{\textbf{New \texorpdfstring{$Z'$}{Z'} bosons.}} The exchange of a new $Z'$ boson with TeV-scale mass, $M_{Z'}$, can shift $Q_W$ by $\Delta Q_W^{Z'} = -\frac{6\sqrt{2}}{G_F}A\frac{\overline{g_{Z'}}}{M^2_{Z'}}$ \cite{bouchiat1983parity,bouchiat2005constraints,porsev2009precision,diener2012constraining}, where $\overline{g_{Z'}}$ is the average axial-electron vector-(up/down) quark $Z'$-mediated interaction strength.  The $^{133}$Cs measurement \cite{wood1997} sets lower limits $\gtrsim 0.5$~TeV on $M_{Z'}$, for a wide range of BSM scenarios\cite{porsev2009precision,diener2012constraining,tran2022implications}.  Measuring $Q_W$ in several isotopes (e.g., Fr, Yb) at the $0.1\%$ level can push these limits to several TeV \cite{diener2012constraining}.

\paragraph{\textbf{Oblique corrections.}}
Vacuum polarization corrections to the W and Z bosons' self-energies, due to new particles beyond the SM  \cite{marciano1990atomic,rosner2002role,porsev2009precision}, can contribute to the weak charge through the Peskin--Takeuchi $S$ and $T$ parameters~\cite{peskin1990new,marciano1990atomic,peskin1990new,veltman1977limit}. For the case of $^{133}$Cs, this correction can be written as $\Delta Q_W^\text{ST}(\text{Cs}) = -0.800 S - 0.007 T$ \cite{marciano1990atomic,rosner2002role}. While the effect is dominated by the S parameter, other electroweak parameters, such as the mass of the W boson or the leptonic width of the Z boson, are sensitive to both S and T, highlighting the importance of high-precision atomic PV measurement for disentangling the effects of the two parameters \cite{rosner2002role}.

\paragraph{\textbf{Dark bosons.}}
A dark $Z_d$ boson with MeV--GeV mass, coupled weakly to the SM through kinetic mixing can shift $\sin^2\theta_W$ at low momentum transfer, $Q$, by $\Delta\sin^2\theta_W \propto (m_Z/m_{Z_d})\sin\theta_W\cos\theta_W\frac{m_{Z_d}^2}{Q^2+m_{Z_d}^2}$ \cite{davoudiasl2012muon,davoudiasl2012dark,davoudiasl2014muon,davoudiasl2015low}. This shows the enhanced sensitivity of low-energy experiments ($Q\ll m_{Z_d}$) to the $Z_d$ boson effects compared to high-energy ones. Given the low momentum transfer in atomic parity violation experiments ($Q \approx 2.4$ MeV for the $^{133}$Cs measurement \cite{bouchiat1983parity}), such experiments are expected to be very sensitive to the physically motivated search space of $M_{Z_d} \gtrsim 10$ MeV \cite{davoudiasl2012dark}.

\paragraph{\textbf{Cosmic fields.}}
Pseudoscalar fields, such as the axion \cite{peccei1977cp}, can produce time-varying PV amplitudes at the axion Compton frequency~\cite{stadnik2014axion,roberts2014parity}. Unlike the methods described above for measuring static atomic PV, in this case, one would need to measure the PV effect as a function of time and extract the frequency variation of such a signal from the associated power spectral density \cite{stadnik2014axion}.

\subsubsection{The neutron skin from atomic parity violation}
\label{sec:neutron_skin_PV}

As discussed in Sec.~\ref{sec:W_expansion}, the weak charge provides information mainly about the neutrons inside the nucleus.  If the proton, $\rho_p(r)$, and neutron, $\rho_n(r)$, densities differ (i.e. the neutron skin, $\Delta R_{np} \neq 0$), the weak-charge value acquires a correction $\Delta Q_W^\text{NS}$, relative to the $\Delta R_{np} = 0$ case, proportional to $1 - q_n/q_p$, where $q_{n,p} \equiv \int \rho_{n,p}(r) f(r)\, d^3r$, with $f(r)$ being the radial electronic wave function inside the nucleus~\cite{fortson1990nuclear,pollock1992atomic,brown2009,sahoo2021new}.  For heavy elements, such as Ra or Fr, this correction can reach $\sim 0.6\%$~\cite{brown2009}, making it a limiting factor for future precision SM tests but also an opportunity: by assuming there are no BSM corrections to the weak charge, atomic PV measurements can be used to determine $\Delta R_{np}$, providing information about the nuclear equation of state in addition to parity-violating electron scattering (PREX/CREX \cite{adhikari2021accurate,adhikari2022precision}). Isotopic ratio measurements of the weak charge can also be used to reduce the sensitivity of atomic PV experiments to the neutron skin, by probing differences between isotopes, where systematic errors partially cancel.  This could make isotopic atomic PV a clean probe of new physics \cite{fortson1990nuclear,pollock1992atomic,brown2009,antypas2019isotopic,ramsey1999low,derevianko2002reevaluation,diener2012constraining}.

\subsection{Nuclear spin-dependent parity violation}
\label{sec:NSD_PV}

\subsubsection{Three contributions to NSD PV}

\begin{figure}[t]
    \centering
    \includegraphics[width=\linewidth]{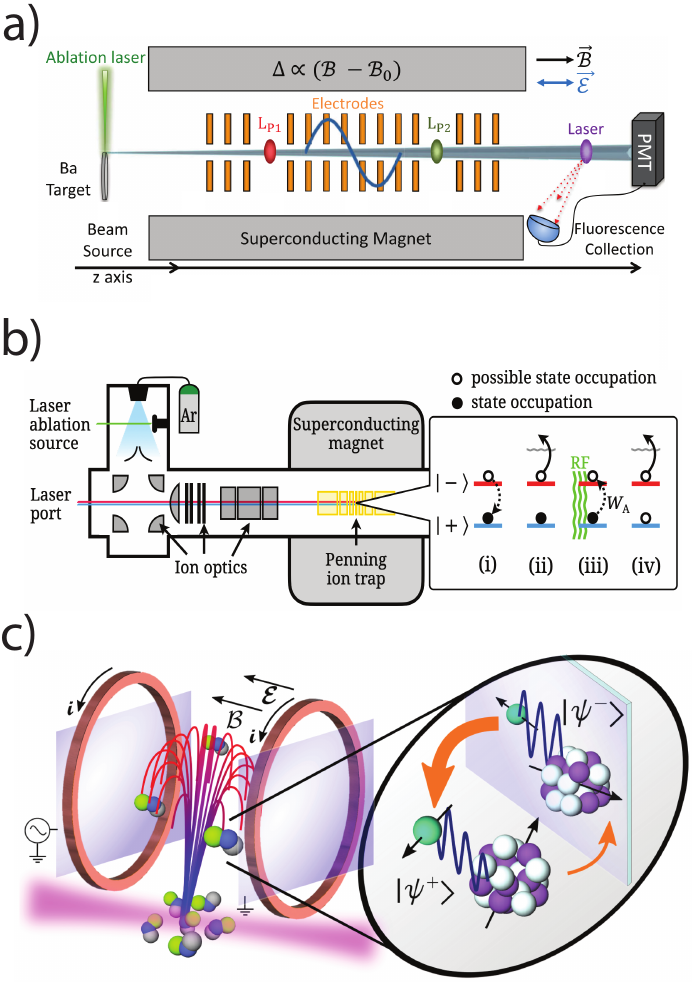}
    \caption{Experimental approaches for measuring NSD PV effects in molecules.  a) Neutral beam of diatomic molecules (Figure adapted from Ref. \cite{altuntacs2018demonstration}, doi:10.1103/PhysRevLett.120.142501. Copyright 2018 American Physical Society.). b) Trapped diatomic molecular ions (Figure adapted from Ref. \cite{karthein2023electroweak}, under CC BY 4.0). c) Molecular fountain using polyatomic neutral molecules (Figure adapted from Ref. \cite{hao2020nuclear}, 10.1103/PhysRevA.102.052828. Copyright 2020 American Physical Society.). In all cases, the experimental technique exploits the small opposite-parity splittings present in molecules to enhance the PV signal by many orders of magnitude relative to atoms.}
    \label{fig:PV_anapole_experiments}
\end{figure}

The dominant atomic NSD PV Hamiltonian, acting in the electronic space, can be written as \cite{flambaum1984nuclear,flambaum1985enhancement,ginges2004violations} (Sec.~\ref{sec:anapole}):
\begin{equation}
H_\text{NSD}^\text{PV} = \frac{G_F}{\sqrt{2}}\,\eta\,(\boldsymbol{\alpha}\cdot\mathbf{I})\,\rho(r)\,,
\label{eq:HNSD}
\end{equation}
with $\eta = \eta_\text{anapole} + \eta_\text{axial} + \eta_\text{hf}$, where:

\paragraph{\textbf{$\eta_\text{anapole}$}} The dominant contribution in heavy nuclei ($\propto A^{2/3}$), arising from the nuclear anapole moment (Sec.~\ref{sec:anapole}).  It probes hadronic PV interactions inside the nucleus.

\paragraph{\textbf{$\eta_\text{axial}$}} The direct NSD weak interaction (Sec.~\ref{sec:W_expansion}) between the electron vector current and nucleon axial current, mediated by the $Z^0$ boson. This probes some of the least well-constrained parameters of the SM, $C_{2u}$ and $C_{2d}$ \cite{langacker1992high,buckley2012precision}. The complementarity of atomic and molecular experiments to other past and proposed measurements aiming to extract linear combinations of the $C_{2u}$ and $C_{2d}$ parameters is clearly displayed in Fig. \ref{fig:NSD_constraints} a).
 The $C_{1q}$ build up coherently over the
nucleus and give rise to the weak charge discussed in
Sec.~\ref{sec:PV}. At tree level,
\begin{equation}
C_{2u}=-\tfrac{1}{2}+2\sin^{2}\thetaW=-C_{2d},
\label{eq:c2_tree}
\end{equation}
so that both couplings are proportional to $1-4\sin^{2}\thetaW$ and
inherit the same accidental suppression as the proton weak charge~\cite{erler2014weak};
including electroweak radiative corrections, the SM predicts
$2C_{2u}-C_{2d}=-0.0949(4)$~\cite{Wan15a},
with theoretical uncertainties far below present experimental
sensitivity. Because accessing the $C_{2q}$ requires the
axial quark current, they are constrained only through the cross term
in parity violating deep inelastic scattering and through NSD PV, and
present measurements determine them to roughly 50\% precision. This
uncertainty refers to the measurement of the low energy effective
couplings, not to the SM prediction. The quantity probed by
$\eta_{\mathrm{axial}}$ is the corresponding nucleon level coupling,
obtained by weighting the quark couplings with the quark spin content
of the nucleon \cite{musolf1994intermediate,erler2013weak},
\begin{equation}
C_{2p}=\Delta u\,C_{2u}+\Delta d\,C_{2d},
\qquad
C_{2n}=\Delta d\,C_{2u}+\Delta u\,C_{2d},
\label{eq:c2_nucleon}
\end{equation}
where $\Delta u\approx0.84$ and $\Delta d\approx-0.43$ denote the
fractions of the nucleon spin carried by up and down quarks. Their
difference $\Delta u-\Delta d=g_{A}=1.2754$ is fixed by neutron beta
decay~\cite{pdg2024review}, while the individual flavors are
determined consistently by polarized deep inelastic
scattering~\cite{adolph2016spin} and by lattice QCD calculations of
the flavor diagonal axial charges~\cite{gupta2018flavor}. Nuclei with an
unpaired proton and with an unpaired neutron therefore constrain
nearly orthogonal combinations of $C_{2u}$ and $C_{2d}$.

\paragraph{\textbf{$\eta_\text{hf}$}} The interference between the NSI weak charge and the hyperfine interaction. This is typically much smaller than the other two effects, and it can be calculated at the $< 10\%$ relative uncertainty \cite{bouchiat1991nuclear,johnson2003combined}, which is enough for future measurements of the NSD PV effects at the percent level.

While disentangling the different contributions to $\eta$ can be challenging from a single measurement, the different $A$-scaling of $\eta_\text{anapole}$ ($\propto A^{2/3}$) and $\eta_\text{axial}$ ($A$-independent) should allow this, in principle, from NSD PV measurements across a range of nuclear masses~\cite{demille2008using}.

\subsubsection{The \texorpdfstring{$^{133}$Cs}{133Cs} anapole moment measurement}

\begin{figure}[t]
    \centering
    \includegraphics[width=0.8\linewidth]{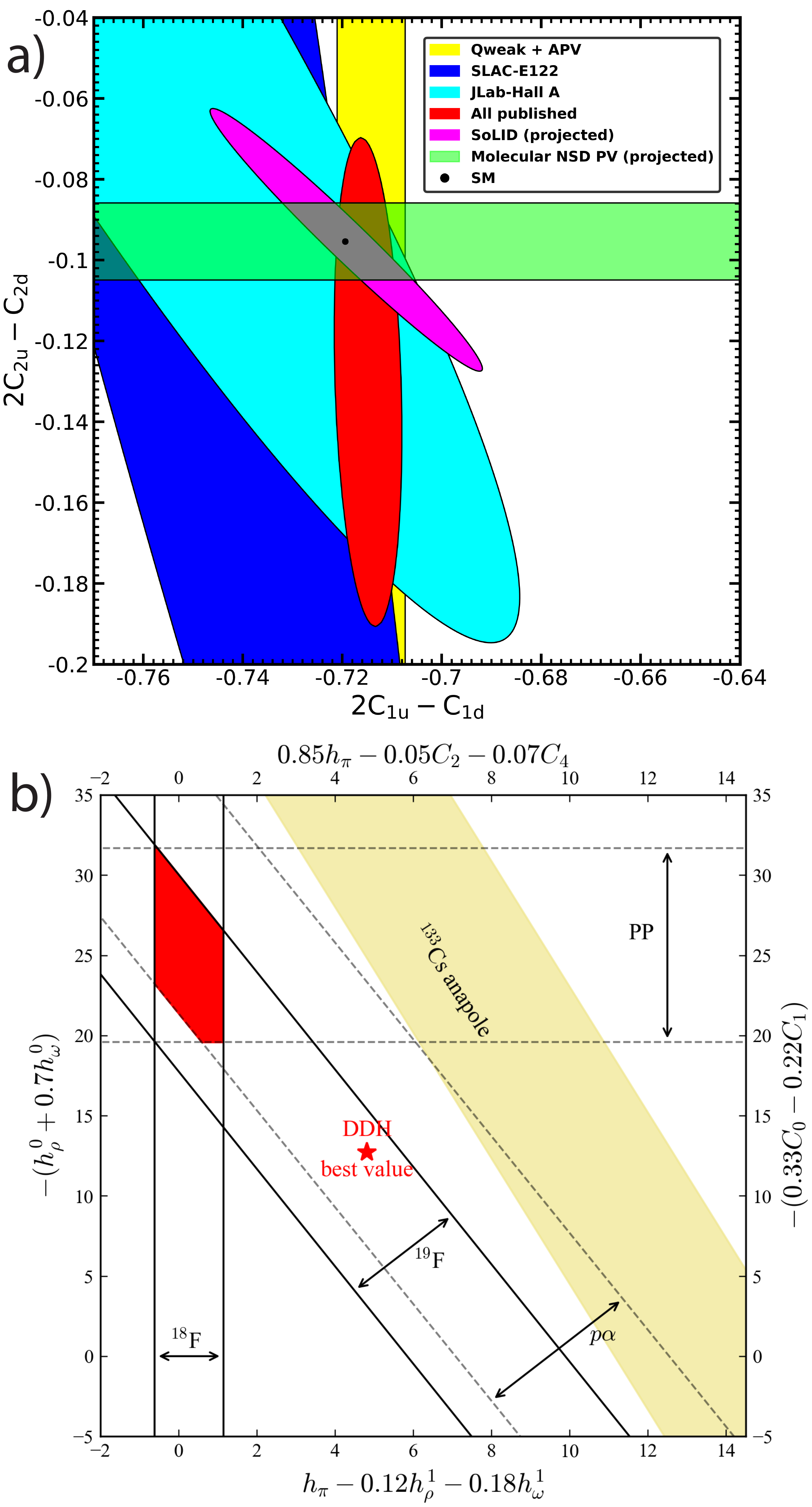}
    \caption{a) Constraints on the SM parameters $C_{1q}$ and $C_{2q}$ from existing and proposed experiments. The combined ellipse of existing measurements is shown in red. Expected bounds from the deep inelastic scattering SoLID experiment are shown in magenta \cite{moller_solid_2022_jlab}. In green, we show expected results from future molecular NSD PV measurements. It can be clearly seen that such measurements would be highly important, having the potential to significantly reduce the uncertainties on the $C_{2u}$ and $C_{2d}$ parameters of the SM. Figure adapted from Ref. \cite{moller_solid_2022_jlab}.  b) Constraints on the PNC meson couplings of the DDH potential ($\times 10^7$) obtained from different experimental constraints. Corresponding combination of parameters in the chiral potentials obtained with the equivalence relations from~\cite{deVries2014study}. The DDH potential best values are in mild tension with current experimental constraints, although nuclear theoretical uncertainties have not been included. Figure reproduced using data available from~\cite{haxton2013hadronic}.}
    \label{fig:NSD_constraints}
\end{figure}

So far, the only non-zero NSD PV measurement in an atom was performed in the $^{133}$Cs experiment~\cite{wood1997}, where the dominant contribution comes from the anapole moment \cite{flambaum1997anapole}. Despite its much smaller effect relative to the weak charge, the anapole moment was cleanly extracted as the difference between the $E_1$ induced PV amplitude between 2 different pairs of hyperfine levels: $6S_{1/2}(F=3)$ $\rightarrow$ $7S_{1/2}(F=4)$ and $6S_{1/2}(F=4)$ $\rightarrow$ $7S_{1/2}(F=3)$, with a value of \cite{wood1997}:

\begin{equation}
\text{Im}(E_\text{PV}/\beta)_\text{NSD} = 0.077 \pm 0.011~\text{mV/cm}\,,
\end{equation}
from which it was obtained that $\eta_\text{anapole}(^{133}\text{Cs}) = 0.090 \pm 0.016$~\cite{flambaum1997anapole,dzuba1987calculation,kraftmakher1988hartree,bouchiat1991nuclear,haxton2001anapole,haxton2001atomic,johnson2003combined}.  Together with measurements performed in Tl~\cite{vetter1995,bouchiat1991nuclear,haxton2001anapole,haxton2001atomic}, $(\eta_{\mathrm{anapole}}(Tl) = 0.376 \pm 0.400)$, this constrains combinations of the hadronic PV meson-nucleon couplings (Fig.~\ref{fig:NSD_constraints} b)). However, tensions still exist between different experiments~\cite{haxton2001atomic,haxton2013hadronic}, motivating further NSD PV measurements, particularly in nuclei with an unpaired neutron, which would provide a band roughly perpendicular to the Cs constraint in Fig.~\ref{fig:NSD_constraints} b).

\subsubsection{Molecular enhancements for NSD PV}

As derived in Sec.~\ref{sec:mol_enhance}, molecules provide enhancements of $10^5$-$10^{7}$ for NSD PV effects relative to atoms, by displaying small opposite-parity splittings \cite{brown2003rotational,safronova2018}, while in some cases, these can reach $10^{11}$-$10^{12}$ under proper experimental conditions \cite{altuntacs2018demonstration,altuntacs2018measuring,karthein2023electroweak}. In addition, in most molecules, the NSI weak-charge contribution is significantly suppressed, being identically zero to first order in perturbation theory \cite{kozlov1995} (see Sec. \ref{sec:appendix_NSI}).  Thus, any non-zero PV signal in such a molecule will be dominated by the NSD effects, making molecules ideal laboratories for exploring these phenomena.

In molecules with $^2\Sigma_{1/2}$ or $^2\Pi_{1/2}$ electronic states, the NSD PV Hamiltonian can be written as an effective spin-rotational and hyperfine operator \cite{flambaum1985enhancement,kozlov1995,demille2008using}:

\begin{equation}
H_\text{NSD}^{\text{PV},\text{eff}} = \eta\, W_\text{PV}\,(\hat{\mathbf{n}}\times\mathbf{S}_\text{eff})\cdot\mathbf{I}\,,
\label{eq:HNSI}
\end{equation}
where $\hat{\mathbf{n}}$ is the unit vector along the internuclear axis, $\mathbf{S_{\mathrm{eff}}}$ is the effective nuclear spin, and $W_\text{PV}$ is an electronic form factor which can be calculated to percent-level accuracy using relativistic coupled-cluster methods.

Beyond testing the electroweak sector of the SM, NSD PV measurements in
molecules probe parity violating interactions mediated by hypothetical
new vector bosons across an exceptionally wide mass range. The isotope
shift searches discussed in Sec.~\ref{sec:IS_newphysics} constrain parity
conserving products of vector couplings, $g_V g_V$; parity violating
observables are the only laboratory probes of the complementary axial
vector products $g_A g_V$, to which fifth force and exotic spin
dependent force searches are blind~\cite{safronova2018thoriumradii,
cong2025spin}. A recent reinterpretation of the $^{138}$Ba$^{19}$F
experiment~\cite{altuntacs2018demonstration} set the first direct
constraints on axial vector nucleus-electron interactions, a channel
previously unconstrained for boson masses above
$\sim$$10^{-4}$~eV$/c^{2}$~\cite{gaul2026constraints}, while at large
mediator masses the corresponding atomic bounds from
$^{133}$Cs~\cite{dzuba2017probing} reach effective scales of several
TeV, competitive with collider contact interaction searches while
retaining a distinct chirality structure. In contrast to Cs, where the
extraction depends on the nuclear anapole moment calculation with an
uncertainty of at least 30\%, present molecular extractions are not
limited by nuclear theory. Projected experiments with $^{29}$SiO$+$, $^{137}$BaF
and $^{225}$RaF have the potential to improve the coupling sensitivity by at least two orders of magnitude~\cite{gaul2026constraints}. Moreover,
comparisons across isotopologue chains can
disentangle SM from beyond SM contributions through their different
$Z$ and $A$ scalings, a handle unavailable to single isotope atomic
measurements~\cite{gaul2026constraints}.
\\

Several experimental techniques are currently being developed to measure NSD PV effects in molecules (Fig.~\ref{fig:PV_anapole_experiments}):
\paragraph{\textbf{Neutral beam experiments.}}
The first proof-of-principle molecular NSD PV experiment was performed in $^{138}$Ba$^{19}$F \cite{altuntacs2018demonstration,altuntacs2018measuring}, using the AC Stark interference technique \cite{demille2008using} in a supersonic beam. The experiment used a static magnetic field in order to bring opposite-parity levels to within $\sim 1$ kHz of each other, which provided an enhancement of the PV mixing of more than $11$ orders of magnitude compared to the atomic case \cite{cahn2014zeeman,altuntacs2018demonstration,altuntacs2018measuring}. In addition, a sufficient control of systematics was demonstrated to allow future measurements of the NSD PV in $^{137}$Ba$^{19}$F, where the effect due to the $^{137}$Ba is large enough to be observed experimentally. Optical cycling and transverse laser cooling of $^{137}$BaF have recently been demonstrated~\cite{kogel2025laser, kogel2025isotopologue}, paving the way for future higher precision measurements. A new approach using TlF molecular beams has been proposed to measure the Tl anapole moment via PV spin-spin coupling between the Tl and F nuclei \cite{blanchard2023using}, leveraging the CeNTREX experiment infrastructure \cite{grasdijk2021centrex}.

\paragraph{\textbf{Trapped molecular ions.}}
A Penning-trap approach for NSD PV in molecular ions has been proposed in Ref. \cite{karthein2023electroweak}, taking advantage of the trap's strong magnetic field, able to tune opposite-parity levels close to degeneracy and a quadrupolar electric field, which can be used for AC Stark interference.  Predicted enhancements of over $10^{12}$ are expected relative to atoms, with much longer coherence times relative to molecular beam experiments envisioned \cite{karthein2023electroweak}.  This method applies to nuclei across the nuclear chart, including short-lived radioactive species produced at RIB facilities.

\begin{figure*}[t]
    \centering
    \includegraphics[width=\textwidth]{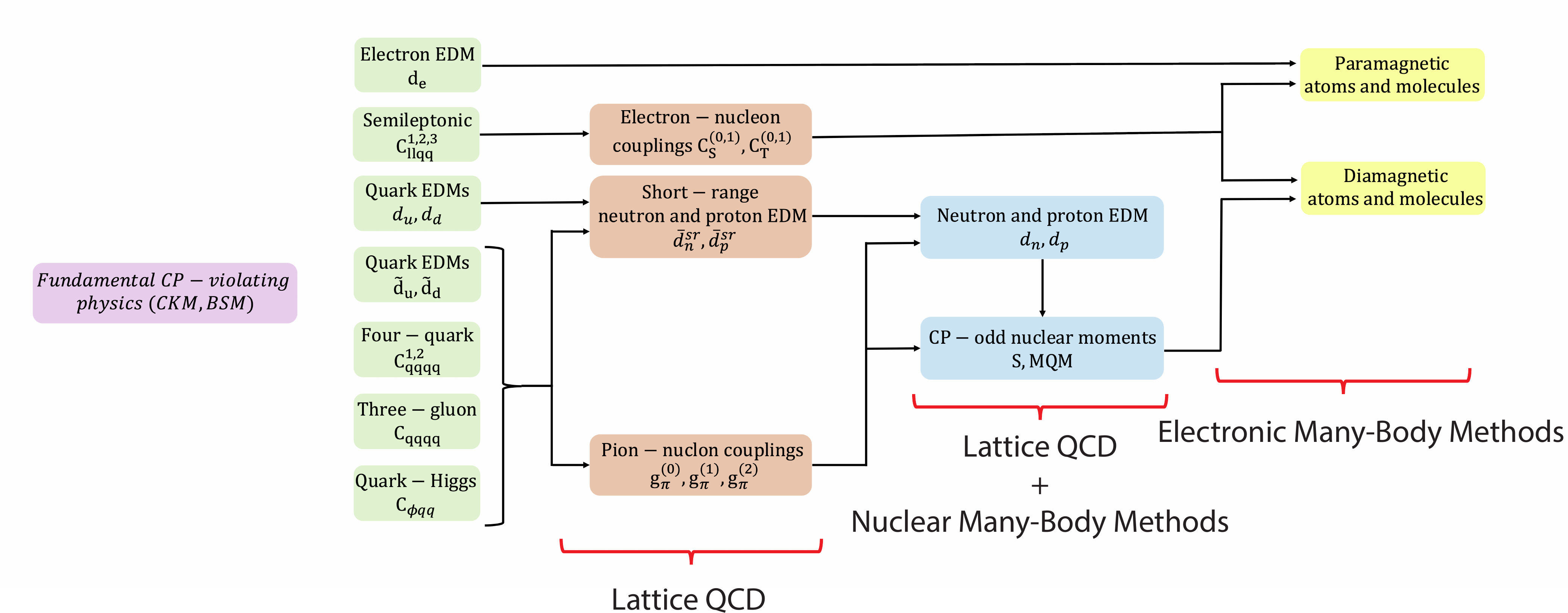}
    \caption{Manifestation of CP violating effects at different energy scales.  BSM physics (purple box) at $\Lambda \gg v$ generates Wilson coefficients in the SMEFT (green boxes), which are matched through the electroweak and QCD scales onto hadronic parameters: $d_e$, $d_{n,p}$, $C_S$, $C_T$, $\bar{g}_\pi^{(i)}$ (orange and blue boxes), and finally onto atomic/molecular observables (yellow boxes). Several computational methods are used for this matching, such as lattice QCD and nuclear and electronic many-body methods (see main text for details).}
    \label{fig:CP_EFT}
\end{figure*}

\paragraph{\textbf{Polyatomic molecules.}}
Certain polyatomic molecules such as YbOH, BeCN, and MgNC offer closely spaced energy levels of opposite parity, called $\ell$-doublets \cite{merer1971rotational}, that can be tuned to degeneracy in magnetic fields $\sim 100\times$ smaller than in the case of diatomics \cite{norrgard2019nuclear,hao2020nuclear}.  Combined with laser cooling and trapping techniques, which can allow for long coherence times, these systems are expected to enable NSD PV measurements even in light nuclei (e.g., Be, C, N, Mg), where \textit{ab initio} nuclear theory is more reliable and can provide quantified uncertainties on the anapole moment \cite{hao2020nuclear} (Sec.~\ref{sec:anapole_theory}).

\subsubsection{The nuclear weak quadrupole moment}

The rank-2 term in the weak multipole expansion (Sec.~\ref{sec:W_expansion}) gives rise to the nuclear weak quadrupole moment $Q_{W}^{(2)}$ (Fig. \ref{fig:weak_charge_interaction} b)), which probes the quadrupole deformation of the neutron distribution inside nuclei \cite{flambaum2016enhancing,flambaum2017effect,lackenby2018weak}.  This NSD PV effect requires isotopes with spin $I \geq 1$, and it is enhanced in heavy, deformed nuclei.  Certain electronic configurations in molecules (e.g., $^3\Delta_1$ electronic states) are expected to be particularly suited for measurements, as in these cases the weak quadrupole effect can be isolated from the anapole and weak charge contributions, enabling a background-free measurement of the neutron quadrupole moment \cite{skripnikov2019hff,penyazkov2022effect}.  Candidate molecules include HfF$^+$ \cite{skripnikov2019hff}, TaN \cite{skripnikov2019hff}, ThO \cite{skripnikov2019hff}, ThF$^+$ \cite{skripnikov2019hff} and  TaO$^+$ \cite{penyazkov2022effect} where the effect expected to be $\sim 1$ mHz \cite{skripnikov2019hff,penyazkov2022effect}.

\subsection{CP violating nuclear properties}
\label{sec:CPV_nuclear}

\subsubsection{Motivation: baryon asymmetry and the strong CP problem}

The observed dominance of matter over antimatter in the Universe requires, through the Sakharov conditions~\cite{sakharov1998violation}, new sources of CP violation beyond what the CKM matrix of the SM can provide.  Searches for permanent electric dipole moments (EDMs) of atoms and molecules are among the most sensitive probes of such new phenomena, given that a non-zero EDM requires simultaneous P and T symmetry violations, which is equivalent to CP violation under the assumption that the combined CPT symmetry of the Universe is conserved \cite{purcell1950possibility,streater2000pct,safronova2018,arrowsmith2023opportunities}.

An additional motivation for ongoing searches for CP violation comes from the strong CP problem \cite{weinberg1976gauge}. Rotating the axial phase of the quark mass matrices, the QCD Lagrangian consist of a CP violating term $\propto \bar{\theta}_\text{QCD}\, G^{\mu\nu}\tilde{G}_{\mu\nu}$, where $G^{\mu\nu}$ is the gluon field, $\tilde{G}_{\mu\nu} \equiv \epsilon_{\mu\nu\alpha\beta}G^{\alpha\beta}/2$ is its dual, with $\epsilon_{\mu\nu\alpha\beta}$ being the completely antisymmetric tensor and $\bar{\theta}_\text{QCD}$ quantifies the strength of this CP violating effect \cite{weinberg1976gauge,navas2024review}. However, the best experimental limits on $\bar{\theta}_\text{QCD}$, coming from the $^{199}$Hg \cite{graner2016reduced} and ultracold neutron \cite{abel2020measurement} measurements (see Fig. \ref{fig:CP_limits}), imply $\bar{\theta}_\text{QCD} \lesssim 10^{-10}$, while the CKM phase, the other CP violating parameter in the SM, is of order unity, leading to the celebrated strong CP problem. A leading proposed explanation for this observation is provided by the Peccei-Quinn mechanism and the existence of the axion \cite{peccei1977cp,wilczek1978problem,kim2010axions}.  

A wide variety of QCD axion models have been proposed in the literature, differing in their ultraviolet realization and in the pattern of axion couplings to SM fields. For simplicity, and because the phenomenology discussed below is largely insensitive to these details, we restrict our attention to the minimal low-energy description in which the axion couples directly only to the QCD topological density:
\begin{equation}
\mathcal L \supset \frac{\alpha_s}{8\pi}\frac{a}{f_a}G^a_{\mu\nu}\tilde G^{a\mu\nu},
\end{equation}
where $f_a$ denotes the axion decay constant. Additional couplings to quarks, leptons, and photons may be present depending on the ultraviolet completion, but are not essential for our discussion. The relation between ultraviolet axion models and the resulting infrared effective theory, including the determination of the axion mass and low-energy couplings after QCD confinement, has been studied extensively in the literature (see, for example, Refs. \cite{GrilliDiCortona:2015jxo,DiLuzio:2020wdo,Marsh:2015xka}).

Planck-suppressed operators that explicitly violate the Peccei--Quinn (PQ) symmetry shift the axion vacuum away from the CP conserving minimum and regenerate a non-zero strong CP phase
\cite{Barr:1992qq,Kamionkowski:1992mf,Holman:1992us,Ghigna:1992iv}. Taking as a benchmark
\begin{equation}
f_a = 10^{10}\ {\rm GeV},
\end{equation}
which lies close to the astrophysically preferred lower end of the allowed QCD axion window \cite{Caputo:2024oqc}, the induced strong CP phase generated by the leading PQ-violating operator of dimension $n-4$ may be conveniently estimated as
\begin{equation}
\bar\theta_{\rm axion}
\simeq
10^{-10}\,
10^{-8.4\,(n-n_c)}
\left(
\frac{f_a}{10^{10}\ {\rm GeV}}
\right)^n ,
\end{equation}
where $n_c\simeq10.5$ denotes the operator dimension for which the induced $\bar\theta$ saturates the current neutron EDM bound for $f_a=10^{10}\,$GeV. The exponential dependence on $n$ illustrates the well-known sensitivity of axion models to higher-dimensional PQ-violating operators.

In minimal Nelson--Barr (NB) models, the quality problem is considerably milder. The dominant contribution from Planck-suppressed operators typically arises from the leading dimension-five operator that violates the symmetry protecting the Nelson--Barr structure
\cite{Nelson:1983zb,Barr:1984qx,Dine:1993qm,Perez:2020cwa}. Combining the flavor dependence of the minimal Nelson--Barr construction with existing estimates of the leading dimension-five quality-violating operators (see, for example,
Refs.~\cite{Dine:2024bxv,Bai:2022nat,Asadi:2022vys,Valenti:2021rdu}), we estimate
\begin{equation}
\bar\theta_{\rm NB}
\simeq
10^{-10}\,
\frac{m_c}{m_{q^u}},
\end{equation}
where $m_{q^u}=m_u,m_c,m_t$ is determined by the up-type flavor direction selected by the vector-like quark couplings. This implies the characteristic range (see Fig. \ref{fig:CP_limits}):
\begin{equation}
10^{-12}
\lesssim
\bar\theta_{\rm NB}
\lesssim
10^{-7},
\end{equation}
demonstrating that, in minimal Nelson--Barr constructions, the residual strong CP phase is controlled primarily by the ultraviolet flavor structure. In contrast, for the QCD axion, the quality problem is governed predominantly by the dimension of the leading PQ-violating operator.

\begin{figure}[t]
    \centering
    \includegraphics[width=\linewidth]{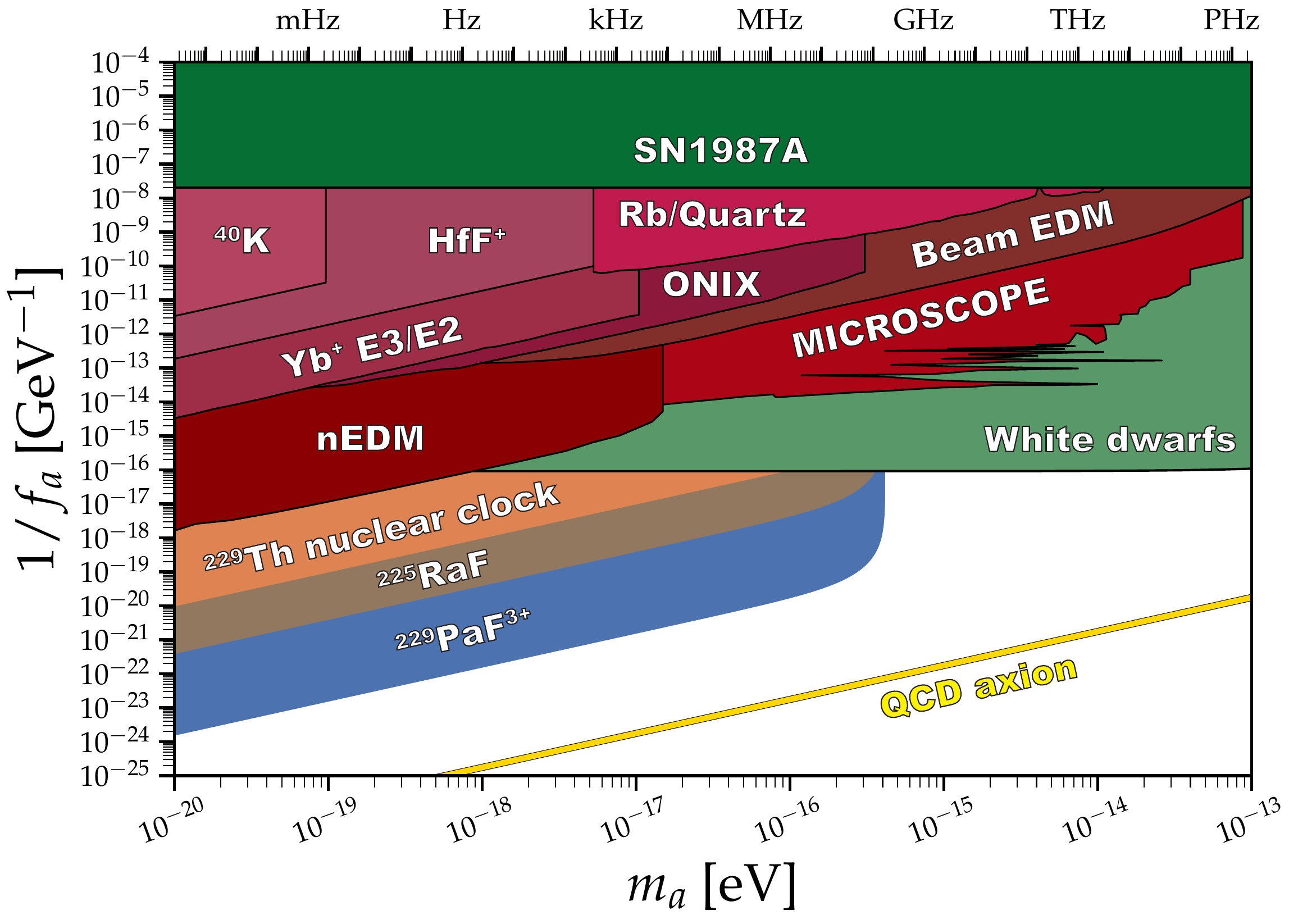}
    \caption{Exclusion plot for the axion coupling $1/f_a$ as a function of axion mass.  The QCD axion band (yellow) shows the predicted relationship between mass and coupling; axion-like particles (ALPs) can exist anywhere in the parameter space.  Bounds for several lab-based experiments and astrophysical observations are shown (figure adapted from Ref. \cite{OHare:2020wah}, under CC BY 4.0). Projected limits using radioactive molecules ($^{225}$RaF and $^{229}$PaF$^{3+}$) and the $^{229}$Th nuclear clock are shown. We assume for RaF: $\tau = 10$ s coherence time, $N = 10^4$ molecules per measurement and $1$ month of integration time; for PaF$^{3+}$: $\tau = 10$ s, $N = 10^3$ and $1$ month of integration time; for the $^{229}$Th nuclear clock we assume a relative measurement precision of $df/f = 10^{-19}$ and an enhancement factor $K = 10^5$ (see Sec. \ref{sec:nuclear_transitions} for details).}
    \label{fig:axion_exclusion}
\end{figure}

Besides the QCD axion (yellow band in Fig.~\ref{fig:axion_exclusion}), more general types of axion-like particles (ALP) exist. Searching for such general ALP is further motivated by the possibility of them making up a significant amount of the dark matter in the Universe \cite{navas2024review}. Such ALP can in principle be located anywhere in the parameter space displayed in Fig. \ref{fig:axion_exclusion}, where we show existing limits for various lab-based and astrophysics-based experiments. We also show projected limits using the radioactive molecules RaF and PaF$^{3+}$, as well as the new $^{229}$Th nuclear clock platform (see Sec. \ref{sec:nuclear_transitions} for details). Assuming experimental parameters which have already been achieved in other systems (e.g., stable atoms and molecules and atomic clocks) \cite{robichaud2026parity,roussy2023improved,campbell2017fermi}, these species can significantly push the limits on the existence of axions with masses below $\sim 10^{-16}$ eV. 

\subsubsection{The EFT framework: from BSM physics to laboratory observables}

The connection between high-energy CP violating physics, which is expected in BSM theories, and low-energy atomic and molecular EDM measurements is made through an EFT matching chain as shown in Eq.~(\ref{eq:EFT_chain}) and Fig. \ref{fig:CP_EFT} \cite{chupp2015electric,chupp2019electric,degenkolb2026global}. In this formalism, the effects of new heavy particles at a scale $\Lambda \gg v$ (the electroweak symmetry-breaking scale, $v = 246$~GeV \cite{chupp2019electric}) are parameterized by higher-dimensional operators in the Standard Model Effective Field Theory (SMEFT), with Wilson coefficients whose effect scales as $(v/\Lambda)^2$ \cite{chupp2019electric,degenkolb2026global}.  At the electroweak scale, BSM effects are parameterized by 12 Wilson coefficients, such as the quark EDMs, quark chromo EDMs, or three-gluon operators (see Ref. \cite{chupp2019electric} for more details of this parameterization).  These are then matched onto hadronic-scale parameters: the electron EDM $d_e$, the nucleon EDMs $d_{n,p}$, the scalar ($C_S$) and tensor ($C_T$) electron-nucleon couplings, and the pion-nucleon couplings $\bar{g}_\pi^{(0,1,2)}$ (Fig.~\ref{fig:CP_EFT}), which in turn contribute to the CP violating nuclear moments - the Schiff moment $\mathbf{S}$ (Sec.~\ref{sec:CPV_expansion}) and the magnetic quadrupole moment $M_{ij}$ (Fig.~\ref{fig:multipoles}f) - and to the electron-nucleus CP violating contact interactions $H_S$ and $H_T$ (Eqs.~(\ref{eq:HS}) and~(\ref{eq:HT})). Finally, these effects can be measured as atomic and molecular observables (see Fig. \ref{fig:CP_EFT}).

At each step, a factorization principle applies, similar to the one described in Sec. \ref{sec:ch2}: the physics above the matching scale is absorbed into the coefficients, while the physics below is encoded in the matrix elements.  For the case of CP violating searches, this chain takes the explicit form:
\begin{equation}
\begin{aligned}
& \underbrace{C_i(\Lambda)}_{\text{BSM Wilson coeff.}} \\
& \quad \xrightarrow{\text{EW matching}} \underbrace{d_q,\, \tilde{d}_q,\, C_S,\, C_T,\, \ldots}_{\text{quark-level couplings}} \\
& \quad \xrightarrow{\text{QCD matching}} \underbrace{d_n,\, d_p,\, \bar{g}_\pi^{(i)}}_{\text{hadronic couplings}} \\
& \quad \xrightarrow{\text{nuclear many-body}} \underbrace{S,\, M,\, d_A}_{\text{nuclear moments}} \\
& \quad \xrightarrow{\text{atomic/molec.\ theory}} \underbrace{\delta E,\, d_\text{atom},\, d_\text{mol}}_{\text{laboratory observables}}
\end{aligned}
\label{eq:EFT_chain}
\end{equation}
Each arrow represents a factorization: the physics $above$ is encoded in the coefficients passed down. The physics $below$ is computed independently at each scale.  For example, the CP violating nuclear moments $\Mn^{(\lambda)}$ in our multipole expansion in Sec. \ref{sec:ch2} sit at the penultimate step of this chain, receiving input from hadronic and nuclear physics and feeding into the atomic/molecular observables through the electronic factors $\Fn_e^{(\lambda)}$.

Different atomic and molecular systems are sensitive to different CP violating parameters \cite{chupp2015electric,chupp2019electric,degenkolb2026global}:
\begin{itemize}
\item \textbf{Paramagnetic systems} (unpaired electron spin): dominated by $d_e$, $C_S$, and MQM (for systems with non-zero nuclear spin).
\item \textbf{Diamagnetic systems} (zero electronic spin, non-zero nuclear spin): dominated by the nuclear Schiff moment and $C_T$.
\end{itemize}
Therefore, constraining all 12 Wilson coefficients requires a program of measurements in multiple systems.


This EFT chain is directly analogous to the factorization used in collider physics to predict cross sections: PDFs absorb the non-perturbative hadron structure, hard cross sections encode the short-distance partonic physics, and the two are convolved to produce the observable \cite{navas2024review}.  The key difference is that, in the atomic/molecular program, the ``beam energy'' cannot be tuned---it is fixed by the electronic structure---but the ``target'' can be changed by choosing different atoms, molecules, isotopes, electronic states, and rovibrational levels.  Each choice provides a different linear combination of the nuclear moments $\Mn^{(\lambda)}$, and a sufficiently diverse set of measurements can disentangle the individual contributions, just as measurements at different $Q^2$ and $x$ in DIS disentangle the individual parton flavors \cite{navas2024review}.

This perspective---atoms and molecules as fixed-target, variable-probe experiments, analyzed within the same factorization and EFT framework used at colliders---unifies the low-energy precision program with the high-energy frontier and makes explicit why the two approaches are complementary rather than competing (see Sec. \ref{sec:appendix_form_factors}).

\subsubsection{Current experimental limits}

\paragraph{ Electron EDM and \texorpdfstring{$C_S$}{CS}.}
The most precise limits for \texorpdfstring{$d_e$}{de} and \texorpdfstring{$C_S$}{CS} come from molecular experiments: \textbf{ThO} (ACME collaboration)~\cite{acme2018improved}: a cryogenic beam of ThO molecules in a metastable electronic state, which can be fully polarized at $E \approx 100$~V/cm and with a $\sim 1$~ms precession time; \textbf{HfF$^+$} (JILA)~\cite{roussy2023improved}: the first EDM measurement using trapped molecular ions, with full polarization at $\sim 58$~V/cm and variable precession times, which can be on the second scale.

The current best limits, from a global fit to both experiments (90\% C.L.), are $|d_e| < 2.1 \times 10^{-29}$~$e\cdot$cm and $|C_S| < 1.9 \times 10^{-9}$~\cite{roussy2023improved, acme2018improved}.  These measurements are already able to probe BSM CP violation physics at energy scales exceeding $10$'s of TeV. 

\paragraph{ Schiff moment and \texorpdfstring{$C_T$}{CT}.}
The most strict limit in this case comes from $^{199}$Hg~\cite{graner2016reduced}. Despite a much smaller achievable polarization compared to molecular systems, the macroscopic quantity of $^{199}$Hg used ($\sim 0.5$ mg) and the long precession times achieved in the presence of nominally uniform electric and magnetic fields ($\tau \approx 170$ s) allowed an upper limit on the mercury EDM to be set $|d_\text{Hg}| < 7.4 \times 10^{-30}$~$e\cdot$cm (95\% C.L.), after $\sim 300$ days of data taking. Other limits exist from $^{129}$Xe~\cite{sachdeva2019new, allmendinger2019measurement}, $^{225}$Ra~\cite{bishop2016improved}, and $^{171}$Yb~\cite{lu2023measurement}. While weaker than the ones from the $^{199}$Hg measurement, these experiments are very useful in setting limits on various Wilson coefficients in a global analysis framework \cite{chupp2015electric,chupp2019electric,degenkolb2026global}.  The $^{225}$Ra experiment, using laser-cooled atoms in an optical dipole trap, is particularly promising due to the expected $\sim 10^3$-fold octupole enhancement of the Schiff moment effect relative to $^{199}$Hg (Sec.~\ref{sec:CPV_expansion}); the current limit, $|d_\text{Ra}| < 1.4 \times 10^{-23}$~$e\cdot$cm, is statistics-limited, with at least two orders of magnitude improvement expected \cite{parker2015first,bishop2016improved}.

A limit on the Schiff moment of the stable
nucleus $^{153}$Eu, $|\mathcal{S}(^{153}\mathrm{Eu})| < 1.7\times10^{-8}~e\,\mathrm{fm}^3$
(95\% C.L.), was recently obtained through nuclear-spin resonance
spectroscopy of $^{153}$Eu$^{3+}$ ions doped into a Y$_2$SiO$_5$ crystal~\cite{Nima2026}.
The measurement introduces a novel solid-state route to Schiff-moment
searches, and the authors
identify no fundamental obstacle to reaching substantially higher
precision in future implementations. Translating such a bound into
constraints on the CP violating couplings relies on the
assumed octupole enhancement of the $^{153}$Eu Schiff moment. A robust
interpretation of this measurement will require microscopic calculations
of the relevant nuclear matrix elements with quantified
uncertainties.

 A current experiment,  CeNTREX \cite{grasdijk2021centrex}, is being developed to measure the Schiff moment of $^{205}$Tl using $^{205}$TlF molecules. By using a cryogenic beam of TlF and increasing the available flux using electrostatic focusing and rotational cooling~\cite{grasdijk2021centrex}, the experiment is expected to improve on the previous best measurements of the $^{205}$Tl nuclear Schiff moment by a factor of $2500$ \cite{cho1991search,grasdijk2021centrex}, corresponding to an improvement in the existing bounds on certain underlying CP violation sources of up to two orders of magnitude \cite{grasdijk2021centrex}.
Efforts towards measuring the nuclear MQM using paramagnetic molecules are also underway \cite{takahashi2025engineered}. Such searches could set bounds on nuclear CP violation physics complementary to the ones coming from searches for the nuclear Schiff moment \cite{arrowsmith2023opportunities}.

\subsubsection{New physics searches and the role of radioactive molecules}
\label{sec:radioactive_molecules_CPV}

Neutron EDM measurements and other hadronic CP violation tests strongly constrain the flavor-diagonal quark, chromoelectric dipole operators. These typically probe new-physics scales on the order of $\Lambda \sim 10^2~{\rm TeV}$ in a model-independent EFT analysis. In particular, following Ref.~\cite{Kley:2021yhn}, the  operators  are defined as:

\begin{equation}
    \begin{split}
        O_{uG}^{ij} &\equiv (\bar Q_L^i \sigma^{\mu\nu} T^A u_R^j)\widetilde HG^A_{\mu\nu} \\
        O_{dG}^{ij} &\equiv (\bar Q_L^i \sigma^{\mu\nu} T^A d_R^j)HG^A_{\mu\nu},
    \end{split}
\end{equation}
where $Q_L$ denotes the left-handed quark doublet, $u_R$ and $d_R$ the right-handed quark singlets, $T^A$ is the $SU(3)_c$ generator, and $\widetilde H=i\sigma_2H^\ast$. After electroweak symmetry breaking, the imaginary parts of the Wilson coefficients $C_{uG}^{11}$ and $C_{dG}^{11}$ generate the up- and down-quark chromoelectric dipole moments, respectively. Using the neutron EDM constraint and assuming a single operator at a time, Kley {\it et al.} obtain lower bounds on the effective scale of CP violating new physics of $ \Lambda \gtrsim 86\ {\rm TeV}$ and $\Lambda \gtrsim 170\ {\rm TeV}$ for $\mathrm{Im}\left(C_{uG}^{11}\right)$ and $\mathrm{Im}\left(C_{dG}^{11}\right)$, respectively, when the Wilson coefficients are normalized according to their natural NDA scaling and one includes both RG running and finite one-loop contributions.

\begin{figure}[t]
    \centering
    \includegraphics[width=\linewidth]{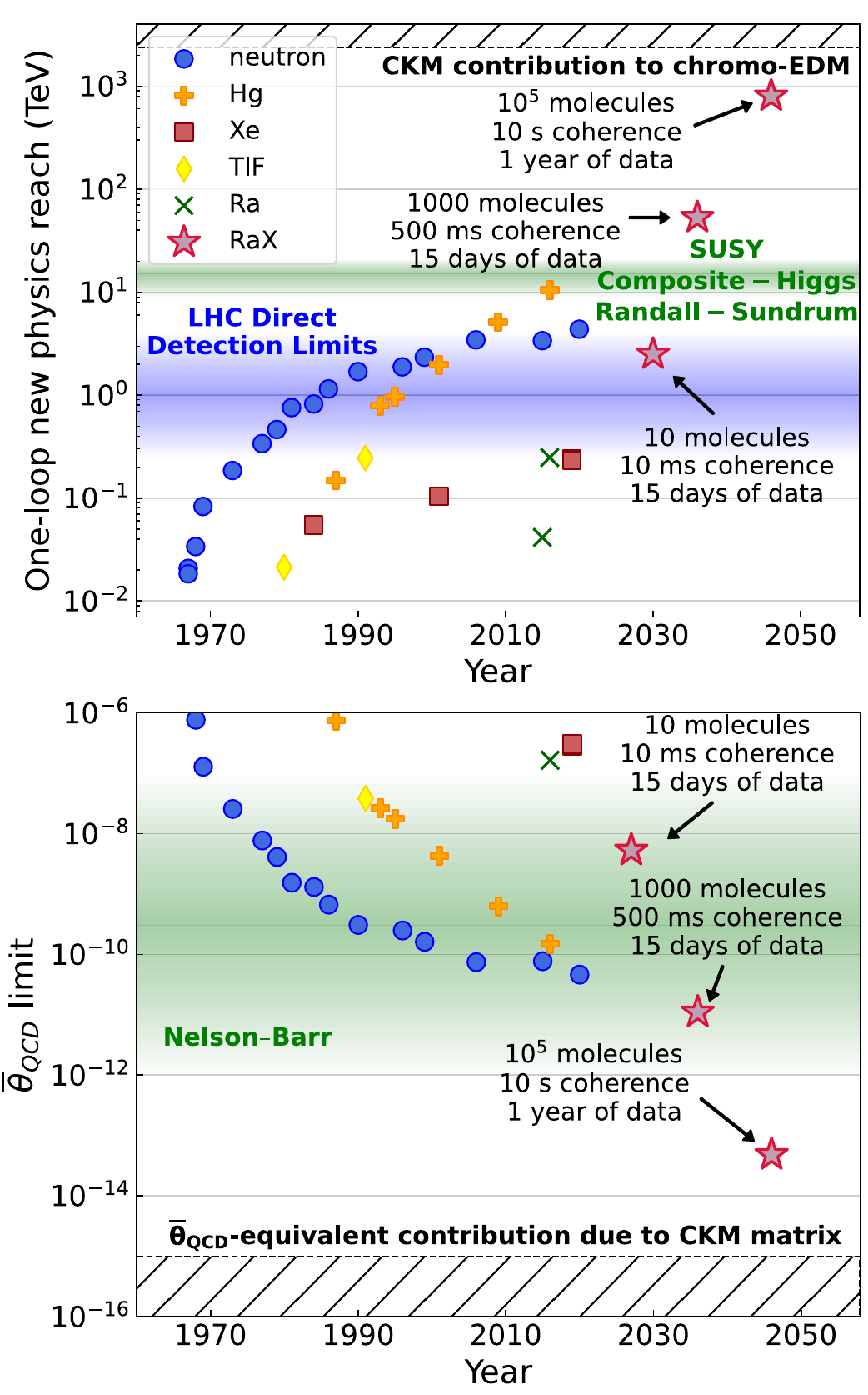}
    \caption{Top: Current one-loop limits on new particle masses inducing nuclear CP violation phenomena \cite{chupp2015electric,chupp2019electric,Engel2013,alarcon2022electric}. Projections from measurements using radium-containing molecules (e.g., $^{225}$RaF, $^{225}$RaOH) \cite{udrescu2024precision,wilkins2023observation,arrowsmith2023opportunities,isaev2017laser} are shown, for various numbers of molecules, coherence time, and total measurement time. Limits from LHC are shown as the blue band, while SM contributions from the CKM matrix are shown as the hatched region \cite{alarcon2022electric,navas2024review}. In green we show expected ranges for several BSM models: SUSY, Composite-Higgs, and Randall-Sundrum \cite{Agashe:2004cp,Agashe:2008uz,Azatov:2009na,Konig:2014iqa,Abel:2001vy,Maekawa:2017kfx}. Bottom: Similar to the top plot, but displaying limits on $\overline{\theta}_{\mathrm{QCD}}$. The parameter space covered by Nelson-Barr models \cite{Nelson:1983zb,Barr:1984qx,Dine:1993qm,Perez:2020cwa} is shown in green.  Figure generated using Ref. \cite{JayichLabEDMLimits}.}
    \label{fig:CP_limits}
\end{figure}

Bounds have also been derived in specific, well-motivated frameworks, such as supersymmetry (SUSY), and Randall–Sundrum/composite-Higgs. In explicit ultraviolet completions, the chromodipole operators are typically generated at one loop, so the corresponding bounds on the underlying particle masses are weaker by the expected loop factors. In anarchic Randall--Sundrum and composite Higgs models, neutron EDM constraints on the induced quark CEDMs generally require KK or fermion-resonance masses $ \gtrsim 10\text{--}20$ TeV for $\mathcal O(1)$ CP violating phases \cite{Agashe:2004cp,Agashe:2008uz,Azatov:2009na,Konig:2014iqa} (see Fig. \ref{fig:CP_limits}). Similarly, in supersymmetric models, the gluino--squark contribution to the quark CEDMs implies characteristic bounds on degenerate squark and gluino masses of order $m_{\tilde q}\sim m_{\tilde g}\gtrsim 10$ TeV, with the precise limit depending on the flavor structure and CP phases \cite{Abel:2001vy,Maekawa:2017kfx}. 

The combination of molecular enhancement with nuclear octupole enhancement (Sec.~\ref{sec:CPV_expansion}) makes molecules containing octupole-deformed nuclei the most sensitive systems for probing such well-motivated new physics nuclear CP violation scenarios.  Several experiments are underway or planned in this direction \cite{arrowsmith2023opportunities} (Fig.~\ref{fig:CP_limits}):
 
\paragraph{\textbf{\texorpdfstring{$^{225}$RaF}{225RaF}.}}
Despite being paramagnetic, RaF containing the octupole-deformed $^{225}$Ra nucleus is expected to improve on existing limits on hadronic Wilson coefficients by several orders of magnitude \cite{arrowsmith2023opportunities}.  Among diatomic fluorides, RaF, remains the most developed case. This molecule was proposed more than a decade ago as a laser-coolable probe of molecular symmetry-violation searches
\cite{isaev2010laser}, and a recent characterization of RaF energy level structure has been achieved during the last few years at ISOLDE \cite{udrescu2021isotope,udrescu2024precision,garcia2020spectroscopy,athanasakis2025electron}, including the demonstration of a laser-cooling scheme suitable for slowing and trapping, established the experimental foundation for future CP violation searches using this species.  Parallel efforts with the polyatomic RaOH molecule are also underway. This molecule offers the extra advantage of being fully polarizable in much smaller electric fields compared to RaF, thus allowing a better control on various sources of systematic uncertainties \cite{arrowsmith2023opportunities,conn2025production}.

\paragraph{\textbf{\texorpdfstring{$^{227,229}$ThF$^+$}{227and229ThF+}.}}
Extending the JILA HfF$^+$ technique to thorium isotopes would combine the molecular-ion trapping platform with nuclei predicted to have large Schiff moments. An implementation of this experiment using quantum logic protocols has been suggested \cite{zhou2024quantum}.

\paragraph{\textbf{FrAg.}}
A francium--silver molecule, assembled from laser-cooled Fr and Ag atoms at
ultracold temperatures, has been proposed for CP-violation searches
\cite{Fleig_2021}. Although FrAg is not expected to be directly laser coolable
as a molecule, it could be produced at the temperatures required for precision
measurements through Feshbach magnetoassociation
\cite{Kohler2006,Chin2010}. In this method, ultracold Fr and Ag atoms are
brought near a magnetically tunable scattering resonance, where the free-atom
threshold is coupled to a weakly bound molecular state. A magnetic-field sweep
across the resonance converts atom pairs into weakly bound FrAg molecules,
which may subsequently be transferred to more deeply bound states by optical
Raman techniques such as stimulated Raman adiabatic passage
\cite{Ni2008}.

\paragraph{\textbf{d. Other proposed species.}}
In addition to the molecules discussed above, a substantial body of theoretical work has proposed different radioactive molecules for symmetry violation searches across several
molecular classes. Recent spectroscopy results have been achieved for AcF \cite{AthanasakisKaklamanakis2025AcF}. Oxide molecules such as $^{225}$RaO were among the
earliest proposals suggested for Schiff moment searches
shortly after the octupole enhancement was recognized
\cite{flambaum2008electric}, with the molecular sensitivity factor later
computed with relativistic coupled-cluster methods
\cite{kudashov2013calculation}, and enhanced CP violating effects were
identified in $^{229}$Th-containing systems, including $^{229}$ThO in
its $^{1}\Sigma$ ground and metastable $^{3}\Delta_1$ states,
$^{229}$ThOH$^{+}$, and $^{229}$ThF$^{+}$ \cite{flambaum2019}.
Polyatomic species such as  RaOH and related hydroxides offer parity doublets in the bending mode, full
polarization in fields of a few V/cm, and internal comagnetometer
states
\cite{kozyryev2017precision,conn2025production}. Molecular
ions RaOH$^{+}$ and RaOCH$_3^{+}$, have already been produced and identified
alongside laser-cooled Ra$^{+}$ \cite{fan2021optical,isaev2022radium}. Alternative diatomics
include RaCl \cite{isaev2021ab} and molecules assembled atom-by-atom
from laser-cooled species, such as RaAg and FrAg
\cite{Fleig_2021,smialkowski2021highly,marc2023candidate}. Highly
charged molecular ions extend the family toward few-valence-electron
systems, with $^{229}$PaF$^{3+}$ as a compelling candidate that could take advantage the exceptionally small
parity-doublet splitting suggested for the nucleus $^{229}$Pa \cite{Ahmad2015Pa229}.  The recent formation of the chemical homolog CeF$^{2+}$ represents an
experimental step in this direction~\cite{nes2026formation}. In addition,
molecules containing the heaviest actinides have been identified as promising
systems with very large effective internal electric fields, for example
$\mathcal{E}_{\rm eff}\sim 200~\mathrm{GV/cm}$ in NoF, NoOH, LrO, and
LrOH$^{+}$~\cite{zhang2021calculations,mitra2021towards}. Common challenges
for all of these proposals include molecule formation from sub-nanogram samples
of radioactive material; efficient deceleration, cooling, and state
preparation; and accurate electronic-structure and nuclear calculations to guide
experimental searches and the interpretation of measurements.

\subsubsection{Nuclear theory as a main challenge.}
As emphasized in Sec.~\ref{sec:schiff_theory}, the nuclear Schiff moment coefficients $a_{0,1,2}$ (Eq.~(\ref{eq:SchiffCoefficient})) are the limiting source of theoretical uncertainty in extracting hadronic CP violating couplings from experimental limits.  The spread of phenomenological calculations (Fig.~\ref{fig:SchiffMoment_theory}) is large, and \textit{ab initio} methods have only recently begun to address this observable ($^{19}$F in the NCSM~\cite{ng2025nuclear} and $^{129}$Xe with the PV-IMSRG~\cite{belley2026abinitio}).  Extending \textit{ab initio} Schiff moment calculations to the heavy, octupole-deformed nuclei of experimental interest ($^{199}$Hg, $^{225}$Ra, $^{229}$Pa) is one of the central challenges for nuclear theory (Sec.~\ref{sec:outlook_theory}).

\subsection{Open challenges and outlook}
\label{sec:outlook_symm}

\subsubsection{Resolving the anapole moment tension.}
The existing constraints on hadronic PV couplings from the Cs anapole moment show tensions with other PV data~\cite{haxton2001atomic,haxton2013hadronic} as shown in Fig.~\ref{fig:NSD_constraints} b).  New measurements in nuclei with unpaired neutrons (e.g., in $^{137}$BaF, Yb isotopes, or radioactive species in ion traps) would provide orthogonal constraints which could help resolve these inconsistencies.

\subsubsection{Separating \texorpdfstring{$\eta_\text{anapole}$ from $\eta_\text{axial}$}{eta anapole from eta axial}.}
In heavy nuclei, NSD PV effects are dominated by the anapole moment, making it difficult to extract the direct $C_{2u,d}$ contribution.  Measurements in light nuclei (Be, C, N in polyatomic molecules) where $\eta_\text{axial}$ dominates, combined with \textit{ab initio} calculations of the anapole moment (Sec.~\ref{sec:anapole_theory}), offer a path to disentangling the two effects and constraining $C_{2u,d}$ - some of the least well-constrained parameters of the SM.

\subsubsection{Global analysis of CP violating experiments.}
Extracting bounds on individual Wilson coefficients requires a global fit to multiple experiments~\cite{degenkolb2026global}, which in turn requires reliable nuclear theory for the CP odd nuclear moment coefficients.  Currently, nuclear theory uncertainties significantly weaken the extracted limits from existing experiments \cite{degenkolb2026global}. The development of \textit{ab initio} methods for heavy nuclei (Sec.~\ref{sec:outlook_theory}) is critical for the full exploitation of the next generation of nuclear EDM experiments.

\subsubsection{Radioactive molecules at RIB facilities.}
The production, cooling, and trapping of molecules containing short-lived nuclei opens a path to precision measurements of CP violating effects in systems where both the molecular and nuclear enhancements are maximized.  Scaling this approach to other species (RaOH, ThF$^+$, FrAg) and to higher production rates at next-generation facilities (FRIB, FAIR) is a central goal.

\subsubsection{Connection to cosmology and high-energy physics.}
EDM experiments are among the very few ways to probe new sources of CP violation, beyond the SM, needed to explain the baryon asymmetry of the Universe.  The projected sensitivity of next-generation molecular experiments, able to explore physics at energy scales of $\sim 1000$~TeV and constraining $\bar{\theta}_\text{QCD}$ at the $<10^{-13}$ level (Fig. \ref{fig:CP_limits}), makes this program a vital complement to high-energy collider searches and cosmological observations.

\section{Nuclear Excitations and Decay}
\label{sec:nuclear_transitions}

The preceding section exploits the imprint of \emph{static} nuclear properties on atomic and molecular energy levels: the nucleus sits in its ground state while the electrons report on the charge radius, the moments, and the symmetry-violating properties through isotope shifts, hyperfine structure, and PV amplitudes.  In this section, we consider instead processes in which the nucleus
itself changes state. We focus the discussion on the excitation of the $^{229}$Th isomer, and the weak transformation of a nucleon in $\beta$ decay. The measured quantity is a nuclear transition rate or frequency rather than an
electronic shift. The two cases probe complementary sectors. The $^{229}$Th nuclear clock gives access to the strong and electromagnetic structure responsible for the isomer's anomalously small excitation energy, and to its extreme sensitivity to variations of fundamental constants and to ultralight dark matter. Precision $\beta$ decay probes the weak sector through the Fermi constant and the CKM matrix element $|V_{ud}|$, providing the most demanding test of first-row CKM unitarity. As with the static observables, the interpretation of both rests on the nuclear theory (Sec.~\ref{sec:nuclear_theory}), benchmarked against the electromagnetic properties of Sec.~\ref{sec:EM_properties}. The same Hamiltonians and many-body methods that reproduce charge radii and moments are required to compute the isomer energy, the change in quadrupole moment across the transition, and the isospin-breaking and radiative corrections to $\beta$ decay.

\subsection{The \texorpdfstring{$^{229}$Th}{229Th} nuclear clock}
\label{sec:Th229}

Among all known nuclides, $^{229}$Th is unique: its first excited nuclear state (the isomeric state $^{229m}$Th, $I^\pi = 3/2^+$) lies only $\sim 8.4$~eV above the ground state ($I^\pi = 5/2^+$)~\cite{kroger1976,reich1990,Wense2016}, placing it in the vacuum ultraviolet (VUV) at $\lambda \approx 148$~nm---within reach of state-of-the-art laser sources.  No other nucleus has a confirmed transition below $50$ eV.  This extraordinary accident of nuclear structure---a near cancellation between electromagnetic and strong-interaction contributions to the nuclear level spacing---opens the door to a ``nuclear clock'': a frequency standard based on a nuclear rather than electronic transition~\cite{peik2003,campbell2012}.  Because the nucleus is $\sim 10^5$ times smaller than the atom, it is intrinsically less sensitive to external electromagnetic perturbations, promising fractional frequency uncertainties at or below $10^{-19}$~\cite{peik2003,campbell2012,beeks2021}.

\subsubsection{The nuclear isomer: from discovery to laser excitation}
\label{sec:Th229_history}

The existence of a low-energy isomeric state in $^{229}$Th was first inferred in 1976 from $\gamma$-ray spectroscopy of the $^{233}$U decay chain at the Idaho National Engineering Laboratory~\cite{kroger1976}.  For nearly three decades, the isomer energy was known only at the level of a few eV, with large uncertainties that hindered direct experimental access.  The key milestones in the progression toward laser excitation were:

\paragraph{Energy determination.}
Successive measurements using $\gamma$-ray spectroscopy~\cite{reich1990,beck2007,beck2009}, internal conversion electron spectroscopy~\cite{seiferle2019}, and microcalorimetry~\cite{sikorsky2020} progressively refined the isomer energy from $3.5 \pm 1.0$~eV to $8.28 \pm 0.17$~eV.  The observation of the radiative decay of the isomer at ISOLDE (CERN) in $^{229}$Th-doped CaF$_2$ crystals~\cite{kraemer2023} determined the wavelength to $\lambda = 148.71(42)$~nm, corresponding to $8.338(24)$~eV, with THz-level precision.

\paragraph{First laser excitation (2024).}
In early 2024, Tiedau et al.\ achieved the first direct laser excitation of the $^{229}$Th nuclear transition using a tunable VUV laser source and $^{229}$Th-doped CaF$_2$ crystals~\cite{tiedau2024}.  The excitation was detected through the fluorescence photons emitted upon nuclear de-excitation.  This landmark result improved the energy precision by a factor of $\sim 800$ over the previous measurement and confirmed that the transition is optically accessible.  Independently, Elwell et al. demonstrated laser excitation in a $^{229}$Th:LiSrAlF$_6$ crystal~\cite{elwell2024laser}, providing a second solid-state platform.

\paragraph{Frequency-comb spectroscopy and the nuclear--atomic clock link (2024).}
The JILA group used a VUV frequency comb---generated by intracavity high-harmonic generation of an infrared comb stabilized to the JILA $^{87}$Sr optical lattice clock---to perform the first frequency-based measurement of the $^{229}$Th nuclear transition~\cite{zhang2024nature}.  This achieved kHz-level precision ($\sim 10^{-12}$ fractional uncertainty), a million-fold improvement over the wavelength-based measurements, and established the first direct frequency link between a nuclear transition and an atomic clock.  The measurement resolved the nuclear electric quadrupole splittings of the isomer, extracting the ratio of the quadrupole moments $\Delta Q_0/Q_0 = 1.791(2)\%$ between the excited and ground states---the first determination of an intrinsic nuclear structure parameter via laser spectroscopy of a nuclear transition.

\paragraph{Excitation in non-transparent materials (2025).}
A collaboration between UCLA, LMU Munich, and JGU Mainz demonstrated laser excitation of $^{229}$Th in a non-transparent host material (thorium electrodeposited on steel), detecting the nuclear signal through internal conversion electrons rather than fluorescence photons~\cite{elwell2025nontransparent}.  This removed the requirement for VUV-transparent crystals, dramatically simplifying sample preparation (1000$\times$ less thorium needed) and opening a new class of materials for nuclear laser spectroscopy.

\paragraph{Temperature sensitivity (2025).}
Higgins et al.\ performed quantum-state-resolved spectroscopy of the nuclear transition at three temperatures (150, 229, and 293~K) in CaF$_2$~\cite{higgins2025temperature}.  They found that one particular quadrupole-split transition shifted by only $\sim 62$~kHz over the full range ($\sim 0.4$~kHz/K), at least 30 times less than other transitions, suggesting the existence of a temperature ``sweet spot'' where the nuclear clock frequency is nearly independent of temperature.

\subsubsection{Connection to nuclear structure}
\label{sec:Th229_structure}

The $^{229}$Th isomeric transition connects naturally to the nuclear structure framework of this review.  Both nuclear states have well-defined multipole moments $\Mn^{(\lambda)}$ (Eq.~(\ref{eq:multipole_master})): the charge radius $\langle r^2\rangle$ (Eq.~(\ref{eq:r2_def})), the magnetic dipole moment $\bm{\mu} = g_I\muN\mathbf{I}$ (Eq.~(\ref{eq:Hmu})), the electric quadrupole moment $Q$ (Eq.~(\ref{eq:HQ})), and in principle the symmetry-violating moments of Secs.~\ref{sec:anapole} and~\ref{sec:CPV_expansion}.  The isomeric transition changes the nuclear quantum numbers ($I^\pi = 5/2^+ \to 3/2^+$), and therefore all these moments take different values in the excited state.  Computing them using the nuclear theory methods of Sec.~\ref{sec:nuclear_theory} is a precise test of our understanding of this nucleus.

The isomer energy arises from a near cancellation between the electromagnetic (Coulomb) and strong-interaction (nuclear) contributions to the splitting of the $5/2^+$ and $3/2^+$ Nilsson levels in the deformed $^{229}$Th nucleus.  The extreme sensitivity of this cancellation to the underlying nuclear force makes the isomer energy a stringent test of nuclear models~\cite{minkov2019,minkov2021}.  The JILA measurement of the quadrupole moment ratio $\Delta Q_0/Q_0$~\cite{zhang2024nature} provides a direct probe of the nuclear deformation change upon isomeric excitation.  Recent analysis~\cite{beeks2025alpha, Caputo:2024doz} found that the measured $\Delta Q_0/Q_0$ challenges the constant-volume approximation commonly used in nuclear models (within the classical geometrical model~\cite{Berengut:2009zz}), indicating that the nuclear volume changes upon excitation---a finding with direct implications for the sensitivity of the nuclear clock to variations of the fine-structure constant. 

Extracting the sensitivity parameter, discussed below, is rather challenging, and at present it is subject to large uncertainties~\cite{Caputo:2024doz}. Currently, beyond the classical geometrical model, a halo model was proposed as a quantum mechanical toy model to estimate the transition energy difference~~\cite{Caputo:2024doz}. The model's predictions are in agreement with the measured charge radius and quadrupole differences, and also with those of the geometrical model. A more methodical approach using multi-reference density functional theory (MR-DFT) was presented in~\cite{Restrepo-Giraldo:2026blh}. However, as mentioned, at the moment, all methods are subject to large theoretical uncertainties. 

We notice that the charge radius change $\delta\langle r^2\rangle$ between the ground and isomeric states is related to the isomer shift in the electronic spectrum of thorium ions through Eq.~(\ref{eq:IS_FS}), with the electronic field-shift factor $F$ (Eq.~(\ref{eq:F_def})) calculable from atomic theory.  This establishes a direct connection between the nuclear clock transition and the isotope-shift formalism of Sec.~\ref{sec:term_FS}.

\subsubsection{The nuclear clock concept}
\label{sec:nuclear_clock}

A nuclear clock based on the $^{229}$Th isomeric transition operates on the same principle as an atomic clock: a laser is stabilized to the narrow nuclear resonance, and the resulting oscillation frequency serves as the time standard.  The nuclear clock offers several potential advantages over atomic clocks:

\begin{itemize}
\item \textbf{Reduced sensitivity to external fields.}  The nucleus is shielded by the surrounding electrons, which act as an effective Faraday cage.  Systematic shifts from stray electric and magnetic fields, which limit atomic clocks, are strongly suppressed~\cite{peik2003,campbell2012}.
\item \textbf{Solid-state implementation.}  Unlike atomic clocks, which require ultracold atoms in vacuum, the nuclear clock can operate with thorium embedded in a solid-state crystal (CaF$_2$, LiSrAlF$_6$) or even electroplated on a metal surface~\cite{elwell2025nontransparent}.  This enables compact, robust devices.
\item \textbf{High transition frequency.}  The VUV transition at $2.02 \times 10^{15}$~Hz provides a large number of oscillation cycles per second, enabling high precision in short averaging times.
\end{itemize}

The projected fractional frequency uncertainty of a $^{229}$Th nuclear clock is $\sim 10^{-19}$~\cite{peik2003,campbell2012,beeks2021}, competitive with the best optical lattice clocks~\cite{aeppli2024}.  Two implementation paths are being pursued: solid-state nuclear clocks using doped crystals~\cite{zhang2024nature,higgins2025temperature}, and single-ion nuclear clocks using trapped $^{229}$Th$^{3+}$ ions~\cite{campbell2012}, where the ion-trap environment provides ultimate control over systematics.

Recent work on frequency reproducibility~\cite{zhang2025reproducibility} has demonstrated that the solid-state nuclear clock transition frequency can be reproduced at the kHz level across different crystals and over timescales of months, an essential prerequisite for a practical frequency standard.

\subsubsection{Probing new physics with nuclear clocks}
\label{sec:Th229_newphysics}

The nuclear clock's sensitivity to new physics arises from a combination of two features: (i)~the transition frequency depends on nuclear structure parameters that couple to the strong interaction, the quark masses, and the QCD scale; and (ii)~the small transition energy results from a near cancellation, so that even tiny shifts in the underlying parameters produce large fractional frequency changes.  This is parameterized by the response coefficient:
\begin{equation}
K \equiv \frac{\partial \ln \omega}{\partial \ln X}\,,
\label{eq:K_sensitivity}
\end{equation}
where $X$ is a fundamental constant (e.g., the fine-structure constant $\alpha$, the quark mass ratio $m_q/\Lambda_\text{QCD}$).  The enhanced sensitivity $K \gg 1$ makes the nuclear clock a uniquely powerful probe of new physics that couples to the nuclear sector.
Both the geometrical model and the halo model are consistent with a central value of $K\sim 10^4$~\cite{beeks2025alpha,Caputo:2024doz}, however, with large theoretical and experimental uncertainties. The more systematic study of
\cite{Restrepo-Giraldo:2026blh}, using MR-DFT, currently yield too large a variation between the different models to allow for any concrete conclusions.

\subsubsection{Ultralight dark matter searches}

Ultralight dark matter (ULDM) can be modeled as a coherently oscillating classical field, $\phi(t) = \phi_0\cos(m_\phi t)$, which induces time-dependent variations of fundamental constants.  These variations shift the nuclear transition frequency with an amplitude amplified by $K$.  Two complementary search strategies have been developed:

\paragraph{Lineshape analysis.}
When the ULDM oscillation period is shorter than or comparable to the spectroscopic interrogation time, the signal manifests as distortions of the resonance profile---broadening and sideband-like structures---rather than a simple frequency shift.  Ref.~\cite{Fuchs:2024xvc} developed this framework and showed that existing laser excitation data already probe ULDM couplings to nucleons, extending sensitivity to higher masses than traditional time-domain searches. The method has been applied to the JILA data in~\cite{Arakawa:2026mls}, improving the existing bounds by around 6 orders of magnitude.

\begin{figure*}[ht]
    \includegraphics{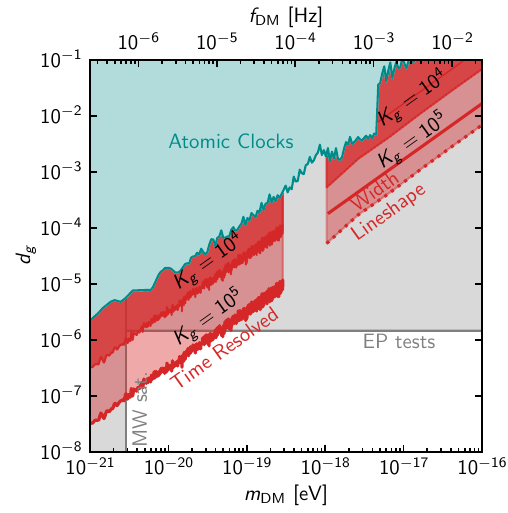}
    \includegraphics{06-nuclearexcitations/Figures/d_g.pdf}
    \caption{%
        Searches for coupling of scalar DM $\phi$ to quark masses (left) and gluons (right). The coloured regions represent constraints from searches for time variations in frequencies, assuming the scalar field constitutes DM. The red region shows the constraint derived under two scenarios for the sensitivity factor $K$ (see text for details on the challenge to calculate $K$). For small masses and correspondingly low frequencies, the bound is inferred from the time-resolved analysis. For larger masses, the lineshape analysis is applied. For the lineshape analysis, an agnostic bound is also shown, relying purely on the recorded width (straight) as well as one where a Lorentzian shape arising from the crystal environment is assumed (dotted). In contrast, the gray regions are excluded by constraints that rely on the scalar being sufficiently massive to support the existence of Milky Way (MW) satellite galaxies~\cite{DES:2020fxi}, as well as from tests of the equivalence principle (EP) violations mediated by the scalar field~\cite{MICROSCOPE:2022doy}. The provisional operation of a nuclear clock already bests the bounds coming from atomic clocks reaching the standard quantum limit~\cite{Hees:2016gop,Kennedy:2020bac,Kobayashi:2022vsf,Sherrill:2023zah,Filzinger:2023zrs,Banerjee:2023bjc} (teal). At masses $m_{DM}\lesssim 10^{-19}$, the nuclear clock already explores yet unconstrained territory. Figure reproduced from Ref. \cite{Arakawa:2026mls}, with permission from the authors.}
    \label{fig:scalar_nuclear_bounds}
\end{figure*}

\paragraph{Time-domain oscillation search.}
Using high-resolution JILA frequency-comb data, Ref.~\cite{Arakawa:2026mls} combined time-domain searches for coherent oscillations in the transition frequency (sensitive to low-mass ULDM) with lineshape analysis (sensitive to higher masses).  The resulting constraints on scalar couplings to nucleons reach effective interaction scales of $\sim 10^6\, M_\text{Pl}$, establishing the strongest laboratory bounds to date for hadronic ULDM couplings over a wide mass range.
In Fig.~\ref{fig:scalar_nuclear_bounds} taken from~\cite{Arakawa:2026mls}, we show the 2$\sigma$-bounds on scalar DM assuming two values for the sensitivity factor. 

\subsubsection{New forces and quintessometers}
Beyond oscillating dark-matter signals, scalar fields coupled to Standard Model operators generate spatial variations of fundamental constants in the presence of macroscopic sources.  Ref.~\cite{perez2025quintessometer} introduces the concept of ``quintessometers'': comparisons between nuclear and atomic clocks, or between spatially separated nuclear clocks, that probe equivalence-principle violation and fifth-force effects.  The projected sensitivities access previously unexplored parameter space, particularly at short distances ($\sim 10^{-7}$~m) in solid-state implementations and at macroscopic scales in transportable configurations.

\subsubsection{Variation of fundamental constants}

The large $K$-factor of the $^{229}$Th transition makes it an ideal system for constraining temporal drifts of fundamental constants.  A comparison between the nuclear clock and an atomic clock over time constrains $\dot{\alpha}/\alpha$ with sensitivity enhanced by $K$ relative to atomic-clock-only comparisons.  As the nuclear and atomic clocks have qualitatively different dependencies on $\alpha$, $m_q/\Lambda_\text{QCD}$, and $m_e/m_p$, their ratio probes specific linear combinations of these drifts that are inaccessible to atomic clocks alone.

\subsubsection{Challenges for the nuclear clock}
\label{sec:Th229_outlook}

The rapid experimental progress of 2023--2025---from the first observation of the isomer's radiative decay to kHz-precision frequency-comb spectroscopy and excitation in non-transparent materials---has established $^{229}$Th as a working experimental platform.  Several challenges and opportunities lie ahead:

\paragraph{Toward a practical nuclear clock.}
Achieving clock-level performance ($\sim 10^{-19}$) requires narrowing the VUV laser linewidth, understanding and controlling crystal-field-induced shifts, and identifying the temperature sweet spot predicted by the data of Ref.~\cite{higgins2025temperature}.  Single-ion implementations in Paul or Penning traps~\cite{campbell2012} offer an alternative route with potentially lower systematics.

\paragraph{Nuclear theory.}
The isomer energy, the quadrupole moment ratio, and the charge radius change are all quantities that challenge nuclear theory.  Computing these observables from first principles using the \textit{ab initio} methods of Sec.~\ref{sec:nuclear_theory}---which have been validated on the electromagnetic properties of lighter nuclei (Sec.~\ref{sec:EM_properties})---would sharpen the new-physics interpretation of frequency measurements and strengthen the connection between the nuclear clock and the broader program of nuclear structure studies reviewed in this article.

\paragraph{Connection to the multipole expansion.}
The nuclear clock transition changes the nuclear quantum numbers ($I = 5/2 \to 3/2$), modifying all the nuclear moments in the multipole expansion of Sec.~\ref{sec:ch2}: the charge radius, the magnetic dipole moment, the electric quadrupole moment, and potentially the anapole and Schiff moments.  Measuring these changes through the electronic response---via the isomer shift, the hyperfine structure of the isomeric state, or the PV amplitude in trapped $^{229}$Th ions---would provide a comprehensive characterization of the nuclear structure change upon isomeric excitation, using the same formalism developed throughout this review.

\paragraph{Broader landscape.}
While $^{229}$Th is currently unique, the search for other low-energy nuclear transitions continues.  Theoretical predictions suggest that other nuclear isomers with sub-keV energies may exist but remain undiscovered.  The experimental techniques developed for $^{229}$Th---VUV frequency combs, solid-state nuclear spectroscopy, internal-conversion detection---constitute a toolkit that could be applied to future discoveries, extending the reach of direct nuclear excitation with photons.

\subsection{Precision \texorpdfstring{$\beta$}{beta} decay and CKM unitarity}
\label{sec:beta_decay}

While the $^{229}$Th nuclear clock probes the nucleus through direct photon excitation, $\beta$ decay probes it through the weak interaction: a neutron converts to a proton (or vice versa), emitting an electron and a neutrino.  In the EFT framework of Sec.~\ref{sec:summary} (Eq.~(\ref{eq:EFT_chain})), this process is governed by the same Fermi constant $\GF$ and quark-mixing matrix elements that enter the weak multipole expansion.  Precision measurements of nuclear $\beta$-decay rates provide the most accurate determination of the CKM matrix element $|V_{ud}|$, which anchors the most demanding test of first-row CKM unitarity: $|V_{ud}|^2 + |V_{us}|^2 + |V_{ub}|^2 = 1$.  Any deviation from unity would signal new physics beyond the Standard Model.

\subsubsection{Superallowed \texorpdfstring{$0^+ \to 0^+$}{0to0} decays}

Superallowed Fermi $\beta$ transitions between $0^+$ isobaric analog states are pure vector transitions, making them the cleanest avenue for extracting $|V_{ud}|$~\cite{hardy2020superallowed,towner2010testing}.  The measured $ft$ value of each transition, corrected for nucleus dependent radiative ($\delta'_R$), nuclear structure ($\delta_{NS}$) and isospin-symmetry-breaking ($\delta_C$) effects, yields a ``corrected'' $\mathcal{F}t$ value:
\begin{equation}
\mathcal{F}t = ft(1+\delta_R')(1+\delta_\text{NS}-\delta_C) = \frac{K}{2G_V^2(1+\Delta_R^V)}\,,
\label{eq:Ft}
\end{equation}
from which $G_V = V_{ud}G_F$, where $G_F$ is the Fermi constant, and hence $|V_{ud}|$ are extracted.  In the above, the factor $K/(\hbar c)^6 = 2\pi^3\hbar \ln(2) / (m_e c^2)^5$~\cite{Workman2022} and $\Delta_R^V$~\cite{seng2018reduced} is a universal radiative correction.  The current world average over 15 transitions with $\sim 0.1\%$ precision gives $|V_{ud}| = 0.97373(31)$~\cite{hardy2020superallowed}, which, combined with current $|V_{us}|$ determinations, shows a $\sim 1.8\sigma$ tension with unitarity \cite{navas2024review}.

The dominant theoretical uncertainties come from two sources:
\paragraph{\textbf{Nuclear corrections ($\delta_C$, $\delta_\text{NS}$):}} Two terms are required from nuclear theory in order to evaluate the ``corrected'' $\mathcal{F}t$ value. The isospin-symmetry-breaking correction $\delta_C$ corrects for the small effects of isospin breaking in the nuclear forces.  Different approaches---the shell model with Saxon-Woods radial wave functions~\cite{towner2010testing}, Hartree-Fock with Skyrme functionals---yield values that differ at the level of the current experimental precision, making $\delta_C$ a limiting factor. Evaluation of this factor from an \textit{ab initio} perspective remains an open challenge~\cite{str21beta}.

The correction $\delta_{NS}$ comes from one-loop radiative corrections to the nuclear matrix elements. This correction has mostly been computed using the nuclear shell-model~\cite{barker1992determination,towner1992nuclear, towner1994quenching}. However, these evaluations relied on a separation of scale argument that was found to be flawed using a dispersion relation~\cite{gorchtein2019gammaw,seng2019dispersive}. This leads to a significant increase in the uncertainty of this correction, making it the biggest source of uncertainty at the moment in the extraction of $|V_{ud}|$~\cite{gorchtein2024superallowed}. Efforts have, however, been made from the \textit{ab initio} community, where this correction has been evaluated using the no-core shell-model for the transition from $^{10}$C to $^{10}$B~\cite{gennari2025abinito} as well as derivation within an EFT framework~\cite{cirigliano2024radiative}. Evaluations in heavier nuclei will be required to really reduce the uncertainty of this term in the future.
\paragraph{\textbf{Electroweak radiative corrections ($\Delta_R^V$):} } Recent re-evaluations using dispersion relations~\cite{seng2018reduced} and lattice QCD~\cite{ma2024lattice} have shifted $|V_{ud}|$ downward and increased the tension with unitarity.

Extending super-allowed measurements to heavier and more exotic nuclei at FRIB, TRIUMF, and ISOLDE tests the consistency of $\delta_C$ calculations across a wider range of nuclear structure, providing essential constraints.

\subsubsection{Mirror and neutron \texorpdfstring{$\beta$}{beta} decays}
Superallowed mixed (Fermi + Gamow-Teller) mirror transitions ($T = 1/2$
nuclei) offer a complementary path to $|V_{ud}|$ that requires measurement
of both the $ft$ value and the Fermi/Gamow-Teller mixing ratio $\rho$, the
latter accessible through $\beta$-asymmetry or $\beta$-neutrino correlation
measurements~\cite{severijns2006tests,gonzalezalonso2019beta}.  Ion traps
such as the St.~Benedict facility at Notre Dame~\cite{brodeur2023stbenedict}
and the TAMUTRAP at Texas A\&M are being developed to measure $\beta$-$\nu$
correlations in mirror decays with $\sim 0.5\%$ precision, sufficient to
impact $|V_{ud}|$.  The interpretation of mirror transitions rests on the
same isospin symmetry arguments as the $0^+ \to 0^+$ decays, and precision
measurements of ground state charge radii along the relevant isotopic
chains provide direct experimental access to the isovector densities that
control the corrections for isospin symmetry breaking: the charge radii of
the mirror $\beta$ emitters $^{36,37}$K were measured with this motivation
at the BECOLA facility~\cite{rossi2015charge}, and differences between the
charge radii of mirror pairs such as $^{54}$Ni--$^{54}$Fe probe the
isovector part of the nuclear density
directly~\cite{pineda2021charge,brown2017mirror,yang2018mirror}.
Free neutron $\beta$ decay avoids nuclear structure corrections entirely
but requires precise knowledge of the neutron lifetime $\tau_n$ and the
axial coupling $g_A$.  The persistent $\sim 4\sigma$ discrepancy between
``bottle'' and ``beam'' measurements of $\tau_n$ remains unresolved and
actively investigated~\cite{ucn2021neutron}.

\subsubsection{Connection to nuclear theory}
The nuclear structure corrections $\delta_C$ and $\delta_\text{NS}$
entering Eq.~(\ref{eq:Ft}) can be computed using the same nuclear many-body
methods discussed in
Sec.~\ref{sec:nuclear_theory}, and the electromagnetic observables of
Sec.~\ref{sec:EM_properties} enter this program at two levels: as inputs
that fix the nuclear densities on which the corrections depend, and as
benchmarks that validate the Hamiltonians,  currents, and many-body methods used to compute
them.  The role of the nuclear charge radius is the most direct.  In the
shell model evaluations that underpin the current $\mathcal{F}t$
survey~\cite{hardy2020superallowed,towner2010testing}, the correction
$\delta_C$ is dominated by the mismatch between the proton radial
wavefunctions of parent and daughter, and the Woods-Saxon potentials from
which these wavefunctions are obtained are constrained by measured charge
radii and separation energies.  A measured radius is therefore not merely
a consistency check but an input whose absence must be compensated by
extrapolation.  

Recent work has sharpened this connection into a
quantitative formalism relating $\delta_C$ to combinations of electroweak
nuclear radii that are sensitive to isospin symmetry
breaking and experimentally
accessible~\cite{seng2023electroweak,gorchtein2024superallowed}, enabling
data driven reevaluations of the superallowed $ft$
values~\cite{seng2024datadriven}.  The impact of such measurements was
demonstrated by collinear laser spectroscopy of the isomeric superallowed
emitter $^{26m}$Al~\cite{plattner2023radius,heylen2021al}. The first
experimental determination of the $^{26m}$Al charge radius differed by 4.5 standard
deviations from the extrapolated value previously used to evaluate
$\delta_C$, shifting the corrected $\mathcal{F}t$ value of the most
precisely known superallowed transition by one standard deviation.  Since
several of the heavier superallowed emitters ($^{62}$Ga, $^{74}$Rb) are
short lived, deformed, and among the cases with the largest and most
model dependent $\delta_C$~\cite{hardy2020superallowed}, extending
precision charge radius and quadrupole moment measurements to these
nuclei with the techniques of Sec.~\ref{sec:EM_properties} is a concrete
experimental path toward reducing the dominant nuclear uncertainty in
$|V_{ud}|$.

Other ground state electromagnetic properties test the spin dependent
part of the problem.  The nuclear structure correction
$\delta_\text{NS}$, which arises from the $\gamma W$ box evaluated within
the nucleus, involves matrix elements of the same spin and isospin
operators that govern magnetic dipole moments and Gamow-Teller
transitions~\cite{towner1992nuclear,towner1994quenching,gorchtein2019gammaw,seng2019dispersive}.
The long standing quenching of these operators has been traced from first
principles to the combined effect of correlations and two-body
currents~\cite{gysbers2019discrepancy}. Two-body currents have also
been shown to be essential for reproducing measured magnetic dipole
moments across the nuclear chart~\cite{miyagi2024impact}.  Precision
measurements of nuclear moments therefore benchmark the development of 
electroweak operators.  

\section{Conclusion and future perspectives}
\label{sec:conclusion}
The past two decades have witnessed a profound transformation in the relationship between nuclear, particle, and atomic and molecular physics. 
Precision measurements in atoms and molecules have become increasingly important
tools for determining the electromagnetic properties of exotic nuclei. In recent
years, they have also emerged as precision probes of electroweak nuclear
structure and as sensitive laboratories for searches for physics beyond the
Standard Model.

In parallel, the field of nuclear theory has undergone its own revolution, with chiral effective field theory and \textit{ab initio} many-body methods enabling increasingly quantitative connections between QCD- and electroweak-scale physics with experimentally accessible observables across the nuclear chart. 
From the convergence of these developments, a unified paradigm has emerged in which nuclei, atoms, and molecules interact collectively as precision probes of fundamental interactions and symmetries.
Complementing \textit{ab initio}  approaches, nuclear DFT has become an
important tool for describing medium-mass and heavy nuclei, where fully
\textit{ab initio} calculations remain challenging. By providing
self-consistent proton and neutron densities, deformations, charge radii, and
electromagnetic moments across long isotopic chains, DFT offers a bridge
between precision measurements and global nuclear-structure systematics, while
also enabling uncertainty quantification and extrapolations toward nuclei far
from stability.

The interaction between nuclei and electrons admits a common multipole
expansion, discussed in detail in Sec.~\ref{sec:ch2}, regardless of whether the
underlying interaction is electromagnetic, weak, or CP violating. Within this
framework, observables as diverse as isotope shifts, hyperfine structure,
parity-violating amplitudes, Schiff moments, weak charges, and nuclear-clock
transitions arise from the same underlying operator structures. This
factorization enables atomic and molecular systems to provide simultaneous
sensitivity to nuclear structure, hadronic weak interactions, and possible new
particles or forces. It also underscores the indispensable role of nuclear
theory in interpreting these observables quantitatively and, should deviations
be observed, in connecting them to possible new physics.

Experimentally, the field has advanced through the simultaneous maturation of
precision quantum control and radioactive-beam science. Laser cooling and
trapping, ion manipulation, high-resolution spectroscopy, and optical frequency
metrology now enable measurements sensitive to energy shifts at the sub-Hz
level, while developments at radioactive-beam facilities worldwide have made it
possible to perform precision studies of exotic isotopes near the limits of
nuclear existence. In these systems, with extreme proton-to-neutron ratios,
distinctive nuclear phenomena emerge, including deformation, shell evolution,
neutron skins, halo structure, and octupole collectivity.
Molecules have emerged as particularly powerful systems to measure symmetry-violating nuclear properties because their closely spaced opposite-parity states can enhance parity- and CP violating effects by more than five orders of magnitude relative to atoms. 
Heavy polar molecules have already produced dramatic improvements in sensitivity to the electron electric dipole moment, and radioactive molecules containing octupole-deformed nuclei are expected to provide similarly transformative sensitivity to CP-violating nuclear properties. 
Additionally, the realization of nuclear-clock spectroscopy in $^{229}$Th has established nuclear transitions themselves as precision sensors of electroweak physics, time variation of fundamental constants, and potentially ultralight dark matter.

The prospects are especially exciting as the next decade is likely to see dramatically rapid progress driven by the growing and essential interplay between experiment and theory. 
New radioactive-beam facilities with higher-intensity isotope production will make increasingly exotic nuclei accessible for such studies. 
Precision isotope-shift measurements across entire isotopic chains will provide detailed information on neutron distributions, nuclear polarizabilities, and higher-order radial moments. 
Experiments combining trapped ions, ultracold molecules, and quantum sensing techniques will extend coherence times and sensitivity far beyond current limits. 
On the theory side, improved chiral interactions, higher-order electroweak operators, exascale computing, and machine-learning-assisted many-body methods are expected to substantially expand the reach and precision of \textit{ab initio} calculations.

More broadly, the field is entering an era in which low-energy precision experiments are becoming fully complementary to high-energy collider searches, long the driving force for discovering new particles and forces. 
Atomic and molecular observables probe virtual processes associated with mass scales extending far beyond those directly accessible at accelerators, while also providing unique sensitivity to symmetry structures and flavor patterns that are difficult to disentangle at high energies. 
At the same time, these measurements illuminate emergent phenomena of the nuclear many-body problem itself: neutron skins, collective deformation, parity violation, and weak-charge distributions become experimentally accessible through the quantum structure of atoms and molecules.

\section{Acknowledgments}
\label{sec:Ack}

Generative AI tools, including ChatGPT and Claude, were used during the preparation of this review to assist with literature searches, organization, and language polishing.

S.M.U was supported by a William H. Miller Fellowship.

R.F.G.R. and A.B. acknowledge support from the US Department of Energy, Office of Science, Office of Nuclear Physics under grants DE-SC0021176 and the Center for Ultracold Atoms (an NSF Physics Frontier Center). A.B. acknowledges the support of the Natural Sciences and Engineering Research Council of Canada (NSERC) [PDF-587464-2024].

J.D.H. was supported by the Natural Sciences and Engineering Research Council of Canada (NSERC) under grant SAPIN-2024-0003 and the Arthur B.~McDonald Canadian Astroparticle Physics Research Institute.

G.P. is supported by the Israel Science Foundation (ISF), Minerva, the NSF-BSF, and the European Research Council (ERC, DM-Dawn, Grant Agreement No. 101199868).


\bibliography{EDM_RMP_v1}

\appendix
 
\section{Nuclear form factors in atomic and molecular spectroscopy}
\label{sec:appendix_form_factors}
 
The factorized structure of the nucleus--electron interaction developed in the preceding section---nuclear moment $\times$ electronic field operator---has a precise and illuminating reformulation in momentum space.  In this language, every atomic and molecular observable discussed in this review can be expressed as a nuclear \emph{form factor} convoluted with an electronic \emph{probe function}, in exact parallel with the theoretical framework of electron-nucleus scattering experiments.  The atom or molecule is, quite literally, a low-energy scattering experiment in which the bound electrons serve as the probe, and the measured energy shift or transition amplitude replaces the scattering cross section.

\subsection{The finite-size energy shift as a charge form-factor integral}
\label{sec:FS_formfactor}
 
Consider the electromagnetic interaction between the nuclear charge distribution $\rho_c(\mathbf{r})$ and the electrons.  In momentum space, the Coulomb potential of the finite-size nucleus is:
\begin{equation}
\tilde{V}_\text{nuc}(q) = -\frac{Ze^2}{\epsilon_0}\,\frac{1}{q^2}\, F_\text{ch}(q^2)\,,
\label{eq:Vnuc_q}
\end{equation}
where $F_\text{ch}(q^2)$ is the nuclear \emph{charge form factor}---the Fourier transform of the charge density, normalized to $F_\text{ch}(0) = 1$:
\begin{equation}
F_\text{ch}(q^2) = \frac{1}{Ze}\int \dd^3 r\, \rho_c(\mathbf{r})\, e^{\ii\mathbf{q}\cdot\mathbf{r}}\,.
\label{eq:Fch}
\end{equation}
The point-nucleus potential corresponds to $F_\text{ch} = 1$.  The finite-nuclear-size correction to the energy of an electronic state $|\psi_e\rangle$ is therefore:
\begin{equation}
\delta E_\text{FS} = -\frac{Ze^2}{\epsilon_0}\int\frac{\dd^3 q}{(2\pi)^3}\, \frac{1}{q^2}\, \Big[F_\text{ch}(q^2) - 1\Big]\, \rho_e(q)
\label{eq:FS_FF}
\end{equation}
where
\begin{equation}
\rho_e(q) = \int \dd^3 r\, |\psi_e(\mathbf{r})|^2\, e^{\ii\mathbf{q}\cdot\mathbf{r}}
\label{eq:rho_e_q}
\end{equation}
is the \emph{electronic form factor}---the Fourier transform of the electron probability density.  Equation~(\ref{eq:FS_FF}) has a transparent physical interpretation: the energy shift is a sum over all momentum transfers $\mathbf{q}$ of the nuclear structure factor $[F_\text{ch}(q^2) - 1]$, weighted by the Coulomb propagator $1/q^2$ and the electronic form factor $\rho_e(q)$.  The electronic form factor acts as a \emph{window function} in momentum space, selecting the range of $q$ that the bound electrons are sensitive to.
 
For $s$-wave electrons, $\rho_e(q)$ is peaked at $q = 0$ and falls off at $q \sim 1/a_0 \sim \alpha m_e c/\hbar \sim 4$~keV/$c$.  The nuclear form factor deviates from unity only at $q \gtrsim 1/R \sim 100$~MeV/$c$.  The overlap is concentrated in the region $q \lesssim 1/R$ where the form factor can be expanded:
\begin{equation}
F_\text{ch}(q^2) - 1 \approx -\frac{q^2}{6}\langle r^2\rangle + \frac{q^4}{120}\langle r^4\rangle - \ldots\,,
\label{eq:FF_expand}
\end{equation}
and the leading term reproduces the familiar position-space result:
\begin{equation}
\delta E_\text{FS} = \frac{Ze^2}{6\epsilon_0}\,\langle r^2\rangle \cdot |\psi_e(0)|^2 + \mathcal{O}(\langle r^4\rangle)\,.
\label{eq:FS_leading_v2}
\end{equation}
The charge radius $\langle r^2\rangle$ is thus the slope of the charge form factor at $q^2 = 0$: $\langle r^2\rangle = -6\, \dd F_\text{ch}/\dd q^2\big|_{q^2=0}$.  Higher moments probe the curvature and higher derivatives of $F_\text{ch}(q^2)$.
 
The isotope shift between isotopes $A$ and $A'$ involves the difference of form factors:
\begin{equation}
\delta\nu_\text{FS}^{A,A'} = -\frac{Ze^2}{\epsilon_0}\int\frac{\dd^3 q}{(2\pi)^3}\, \frac{1}{q^2}\, \Big[F_\text{ch}^{(A')}(q^2) - F_\text{ch}^{(A)}(q^2)\Big]\, \Delta\rho_e(q)\,,
\label{eq:IS_FF}
\end{equation}
where $\Delta\rho_e(q)$ is the difference of electron form factors between the two electronic states of the transition.  This is the most general, approximation-free expression for the field shift.

\subsection{The hyperfine constant as a magnetization form-factor integral}
\label{sec:HFS_formfactor}
 
The same momentum-space formulation applies to the magnetic hyperfine interaction.  The nuclear magnetization distribution $m(\mathbf{r})$ can be characterized by a magnetization form factor (the magnetic form factor):
\begin{equation}
G_M(q^2) = \frac{1}{\mu}\int \dd^3 r\, m(\mathbf{r})\, e^{\ii\mathbf{q}\cdot\mathbf{r}}\,,
\label{eq:GM}
\end{equation}
normalized such that $G_M(0) = 1$.  The magnetic hyperfine constant can then be written as:
\begin{equation}
A_\text{hfs} = A_\text{hfs}^{(\text{point})} \int\frac{\dd^3 q}{(2\pi)^3}\, G_M(q^2)\, \tilde{b}_e(q)
\label{eq:Ahfs_FF}
\end{equation}
where $A_\text{hfs}^{(\text{point})}$ is the point-dipole hyperfine constant and $\tilde{b}_e(q)$ is the Fourier transform of the electronic magnetic field density at the nucleus---the magnetic analog of the electronic probe function $\rho_e(q)$.
 
The Bohr--Weisskopf effect (Sec.~\ref{sec:term_mu}) is precisely the correction due to the deviation of $G_M(q^2)$ from unity at the momentum transfers sampled by the electrons:
\begin{equation}
\epsilon_\text{BW} = \frac{A_\text{hfs} - A_\text{hfs}^{(\text{point})}}{A_\text{hfs}^{(\text{point})}} = \int\frac{\dd^3 q}{(2\pi)^3}\, \Big[G_M(q^2) - 1\Big]\, \tilde{b}_e(q)\,.
\label{eq:BW_FF}
\end{equation}
This is the magnetic analog of Eq.~(\ref{eq:FS_FF}): the finite-size shift probes $F_\text{ch}(q^2) - 1$, the Bohr--Weisskopf correction probes $G_M(q^2) - 1$, and in both cases the bound electrons provide the probe function.

\subsection{The parity-violating amplitude as a weak form-factor integral}
\label{sec:PV_formfactor}
 
The correspondence extends to the weak interaction.  The nuclear weak form factor is the Fourier transform of the weak-charge density $\rho_W(\mathbf{r})$ introduced in Sec.~\ref{sec:weak_densities}:
\begin{equation}
F_W(q^2) = \frac{1}{\QW}\int \dd^3 r\, \rho_W(\mathbf{r})\, e^{\ii\mathbf{q}\cdot\mathbf{r}}\,,
\label{eq:FW}
\end{equation}
with $F_W(0) = 1$.  The parity-violating matrix element becomes:
\begin{equation}
M_\text{PV} = \frac{\GF}{2\sqrt{2}}\,\QW \int\frac{\dd^3 q}{(2\pi)^3}\, F_W(q^2)\, \tilde{\phi}_{W,e}(q)
\label{eq:MPV_FF}
\end{equation}
where $\tilde{\phi}_{W,e}(q)$ is the Fourier transform of the electronic pseudoscalar density $\psi_e^\dagger \gamma_5 \psi_e$ inside the nuclear volume.  This is the \emph{same} weak form factor measured in parity-violating electron scattering (PVES), where it enters through the asymmetry:
\begin{equation}
A_\text{PV}^\text{scatt.} = \frac{\sigma_R - \sigma_L}{\sigma_R + \sigma_L} \propto \frac{\GF q^2}{4\pi\alpha}\,\frac{F_W(q^2)}{F_\text{ch}(q^2)}\,.
\label{eq:APV_scatt}
\end{equation}
PVES tunes $q$ to map out $F_W(q^2)$ and extract the neutron density; atomic PV integrates $F_W(q^2)$ weighted by the bound-electron probe to determine $\QW$ and $\sin^2\thetaW$ at low energy.  Both probe the same nuclear physics---the neutron distribution---through the same form factor.

\subsection{Unified picture: one form factor per interaction, one probe function per measurement}
\label{sec:unified_FF}
 
The results above reveal a single organizing structure.  For each interaction, the observable is a nuclear form factor convoluted with an electronic probe function:
\begin{equation}
\mathcal{O} = \int\frac{\dd^3 q}{(2\pi)^3}\, \underbrace{F_\text{nuc}(q^2)}_{\substack{\text{nuclear}\\\text{form factor}}} \times \underbrace{W_e(q)}_{\substack{\text{electronic}\\\text{probe function}}}\,.
\label{eq:unified_FF}
\end{equation}
 
\begin{center}
\renewcommand{\arraystretch}{1.4}
\begin{tabular}{lccc}
\hline\hline
\textbf{Obs.} & \textbf{Nuclear FF} & \textbf{Probe $W_e(q)$} & \textbf{Scattering analog} \\
\hline
$\delta E_\text{FS}$ & $F_\text{ch}(q^2)$ & $\rho_e(q)/q^2$ & Elastic $e$-$A$ \\
$A_\text{hfs}$ & $G_M(q^2)$ & $\tilde{b}_e(q)$ & Magnetic $e$-$A$ \\
$M_\text{PV}$ & $F_W(q^2)$ & $\tilde{\phi}_{W,e}(q)$ & PV $e$-$A$ (PREX) \\
\hline\hline
\end{tabular}
\end{center}
 
The correspondence between spectroscopy and scattering is:
 
\begin{center}
\renewcommand{\arraystretch}{1.4}
\begin{tabular}{p{2cm}cc}
\hline\hline
& \textbf{Scattering} & \textbf{Spectroscopy} \\
\hline
Observable & $\dd\sigma/\dd\Omega$ or $A_\text{PV}$ & $\delta E$, $A_\text{hfs}$, or $M_\text{PV}$ \\
Nuclear structure & Form factor $F(q^2)$ & Same $F(q^2)$ \\
Probe & Leptonic tensor $L^{\mu\nu}$ & Electronic probe $W_e(q)$ \\
$q$-range & Tunable (beam energy) & Fixed ($\sim 1/R$, by $\psi_e$) \\
Strength & Full $q$ mapping & Extreme precision at low $q$ \\
\hline\hline
\end{tabular}
\end{center}
 
\subsubsection{Molecules as multi-probe experiments.}
In a molecule, the electronic probe function depends on the internuclear distance $R$: $W_e(q) \to W_e(q; R)$.  Different rovibrational states $|v, J\rangle$ sample different expectation values $\langle W_e(q; R)\rangle_{v,J}$, providing a family of independent probe functions within a single molecular species.  Measuring the same observable in different rovibrational levels is therefore analogous to measuring a scattering cross section at different beam energies: each measurement samples the nuclear form factor with a different weighting, and the combination constrains the $q$-dependence of $F(q^2)$---and thus the spatial distribution of the nuclear property---more tightly than any single measurement alone.
 
\subsubsection{The EFT matching chain.}
The form-factor language connects naturally to the effective field theory framework relating BSM physics to laboratory observables.  The nuclear form factors sit at the penultimate stage of the matching chain:
\begin{equation}
\begin{aligned}
& \underbrace{C_i(\Lambda)}_{\text{BSM Wilson coeff.}}
\;\xrightarrow{\;\text{EW}\;}\; \underbrace{d_q,\, \tilde{d}_q,\, C_S\ldots}_{\text{quark-level}}
\;\xrightarrow{\;\text{QCD}\;}\; \underbrace{d_n,\, d_p,\, \bar{g}_\pi^{(i)}}_{\text{hadronic}} \\
& \quad\xrightarrow{\;\text{nuclear}\;}\; \underbrace{F_\text{ch},\, G_M,\, F_W\ldots}_{\text{form factors}}
\;\xrightarrow{\;\text{atomic}\;}\; \underbrace{\delta E,\, A_\text{hfs},\, M_\text{PV},\, d_\text{mol}}_{\text{observables}}
\end{aligned}
\label{eq:EFT_chain_FFactor}
\end{equation}
At each arrow, the physics above the matching scale is absorbed into coefficients, and the physics below is computed independently---the same factorization principle that underlies collider cross-section predictions via parton distribution functions.  The nuclear form factors are the ``PDFs of the atom'': universal, process-independent quantities that encode the nuclear structure and can be constrained by any experiment---scattering or spectroscopy---that probes them.

\newpage
\section{Supplementary derivations}
\label{sec:appendix}

\subsection{Suppression of NSI parity violation in diatomic molecules}
\label{sec:appendix_NSI}

We show that the expectation value of the NSI PV Hamiltonian (Eq.~(\ref{eq:Q_W_formula})) vanishes in states of definite $\Omega$, explaining why NSD effects dominate in diatomic molecules (Sec.~\ref{sec:NSD_PV}).

The projection of the electronic angular momentum on the internuclear axis, $\Omega = \mathbf{J}_e \cdot \hat{\mathbf{n}}$, changes sign under both parity ($P$) and time-reversal ($T$), since $P$ reverses $\hat{\mathbf{n}}$ but not $\mathbf{J}_e$, while $T$ reverses $\mathbf{J}_e$ but not $\hat{\mathbf{n}}$.  The NSI PV Hamiltonian $H_\text{NSI}$ is $P$-odd and $T$-even.  Therefore:
\begin{align}
\langle\Omega|H_{\rm NSI}|\Omega\rangle
&=
\langle \Omega|P^\dagger P H_{\rm NSI}P^\dagger P|\Omega\rangle \nonumber\\
&=
-\langle-\Omega|H_{\rm NSI}|-\Omega\rangle \nonumber\\
&=
-\left[
\langle\Omega|T^\dagger H_{\rm NSI}T|\Omega\rangle
\right]^* \nonumber\\
&=
-\langle\Omega|H_{\rm NSI}|\Omega\rangle^* \nonumber\\
&=
-\langle\Omega|H_{\rm NSI}|\Omega\rangle ,
\label{eq:NSI_PV_suppression}
\end{align}
which implies $\langle \Omega | H_\text{NSI} | \Omega \rangle = 0$. 
Thus, the diagonal NSI PV interaction vanishes to first order within a definite-$\Omega$ electronic manifold, \cite{kozlov1995}, making the NSD effects (anapole moment and $C_2$ coupling, Sec.~\ref{sec:NSD_PV}) the dominant parity-violating effects in diatomic molecules.

\subsection{Optical rotation in atomic PV experiments}
\label{sec:appendix_optical_rot}

We derive the optical rotation effect used in the pioneering APV measurements in Bi, Pb, and Tl \cite{macpherson1991,warrington1993,meekhof1993,phipp1996,edwards1995,vetter1995} (Sec.~\ref{sec:PV}). Near an atomic resonance at frequency $\omega_0$, the index of refraction for right- ($+$) and left- ($-$) circularly polarized light is:
\begin{equation}
n_\pm(\omega) = 1 - 2\pi N\,\overline{|A_\pm|^2}\,D(\omega_0,\omega)\,,
\label{eq:app_npm}
\end{equation}
where $N$ is the atomic density, $D(\omega_0,\omega) \equiv \langle (\omega - \omega_0 - v\omega + \ii\Gamma/2)^{-1}\rangle_v$ is the Doppler-broadened line shape (the average $\langle\cdots\rangle_v$ is over the Maxwell--Boltzmann velocity distribution), and $\overline{|A_\pm|^2}$ is the polarization-averaged transition amplitude squared.

For two atomic levels of the same parity connected by an allowed $M1$ transition, the PV interaction admixes a state of opposite parity into the excited level:
\begin{equation}
|e'\rangle = |e\rangle + \ii\eta\,|e_2\rangle\,,\qquad
\ii\eta \equiv \frac{\langle e|H_\text{PV}|e_2\rangle}{E_e - E_{e_2}}\,.
\end{equation}
This generates a PV-induced $E1$ amplitude $A_\text{PV} = \ii\eta\,A_{E1}$.  Accounting for the $\pi/2$ phase difference between the electric and magnetic fields of the light wave, the total transition amplitude squared is~\cite{khriplovich1991}:
\begin{equation}
|A_\pm|^2 \approx |A_{M1}|^2 \pm 2|A_{M1}|^2\,\text{Im}\!\left(\frac{A_\text{PV}}{A_{M1}}\right).
\label{eq:app_asymmetry}
\end{equation}
For linearly polarized light propagating along the $z$-axis, the electric field is an equal superposition of right and left circular polarizations.  After traversing a vapor column of length $l$, the two components accumulate different phases due to $n_+ \neq n_-$, rotating the polarization plane by~\cite{khriplovich1991,budker2004atomic}:
\begin{equation}
\phi = \frac{\omega l}{2c}\,\text{Re}(n_+ - n_-)\,.
\label{eq:app_rotation}
\end{equation}
For transitions between hyperfine components $F_g$ and $F_e$, the polarization average gives $\overline{|A_{M1}|^2} = K(F_g,F_e)\,\langle A_{M1}\rangle^2$, where $\langle A_{M1}\rangle$ is the reduced matrix element and $K(F_g,F_e)$ is calculable from angular momentum algebra.  Combining Eqs.~(\ref{eq:app_npm})--(\ref{eq:app_rotation}):
\begin{equation}
\phi = -4\pi N\,\frac{\omega l}{c}\,\text{Re}\!\left[D(\omega_0,\omega)\right]\,K(F_g,F_e)\,\langle A_{M1}\rangle^2\,\text{Im}\!\left(\frac{A_\text{PV}}{A_{M1}}\right).
\label{eq:app_phi}
\end{equation}
The measured rotation angle $\phi$, combined with independent knowledge of $A_{M1}$ and the line shape $D$, yields the ratio $R \equiv \text{Im}(A_\text{PV}/A_{M1})$, from which $\QW$ is extracted using atomic theory (Sec.~\ref{sec:PV}).

\subsection{Schiff shielding theorem}
\label{sec:appendix_schiff_theorem}

We prove that the expectation value of the electric field vanishes for a nonrelativistic point particle in a purely electrostatic potential---the Schiff shielding theorem~\cite{schiff1963measurability}---complementing the discussion in Sec.~\ref{sec:CPV_nuclear}.

For a particle of charge $q$ and mass $m$ in a potential $\Phi$, the Hamiltonian is $H_0 = \mathbf{p}^2/(2m) + q\Phi(\mathbf{r})$.  The EDM interaction is $H_\text{EDM} = -\mathbf{d}\mathbf{E}_\text{tot} = \mathbf{d}\boldsymbol{\nabla}\Phi(\mathbf{r})$, where $\Phi(\mathbf{r})$ is the complete electrostatic potential acting on the particle. For an eigenstate $|n\rangle$ of $H_0$:
\begin{equation}
\begin{split}
0 &= \langle n|[\mathbf{p},\,H_0]|n\rangle = \langle n|[\mathbf{p},\,q\Phi]|n\rangle = -\ii\hbar\,q\,\langle n|\boldsymbol{\nabla}\Phi|n\rangle \\
&= \ii\hbar\,q\,\langle n|\mathbf{E}_\text{tot}|n\rangle\,,
\end{split}
\end{equation}
and therefore $\langle n|\mathbf{E}_\text{tot}|n\rangle = 0$, which implies $\langle n|H_\text{EDM}|n\rangle = 0$.  The average electric field on the particle vanishes---physically, because a nonrelativistic charged particle in equilibrium experiences zero net force.

This theorem is evaded in atoms and molecules by (i)~the relativistic motion of electrons near heavy nuclei, which allows the electron EDM to produce a measurable signal (Sec.~\ref{sec:CPV_nuclear}), and (ii)~the finite size of the nucleus, which gives rise to the Schiff moment $\mathbf{S}$ (Eq.~(\ref{eq:Schiff}))---the residual CP-violating effect that survives shielding.

\subsection{The nuclear Schiff moment: derivation and interaction potential}
\label{sec:appendix_schiff_moment}

We derive the Schiff moment operator and its interaction with the electron cloud, starting from the electrostatic Hamiltonian for a nucleus interacting with $Z_e$ electrons in an external field $\mathbf{E}_0$~\cite{khriplovich1991,chupp2019electric}:
\begin{equation}
H = \sum_i\left[K_i - eV_0(\mathbf{R}_i) - e\mathbf{R}_i\cdot\mathbf{E}_0\right] + \sum_{i>k}\frac{e^2}{|\mathbf{R}_i - \mathbf{R}_k|} - \mathbf{d}\cdot\mathbf{E}_0\,,
\end{equation}
where $K_i$ is the kinetic energy of the $i$-th electron, $\mathbf{R}_i$ its position relative to the nucleus, $\mathbf{d}$ is the nuclear EDM, and the nuclear electrostatic potential is $V_0(\mathbf{R}_i) = e\int \rho(\mathbf{r})\,\dd^3 r/|\mathbf{R}_i - \mathbf{r}|$.

Define the auxiliary operator:
\begin{equation}
V \equiv \mathbf{d}\cdot\mathbf{E}_0 - \frac{1}{eZ}\sum_i \mathbf{d}\cdot\boldsymbol{\nabla}_i V_0(\mathbf{R}_i)\,.
\end{equation}
Using $\frac{\ii}{m}[\sum_i \mathbf{p}_i,\,H] = -e\sum_i \boldsymbol{\nabla}_i V_0(\mathbf{R}_i) + Ze\,\mathbf{E}_0$, one shows that $\langle n|V|n\rangle = 0$ for any eigenstate $|n\rangle$ of $H$.  Adding $V$ to $H$ therefore does not change the spectrum to first order in $\mathbf{d}$, and the effective nuclear potential becomes:
\begin{equation}
V(\mathbf{R}_i) = V_0(\mathbf{R}_i) + \frac{1}{eZ}\,\mathbf{d}\cdot\boldsymbol{\nabla}_i V_0(\mathbf{R}_i)\,.
\label{eq:app_Veff}
\end{equation}
The key result is that the direct coupling $\mathbf{d}\cdot\mathbf{E}_0$ has been absorbed, and the $l = 1$ term in the multipole expansion of $V(\mathbf{R})$ vanishes identically---this is the Schiff shielding theorem.

The first surviving CP-violating contribution appears at $l = 3$ in the Cartesian expansion.  Decomposing this into irreducible spherical tensors yields a rank-1 piece (the Schiff moment) and a rank-3 piece (the electric octupole moment).  The Schiff moment is~\cite{khriplovich1991,chupp2019electric}:
\begin{equation}
    \mathbf{S}^\text{ch} = \frac{|e|}{10}\left(\int\rho_\text{ch}(\mathbf{r})\,r^2\,\mathbf{r}\,\dd^3r
    \;-\;\frac{5}{3}\,\langle r^2\rangle_\text{ch}
    \int\rho_\text{ch}(\mathbf{r})\,\mathbf{r}\,\dd^3r\right),
\end{equation}
reproducing Eq.~(\ref{eq:Schiff}).  It points along the nuclear spin: $\mathbf{S} = S\,\mathbf{I}/I$.  In the point-nucleus limit, the interaction with the electron cloud is:
\begin{equation}
V_\text{Schiff}(\mathbf{R}) = 4\pi\,\mathbf{S}\cdot\boldsymbol{\nabla}\delta(\mathbf{R})\,.
\end{equation}
For a finite-size nucleus with radius $R_N$, the regularized form is~\cite{flambaum1986time}:
\begin{equation}
V_\text{Schiff}(\mathbf{R}) \approx -\frac{15\,\mathbf{S}\cdot\mathbf{R}}{R_N^5}\,\Theta(R_N - R)\,,
\label{eq:app_Vschiff}
\end{equation}
where $\Theta$ is the Heaviside step function (smoothed by the nuclear surface).  This describes a uniform electric field $\mathbf{E} \propto \mathbf{S}$ inside the nuclear volume, directed along the nuclear spin---the physical origin of the Schiff moment's CP-violating effect in diamagnetic atoms and molecules (Sec.~\ref{sec:CPV_nuclear}).

\end{document}